\documentclass[12pt]{article}

\usepackage[dvips]{graphicx}

\usepackage{amsmath}
\usepackage{amsfonts}
\usepackage{placeins}
\usepackage{afterpage}

\usepackage{flafter}

\usepackage[nosort]{cite}

\usepackage[a4paper]{typearea} 
\areaset[0cm]{17.0cm}{24cm}

 \newcommand{\PSImagx}[2]{\includegraphics[width=#2]{psplots/#1.ps}}

\newcommand{\BILD}[4]{\begin{figure}[#1]%

     #2

     \centerline{\parbox{15cm}{\caption[.]{#3} \label{#4}}}
     \end{figure} }

\newcommand{\Int}{\int\limits}
\newcommand{\IInt}{\iint\limits}

\newcommand{\ud}{\text{d}}
\newcommand{\ui}{\text{i}}
\newcommand{\ue}{\text{e}}

\newcommand{\Vol}{\operatorname{vol}}
\newcommand{\Ai}{\operatorname{Ai}}

\newcommand{\R}{\mathbb{R}}
\newcommand{\Z}{\mathbb{Z}}
\newcommand{\N}{\mathbb{N}}

\begin{document}

%%%%%%%%%%%%%%%%%%%%%%%%%%%%%%%%%%% title page %%%%%%%%%%%%%%%%%%%%%%%%%%%%%

\vspace*{-0.5cm}

ULM--TP/01--04

HPL--BRIMS--2001--11

April 2001
 
\vspace*{1cm}

\newcommand{\titel}{Autocorrelation function of eigenstates \\[1ex]
in chaotic and mixed systems}

\normalsize

\vspace{0.5cm}

\begin{center}  \LARGE  \bf

      \titel
\end{center}

   \vspace{2.0cm}

\begin{center}

          {\large Arnd B\"acker%
\footnote[1]{E-mail address: {\tt a.backer@bristol.ac.uk}}$^{,2}$
          }
          {\large and Roman Schubert%
\footnote[3]{E-mail address: {\tt roman.schubert@physik.uni-ulm.de}}
           }

    \vspace{1.5cm}

    $^{1)}$ 
 School of Mathematics, University of Bristol\\
 University Walk, Bristol BS8 1TW, UK\\

    \vspace{2ex}

 $^{2)}$ BRIMS, Hewlett-Packard Laboratories\\
 Filton Road, Bristol BS12 6QZ, UK\\

    \vspace{2ex}

    $^{3)}$ Abteilung Theoretische Physik, Universit\"at Ulm,\\ 
    Albert-Einstein-Allee 11, D--89069 Ulm, Germany\\
    \vspace{4ex}

\end{center}

  \vspace{1cm}

%%%%%%%%%%%%%%%%%%%%%%%%%%%%%%%%%%%%%%%%%%%%%%%%%%%%%%%%%%%%%%%%%%%%%%%%%%%%%

\newcommand{\VIP}{\rule{1cm}{1.5ex} }

\newcommand{\BFx}{{\bf x}}
\newcommand{\BFq}{{\bf q}}
\newcommand{\BFe}{{\bf e}}

\newcommand{\noBFx}{x}
\newcommand{\noBFq}{q}

\newcommand{\erf}{\operatorname{erf}}
\newcommand{\BFp}{{\bf p}}
\newcommand{\BFk}{{\bf k}}
\newcommand{\limacon}{lima\c{c}on }
\newcommand{\omm}{s}
\newcommand{\BFn}{\boldsymbol n}
\newcommand{\E}{\operatorname{\mathbb{E}}}

\newcommand{\VARPHI}{\varphi}

\newcommand{\BFa}{{\bf a}}
\newcommand{\diam}{\operatorname{diam}}

\newcommand{\psiRWM}{\psi_{\text{RWM}}}
\newcommand{\psiRWMD}{\psi_{\text{RWM},D}}
\newcommand{\PRWMDpsi}{P_{\text{RWM},D}(\psi)}

\newcommand{\EPS}{\varepsilon}

\newcommand{\rC}{C}
\newcommand{\co}{\xi}

\newcommand{\ort}{\BFx}
\newcommand{\dort}{{\bf \delta x}}

\newcommand{\VAR}{\sigma^2}
\newcommand{\VARcl}{\sigma^2_{\text{cl}}}
\newcommand{\barVAR}{\overline{\sigma^2}}

\newcommand{\cP}{{\cal P}}
\newcommand{\cD}{{\cal D}}
\newcommand{\Bft}{{\bf t}}

%%%%%%%%%%%%%%%%%%%%%%%%%%%%%%%%%%%%%%%%%%%%%%%%%%%%%%%%%%%%%%%%%%%%%%%%%%%%%%%

\vspace*{1cm}

\leftline{\bf Abstract:}

We study the autocorrelation function of different types 
of eigenfunctions in quantum mechanical systems with either chaotic 
or mixed classical limits. We obtain an expansion of the  
autocorrelation function in  terms 
of the correlation length.
For localized states, like bouncing ball modes or states
living on tori, a simple model using only classical 
input gives good agreement with the exact result.
In particular, a prediction for irregular 
eigenfunctions in mixed systems is derived and tested. 
For chaotic systems, the expansion of the autocorrelation function 
can be used to test quantum ergodicity on different length scales.  

\newpage

%%%%%%%%%%%%%%%%%%%%%%%%%%%%%%%%%%%%%%%%%%%%%%%%%%%%%%%%%%%%%%%%%%%%%%%%%%%%%%%
\section{Introduction}
%%%%%%%%%%%%%%%%%%%%%%%%%%%%%%%%%%%%%%%%%%%%%%%%%%%%%%%%%%%%%%%%%%%%%%%%%%%%%%%

The behavior of a quantum mechanical system in the semiclassical limit 
strongly depends on the ergodic properties of the 
corresponding classical system.
In particular, the \mbox{eigenfunctions} semiclassically reflect the 
phase space structure of the classical system 
and therefore they depend 
strongly on whether the classical system is chaotic or regular. 
In this work we are interested in the fluctuations 
of the wavefunctions, and in the correlations between 
the fluctuations in different regions which are induced by the 
classical phase space structures.  
In particular,
we will consider the case of quantum billiards in a
domain $\Omega\subset \R^2$,
which are described by the time independent Schr\"odinger equation
\begin{equation} \label{eq:Helmholtz-equation}
  (\Delta +E)\psi (\BFq)=0  \qquad \quad \text{ for $\BFq\in\Omega \backslash \partial\Omega$}\;\;,
\end{equation}
with Dirichlet boundary conditions, 
$\psi (\BFq)=0$ for $\BFq\in\partial\Omega$.
For compact $\Omega$ one obtains a discrete spectrum $\{E_n\}$
of eigenvalues, 
$0<E_1\le E_2 \le \ldots$,
with associated eigenfunctions $\psi_n\in L^2(\Omega)$,
which we assume to be normalized, i.e.\
$||\psi_n||:= \int_\Omega |\psi_n(\BFq)|^2 \; \ud q = 1$.
The corresponding classical billiard is given by the free motion
of a point particle inside $\Omega$ with elastic reflections
at the boundary $\partial\Omega$.

The amplitude distribution of an eigenfunction of a quantum mechanical 
system whose classical limit is chaotic is conjectured 
to become Gaussian in the semiclassical limit \cite{Ber77b}, 
and numerical studies support 
this conjecture, see e.g.\ \cite{McDKau88,AurSte93,LiRob94}. 
A more sensitive quantity is
the {\it local} autocorrelation function \cite{Ber77b}
which measures correlations between different points of 
an eigenfunction $\psi$,
\begin{equation}\label{eq:def-loc-auto}
  C^{\text{loc}}(\ort,\dort):=\psi^*(\ort-\dort/2)\psi(\ort+\dort/2) \;\; .
\end{equation}
The crucial fact for 
the theoretical analysis of $C^{\text{loc}}(\ort,\dort)$, 
observed by Berry \cite{Ber77b}, 
is that the autocorrelation function can be expressed as the 
Fourier transformation of the Wigner function 
(see eq.~\eqref{eq:def-Wigner} below) of $\psi$, 
\begin{equation}\label{eq:corr-Wigner}
  C^{\text{loc}}(\ort,\dort)
   =\Int  W(\BFp,\ort) \ue^{-\ui\BFp \dort}\;  \ud p\;\; .
\end{equation}
Hence  information on the behavior of the Wigner function can be used 
to predict the behavior of the autocorrelation function, and since 
semiclassical limits of Wigner functions are concentrated on 
invariant sets in phase space, see e.g.~\cite{Rob98}, 
it follows that in the semiclassical limit
autocorrelation functions are 
determined by the classical phase space structure.
For example, if the classical 
system is ergodic, the quantum ergodicity theorem 
\cite{Shn74,Zel87,CdV85,HelMarRob87,GerLei93, ZelZwo96} 
(roughly speaking) states that almost all quantum expectation
values tend to the corresponding classical limit.
One can show \cite{BaeSchSti98} that for ergodic systems
this is equivalent 
to the semiclassical eigenfunction hypothesis \cite{Vor76,Ber77b,Vor77,Ber83},
when restricted to a subsequence of density one.
Using this result in \eqref{eq:corr-Wigner}
one gets Berry's result \cite{Ber77b} that for chaotic billiards 
in two dimensions
\begin{equation}\label{eq:corr-universal}
  C^{\text{loc}}(\ort,\dort)\sim \frac{1}{\Vol (\Omega)}\, 
  J_0(\sqrt{E} |\dort|)\;\; ,
\end{equation}
weakly as a function of $\ort$ (for fixed $\dort$) 
as $E\to\infty$, where $E$ denotes 
the energy of the eigenstate $\psi$ in \eqref{eq:def-loc-auto}. 
Equivalently we have 
\begin{equation}\label{eq:corr-universal-}
  \lim_{E\to\infty}C^{\text{loc}}(\ort,\dort/\sqrt{E})
= \frac{1}{\Vol (\Omega)}\, 
  J_0(|\dort|)\;\; .
\end{equation}

Numerical and experimental tests of this relation have been performed 
for several chaotic systems 
\cite{McDKau88, AurSte93,LiRob94,EckDoeKuhStoe99} and show at finite
energies noticeable fluctuations of the autocorrelation function 
around the high energy limit \eqref{eq:corr-universal}, especially for 
correlation distances larger than a few de Broglie wavelengths.
These fluctuations have been studied further in 
\cite{SreSti96,Sre96b,HorSre98a,HorSre98},
where for small
correlation length $|\dort|$ a random model 
for the eigenfunctions of a chaotic system was used 
to predict the variance of these fluctuations,   
and for larger $|\dort|$ a formula involving closed
orbits of the system has been derived.
In \cite{ShaGoe84,ShaRonBru88} 
the path correlation function, which is an
average of the local correlations along a given trajectory,
has been introduced.
The path correlation function 
is closely related to the autocorrelation function
and for ergodic systems also tends asymptotically to a 
Bessel function \eqref{eq:corr-universal}. 
This path correlation function 
has been studied in \cite{AurSte93} for an hyperbolic octagon, 
and an expansion in terms of Legendre functions has been derived, which 
can be used to determine corrections to the leading Bessel part 
\eqref{eq:corr-universal}. 

The autocorrelation function in non-chaotic systems has attracted 
much less attention. The integrable case has already been discussed 
by Berry \cite{Ber77b}, and the corresponding formula has been successfully
tested for the circle billiard in \cite{McDKau88}. 
For a system with mixed 
classical phase space the autocorrelation function has been studied in 
\cite{VebRobLiu99}, in particular 
for irregular eigenfunctions an expansion of the 
Wigner function in polar coordinates has been used.

In this work we are interested in the question how the universal limit 
\eqref{eq:corr-universal} is reached, and how, in the case of mixed 
systems, further constraints on the classical motion are reflected 
in the autocorrelation function. For instance, if an eigenfunction is 
concentrated on an ergodic component, then by a generalization 
of the quantum ergodicity theorem \cite{Sch2001:PhD},
the Wigner function becomes equidistributed on that component, and  
this will determine the autocorrelation function.    

The paper is organized as follows. 
In  section \ref{sec:examples} we discuss some examples of the autocorrelation 
function for different eigenfunctions in systems with chaotic and  
mixed classical dynamics. In section \ref{sec:expansion}  a general expansion 
of the autocorrelation function for eigenfunctions in billiards is 
derived, which allows a systematic study of their properties. It is an 
expansion in the correlation length $|\dort|$ which reflects the 
fact that the determination of correlations at 
larger distances needs classical information on finer length scales
than for short range correlations. 
In section \ref{sec:application}  it is shown that 
the correlation length 
expansion provides an efficient way to explain the fine structure of 
the autocorrelation functions of the systems studied 
in the first section. Of particular interest is that 
for chaotic systems deviations of the autocorrelation function 
from the quantum ergodic limit \eqref{eq:corr-universal} 
can be related to the rate of quantum ergodicity. In turn
the autocorrelation function 
can be used to study the  rate of quantum ergodicity on  
different classical length scales.

%%%%%%%%%%%%%%%%%%%%%%%%%%%%%%%%%%%%%%%%%%%%%%%%%%%%%%%%%%%%%%%%%%%%%%%%%%%%%%%
\section{Examples of autocorrelation functions}
\label{sec:examples}
%%%%%%%%%%%%%%%%%%%%%%%%%%%%%%%%%%%%%%%%%%%%%%%%%%%%%%%%%%%%%%%%%%%%%%%%%%%%%%%

For numerical computations as well as for theoretical considerations 
it is much more convenient
to consider a smoothed version of the 
local autocorrelation function \eqref{eq:def-loc-auto}. 
Furthermore, as 
the eigenfunctions oscillate on a scale proportional to 
$1/\sqrt{E}$, we rescale the autocorrelation function by this factor. 
Hence we will  study the autocorrelation function in the form
\begin{equation}\label{eq:gen-def-aotocorrel}
\rC_{\rho}(\ort,\dort):=\Int_\Omega \rho(\ort -\BFq)
\psi^*\bigg(\BFq-\frac{\dort}{2\sqrt{E}}\bigg)
    \psi\bigg(\BFq+\frac{\dort}{2\sqrt{E}}\bigg)\; \ud q \;\;,
\end{equation}
where $\rho$ is a positive function which determines the smoothing of the 
local autocorrelation function. In the literature 
(see the papers mentioned in the introduction)
the mean is usually taken over a small disk,
which corresponds
to taking the characteristic function of a disk for 
$\rho$ in \eqref{eq:gen-def-aotocorrel}.
However, nothing prevents one to consider the case 
$\rho \equiv 1$, i.e.\ to take the mean value of the local 
autocorrelation function \eqref{eq:def-loc-auto} over the 
whole position space. 
In terms of the Wigner function 
\begin{equation} \label{eq:def-Wigner}
  W(\BFp,\BFq):=\frac{1}{(2\pi)^2}\Int \ue^{\ui \BFp\BFq'}
  \psi^*(\BFq-\BFq'/2)\psi(\BFq+\BFq'/2)\; \ud q' \;\;,
\end{equation}
one obtains in this case
\begin{align} 
  \rC(\dort) &:=  \Int
    \psi^*\bigg(\BFq-\frac{\dort}{2\sqrt{E}}\bigg)
    \psi\bigg(\BFq+\frac{\dort}{2\sqrt{E}}\bigg)\; \ud q 
   \label{eq:def-autocorrel-rho1} \\
 &= \IInt W(\BFp,\BFq) \ue^{-\ui\BFp \dort/\sqrt{E}}\;\ud q\,\ud p
  =\Int |\widehat{\psi}(\BFp)|^2 \ue^{-\ui\BFp \dort/\sqrt{E}}\; \ud p\;\; .
\end{align}

This is a particularly good choice for 
the numerical computation of the autocorrelation function in billiards
because it can be reduced to boundary integrals, see 
Appendix \ref{sec:appendix-autocorrel-via-u_n}.  
The resulting formula reads
\begin{equation}\label{eq:autocorrel-via-boundary-integral}
  C(\dort)=\frac{1}{8\sqrt{E}}\IInt_{\partial\Omega\times\partial\Omega}
   |\BFq(s)-\BFq(s')+\dort|\, 
    Y_1(\sqrt{E}|\BFq(s)-\BFq(s')+\dort|) \, u^*(s)u(s')
   \; \ud s\,\ud s'\;\; ,
\end{equation}
where $u(s)$ is the normal derivative of the
normalized eigenfunction $\psi$ 
on the billiard boundary.
This relation provides a very efficient method for the numerical 
computation of the autocorrelation function.

\BILD{t}
     {
     \vspace*{-1cm}
     \begin{center}
      \begin{minipage}{4.6cm}
         \PSImagx{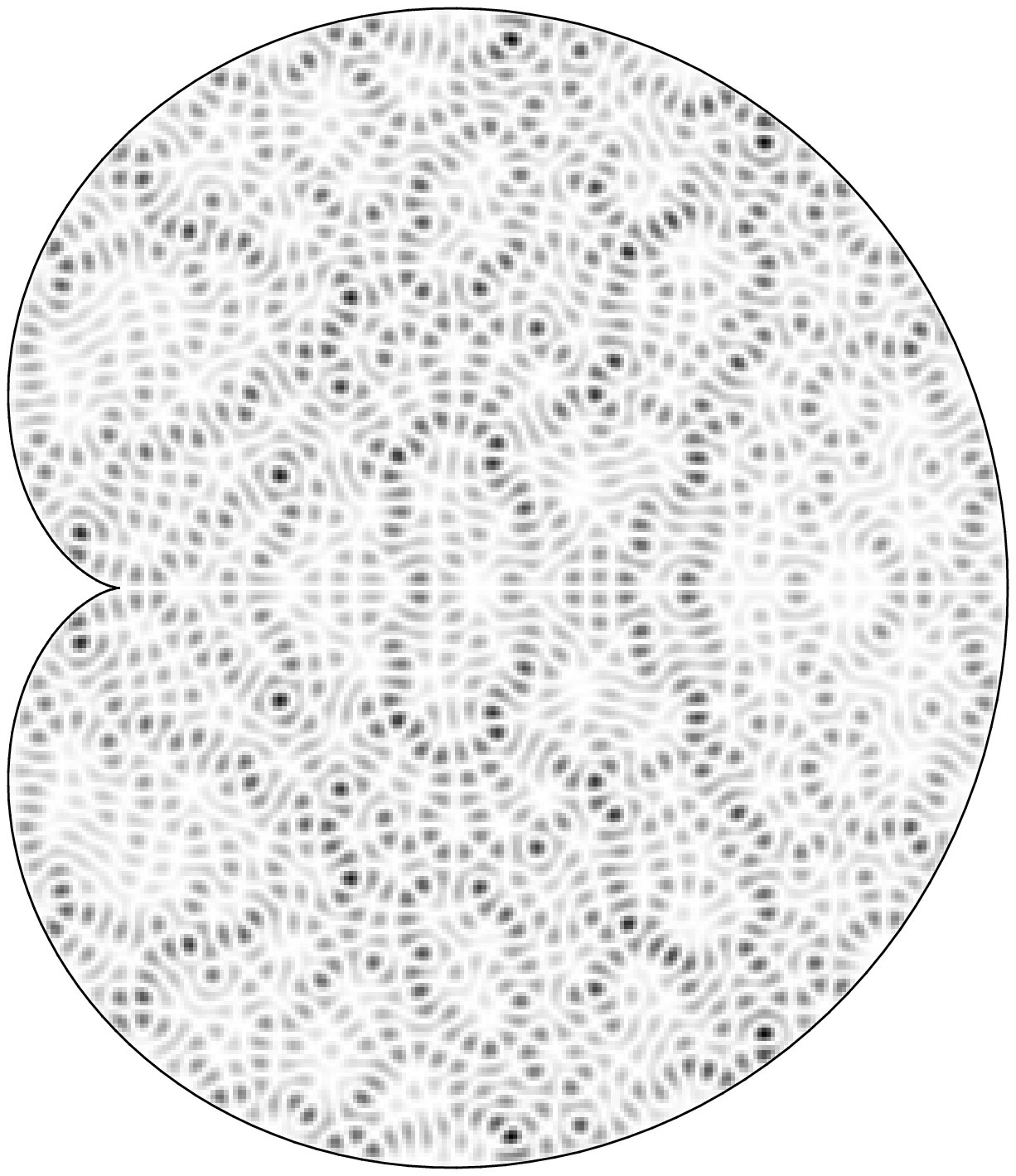}{4.5cm}
      \end{minipage}
      \begin{minipage}{11.6cm}
      \PSImagx{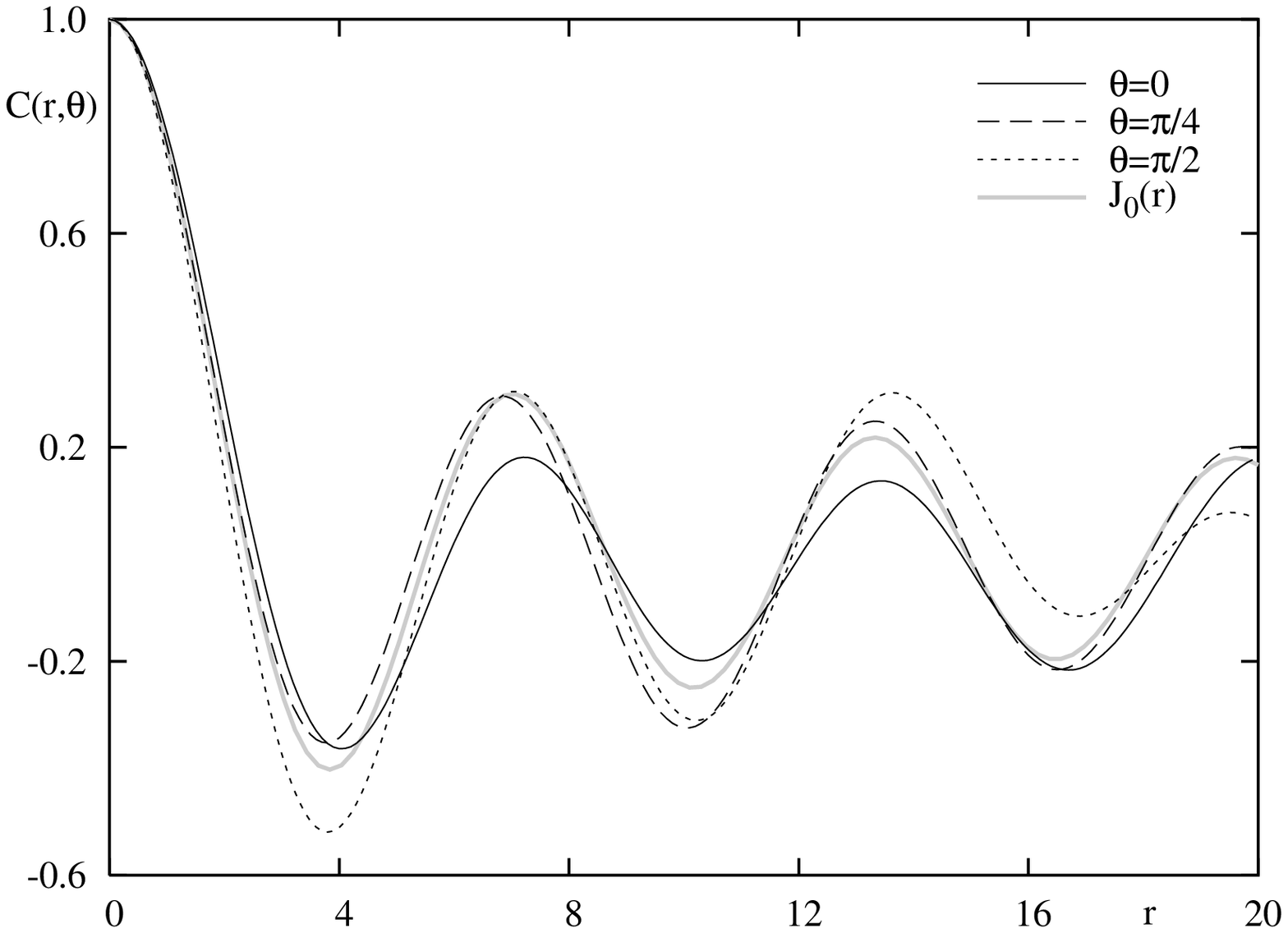}{11.75cm}
      \end{minipage}
     \end{center}
     \vspace*{-0.5cm}
     }
     {Grey scale plot of $|\psi_{1277}(\BFq)|^2 $ 
      in the cardioid billiard with odd symmetry,
      where black corresponds to high intensity. To 
      the right the autocorrelation function 
      $C(r,\theta)$, computed using
      \eqref{eq:autocorrel-via-boundary-integral},
      is shown for three different directions $\theta=0,\pi/4$
      and $\pi/2$. For comparison the asymptotic result $C(r,\theta)=J_0(r)$
      is shown as grey line.
     }
     {fig:cardi-1277}

\BILD{bh}
     {
     \vspace*{-0.5cm}
     \begin{center}
      \PSImagx{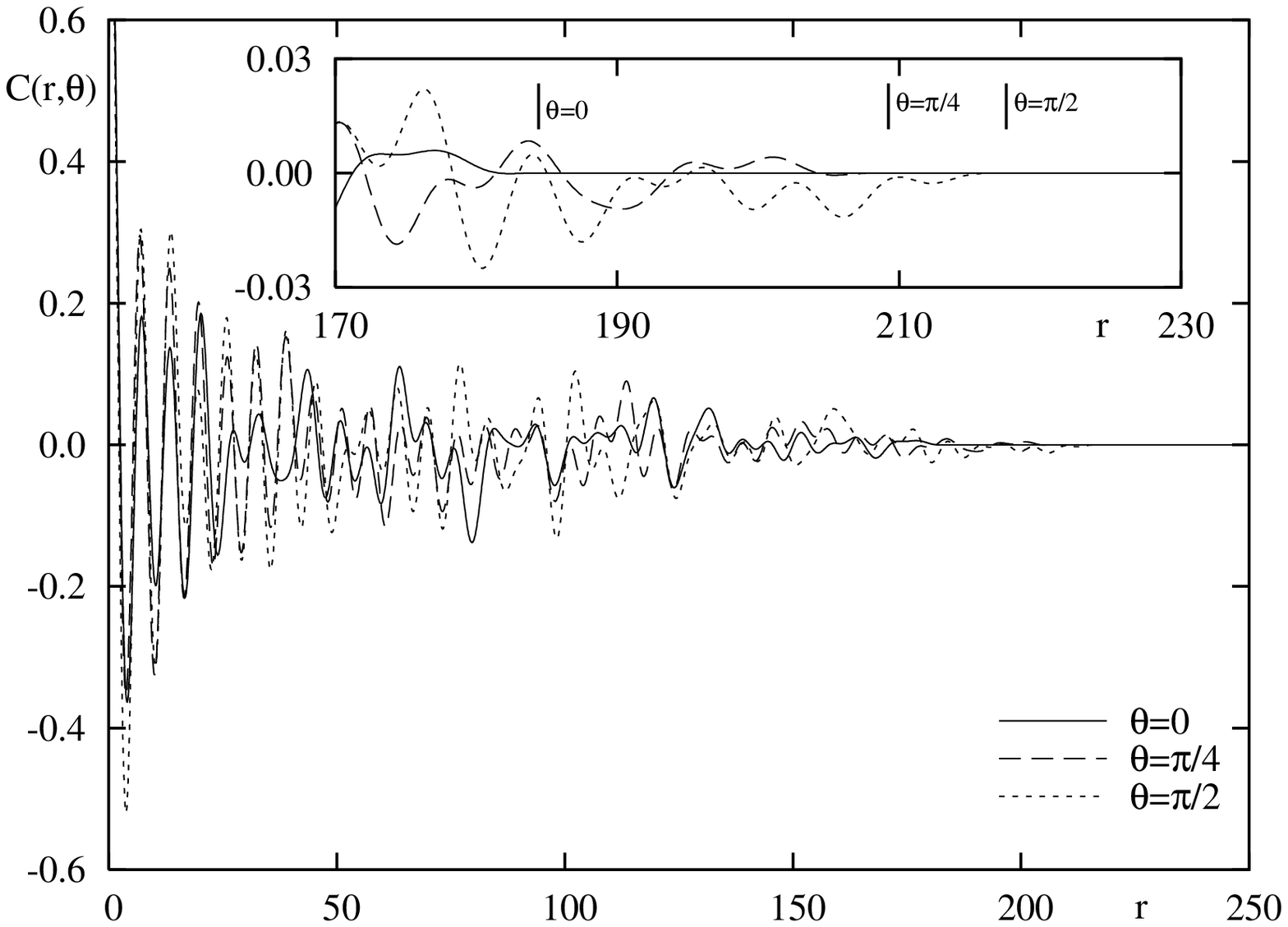}{15cm}
    \vspace*{-0.5cm}
     \end{center}
     }
     {Autocorrelation function for the same state 
      as fig.~\ref{fig:cardi-1277}, but for a larger $r$-interval
      showing the non-universal behavior at larger $r$. 
      The inset shows a magnification and the vertical bars
      indicate the places $r=\sqrt{E_n} \diam(\Omega,\theta)$
      from where on $C(r,\theta)=0$, due to the compactness of the billiard.
     }
     {fig:cardi-1277-large-r}

The systems for which we study the autocorrelation functions are 
the stadium billiard and two members of the family of 
\limacon billiards, namely the cardioid billiard, and a billiard with 
mixed classical phase space. 
The stadium billiard is proven to be strongly
chaotic, i.e.\ it is ergodic, mixing and a $K$-system \cite{Bun74,Bun79}.
The height of the desymmetrized billiard 
is chosen to be 1, and $a$ denotes the length
of the upper horizontal line, for which we have $a=1.8$
in the following.
The family of \limacon billiards is given by
the simplest nontrivial conformal mapping of the unit circle \cite{Rob83,Rob84}
and can be parametrized in polar coordinates by 
$\rho(\varphi) = 1 + \varepsilon\cos(\varphi)$ with $\varphi\in[-\pi,\pi]$
and $\varepsilon\in[0,1]$ denotes the family parameter.
We consider the case $\varepsilon=0.3$ which leads to a mixed 
dynamics in phase space.
For $\varepsilon=1$ one obtains the cardioid billiard,
which is also proven to be strongly chaotic \cite{Woj86,Sza92,Mar93}.
The eigenvalues of the cardioid billiard have been provided
by Prosen and Robnik \cite{PrivComProRob} and were calculated
by means of the conformal mapping technique, see e.g.\ 
\cite{Rob84,ProRob93a}.
For the stadium billiard the eigenvalues and eigenfunctions
have been computed using the boundary element method, 
see e.g.\ \cite{Rid79,BerWil84}, and for the \limacon billiard
the eigenvalues have been computed using the conformal mapping technique
and then the boundary element method has been used 
to compute the eigenfunctions (see \cite{Bae98:PhD} for details).
For the high lying states in the \limacon billiard
the scaling method has been used \cite{VerSar95}.

First we consider a ``typical'' eigenfunction in the cardioid billiard,
see fig.~\ref{fig:cardi-1277}. 
In the plots we show
\begin{equation}
  C(r,\theta)=C(r \hat{\bf e}(\theta)) \;\;,
\end{equation}
where $\hat{\bf e}(\theta)=(\cos \theta,\sin \theta)$, as a function of $r$ 
for three different values of $\theta $.
The quantum ergodicity theorem implies that 
there is a subsequence $\{n_j\}\subset \N$ of density one such 
that $C_{n_j}(r,\theta) \to J_0(r)$ as $n_j\to\infty$ with $r$ fixed.
This convergence is, however, not uniform in $r$.
For the example shown in fig.~\ref{fig:cardi-1277}
$C(r,\theta)$ fluctuates, as expected for a ``quantum ergodic'' state,
around the asymptotic result 
\begin{equation} \label{eq:asymptotic-result-J-0}
  C(r,\theta)\sim J_0(r)\,\, . 
\end{equation}
Actually, for an eigenstate with energy $E_n$
we have $C(r,\theta)=0$
for $r>\sqrt{E_n} \diam(\Omega,\theta)$, where
$\diam(\Omega,\theta)$ is the diameter of $\Omega$ in the direction $\theta$, 
as follows directly from the definition 
\eqref{eq:gen-def-aotocorrel}.
This is illustrated in fig.~\ref{fig:cardi-1277-large-r} which clearly shows
the non-universal behavior for larger $r$.

\BILD{b}
     {
     \begin{center}
      \begin{minipage}{4.6cm}
      \PSImagx{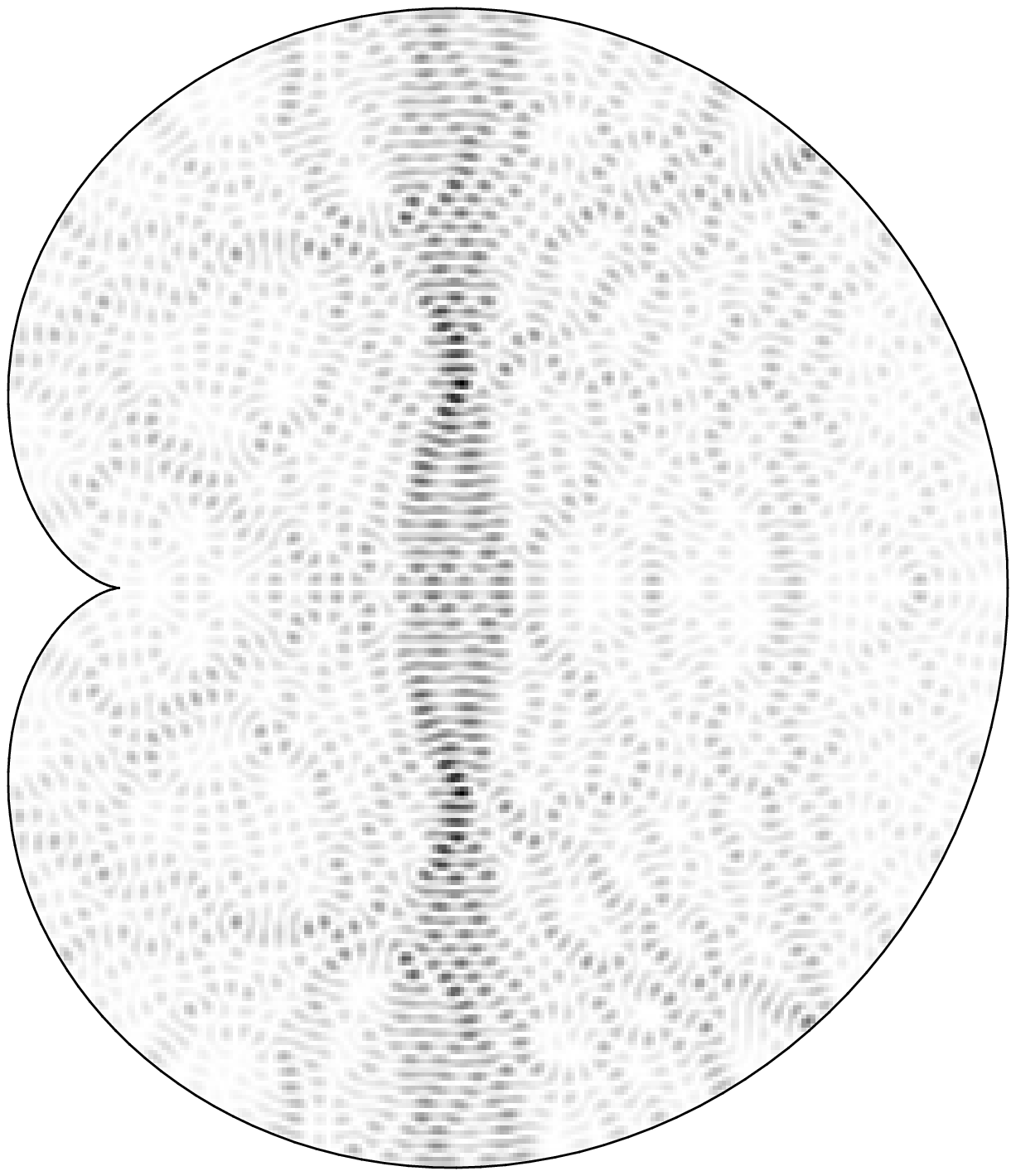}{4.5cm}
      \end{minipage}
      \begin{minipage}{11.6cm}
      \PSImagx{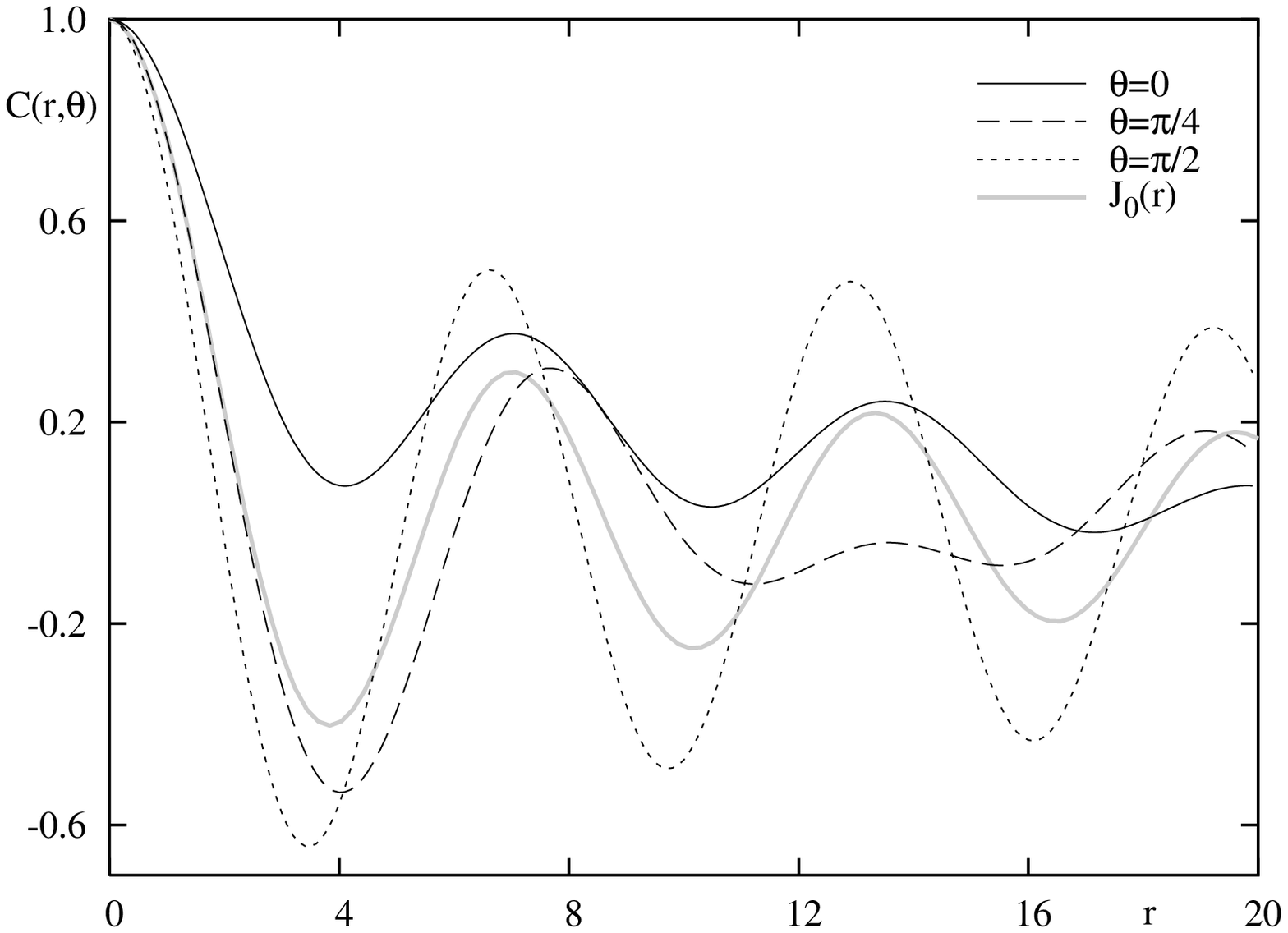}{11.75cm}
      \end{minipage}
     \end{center}
     }
     {Grey scale plot of $|\psi_n(\BFq)|^2 $ with 
      $n=1817$ in the cardioid billiard with odd symmetry. 
      For the autocorrelation function 
      $C(r,\theta)$ one observes clear deviations
      from $C(r,\theta)=J_0(r)$.
     }
     {fig:cardi-1817}

In contrast to the case of quite uniformly distributed eigenfunctions
one expects a stronger directional dependence of the autocorrelation
function for localized eigenfunctions, such as scars \cite{Hel84}.
One example is shown in fig.~\ref{fig:cardi-1817},
where the eigenfunction shows localization along the shortest
unstable periodic orbit in the cardioid.
The corresponding autocorrelation function shows
clear deviations from \eqref{eq:asymptotic-result-J-0}.

\BILD{tbh}
     {
     \begin{minipage}{4.9cm}
      \PSImagx{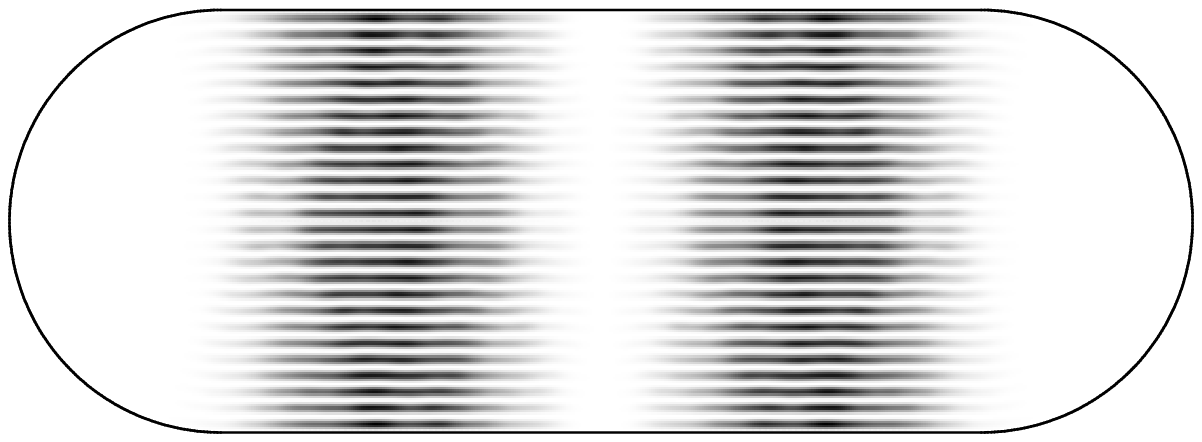}{4.8cm}
     \end{minipage}
     \begin{minipage}{11.6cm}
      \PSImagx{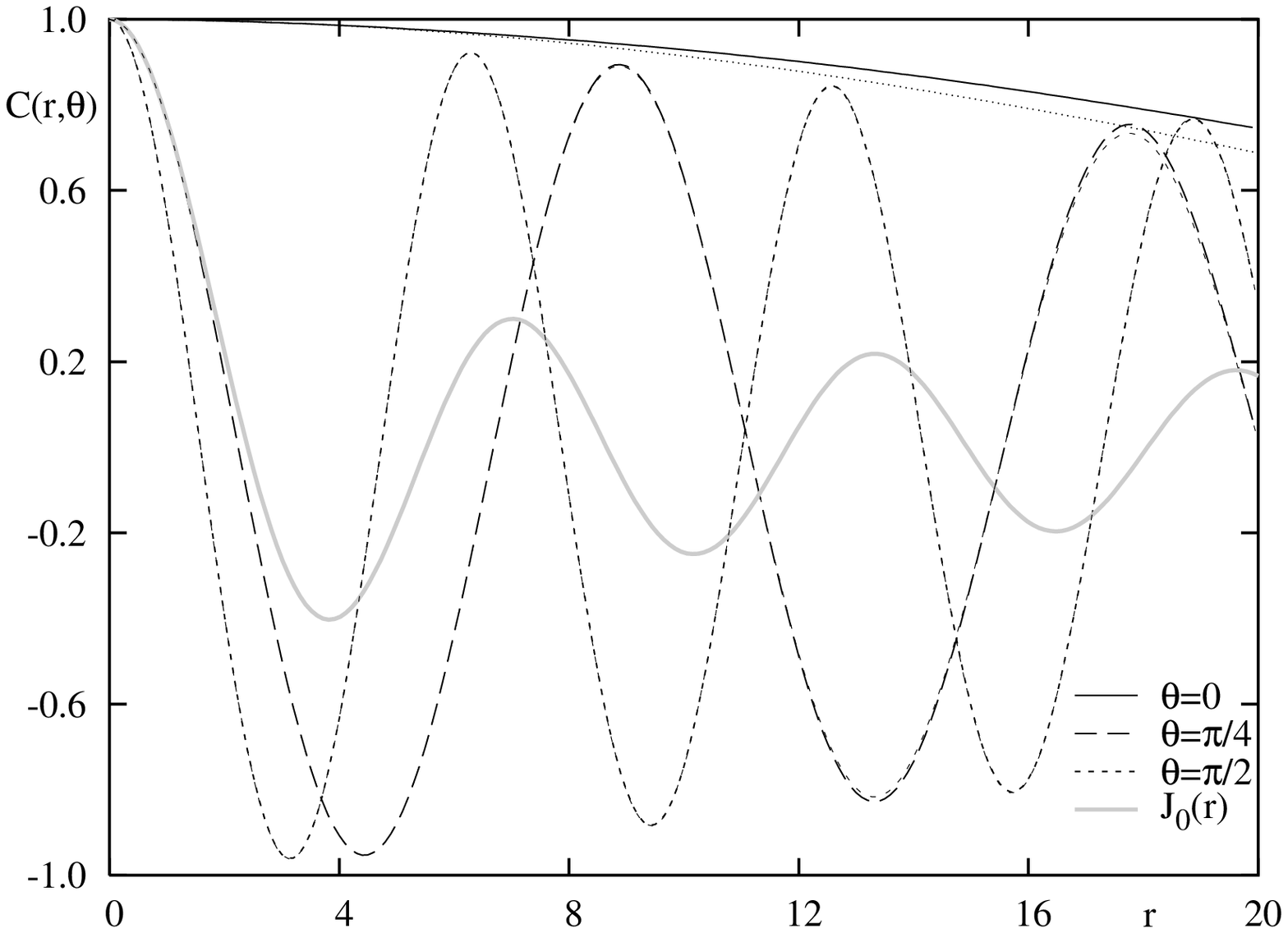}{11.75cm}
     \end{minipage}
     }
     {For the stadium billiard with odd-odd symmetry, $a=1.8$,
      $\psi_{320}(\BFq)$ is a bouncing ball mode.
      The corresponding autocorrelation function is compared 
      with the result $C_{1,13}^{\text{box}}(r,\theta)$, 
     eq.~\eqref{eq:autocorrel-box},
         obtained for a box, shown
      as dotted curves, which 
      follow $C(r,\theta)$.
      Only for $\theta=0$ (full line) and $\theta=\pi/4$ at $r\approx 17$
      are small deviations visible.
     }
     {fig:stad-320}

A class of eigenfunctions which show even stronger
localization are the bouncing ball modes in 
billiards with two parallel walls, see
e.g.\ \cite{BaiHosSteTay85,McDKau88,ConHel88,Tan97,BaeSchSti97a}.
Fig.~\ref{fig:stad-320} shows for the stadium billiard
an example of a bouncing ball mode, which localize
on the so-called bouncing ball orbits having perpendicular
reflections at the parallel walls and thus forming a one-parameter
family. 
The simplest approximation is to consider them as a product
of two sines, one in the $x$ direction and the other in the $y$ direction.
In this case the autocorrelation function can be computed explicitly.
For the odd-odd eigenfunctions 
\begin{equation}
  \psi_{n_x,n_y}(x,y) = 
     \frac{1}{\sqrt{l_x l_y}} \sin(\pi n_x x/l_x) \sin(\pi n_y y/l_y)
\end{equation}
in a box $B:=[-l_x,l_x] \times [-l_y,l_y]$ one gets 
\begin{equation} \label{eq:autocorrel-box}
  C_{n_x,n_y}^{\text{box}}(r,\theta) 
     = F(r \cos(\theta)/\sqrt{E},n_x,l_x) \, 
       F(r \sin(\theta)/\sqrt{E},n_y,l_y) \;\;,
\end{equation}
where 
\begin{align} \label{eq:F-box-for-odd-odd-symmetry}
  F(z,n,l) &:= \chi_{[-l,l]}(z/2)  \frac{1}{l} \Int_{-l+z/2}^{l-z/2} 
                             \sin(\pi n (x-z/2)/l) \sin(\pi n (x+z/2)/l) \; \ud x \\
           &= \chi_{[-l,l]}(z/2)  
      \left[\left(1-\frac{z}{2l}\right) \cos(\pi n z/l)  
           + \frac{1}{2\pi n} \sin(\pi n z/l) \right]     \;\;,          
\end{align} 
and $\chi_{[-l,l]}(z)$ denotes the characteristic function of the interval 
$[-l,l]$. 

In fig.~\ref{fig:stad-320} we compare the autocorrelation function 
$C(r,\theta)$
 for a bouncing ball mode in the stadium billiard
with $C_{1,13}^{\text{box}}(r,\theta)$, 
eq.~\eqref{eq:autocorrel-box},
and observe very good agreement.
Mainly for $\theta=0$ some deviations are visible; these are understandable
from the fact that in this case
only correlations in the $x$--direction are measured, 
where the bouncing ball mode ``leaks''
outside the rectangular region.
To take this into account
one can determine an effective $l_x^{\text{eff}}>2a$, 
by fitting $\sin^2(\pi x/l_x^{\text{eff}})$ to 
\begin{equation}
  \psi_n^{\text{proj}}(x) := \Int_0^{1} |\psi(x,y)|^2 \; \ud y \;\;.
\end{equation}
For the case shown in fig.~\ref{fig:stad-320} this procedure
leads to $l_x^{\text{eff}}\approx 4$ (whereas $2a=3.6$) 
and the corresponding
autocorrelation function gives excellent agreement with the one
for $\psi_{320}$.

\FloatBarrier
%%%%%%%%%%%%%%%%%%%%%%%%%%%%%%%%%%%%%%%%%%%%%%%%%%%%%%%%%%%%%%%%%%%%%%%%%%%%%%%
\section{Expansion of the autocorrelation function}
\label{sec:expansion}
%%%%%%%%%%%%%%%%%%%%%%%%%%%%%%%%%%%%%%%%%%%%%%%%%%%%%%%%%%%%%%%%%%%%%%%%%%%%%%%

In this section we derive an expansion of the 
autocorrelation function which will lead to an understanding of
the directional dependence of 
the autocorrelation function observed in the last section. 
We start from the 
representation of the local autocorrelation function in terms of the 
Wigner function
\begin{equation} \label{eq:starting-point}
\rC_{\rho}(\ort,\dort)=\iint \rho(\ort-\BFq)W(\BFp,\BFq)
       \ue^{-\ui \BFp\dort/\sqrt{E}}\; \ud p\,\ud q \;\;.
\end{equation}
Since the Wigner function is concentrated around the energy shell 
$|\BFp|=\sqrt{E}$, and is furthermore even in $\BFp$ by time 
reversal symmetry,  we get 
\begin{align}
\begin{split}
\rC_{\rho}(\ort,\dort)
&=\Int_0^{\infty}\Int_0^{2\pi}\Int_{\Omega} 
    \rho(\ort-\BFq)W(\BFp,\BFq)\; \ud q
\, \ue^{-\ui |\dort|\cos(\varphi-\theta)}\, r\;
\ud \varphi\, \ud r +O(E^{-1/2})\label{eq:approx_on_energy}\\
&=\Int_0^{\infty}\Int_0^{2\pi}\Int_{\Omega} 
    \rho(\ort-\BFq)W(\BFp,\BFq)\; \ud q\, 
\cos(|\dort|\cos(\varphi-\theta))\, r\; \ud \varphi\, \ud r +O(E^{-1/2}) \;\; ,
\end{split}
\end{align}
where we have used polar coordinates 
$\BFp=(|\BFp|\cos\varphi, |\BFp|\sin\varphi)$, 
$\dort=(|\dort|\cos\theta, |\dort|\sin\theta)$. 
Because of the rescaling by $\sqrt{E}$ the factor 
$\ue^{-\ui \BFp\dort/\sqrt{E}}$ 
is only slowly oscillating for $\BFp$ close 
to the energy shell, on which  
the Wigner function is concentrated.
Therefore  we get that 
the error is of order $1/\sqrt{E}$, 
see Appendix \ref{sec:appendix:estimate} for 
a sketch of the derivation of this remainder estimate.
If we now use that $\cos(r\cos \varphi)$ is a generating 
function for Bessel functions \cite{AbrSte84},  
\begin{equation}\label{eq:gen-function}
\cos( |\dort|\cos\varphi )
=J_0( |\dort|)+2\sum_{l=1}^{\infty} (-1)^l \cos(2l \varphi)
J_{2l}( |\dort|) \;\;,
\end{equation}
we obtain
\begin{equation}
\rC_{\rho}(\ort,\dort)=
\co_0(\ort)\, J_0(|\dort|)+2\sum_{l=1}^{\infty} (-1)^l
\co_{2l}(\ort,\theta)\, J_{2l}( |\dort|)+O(E^{-1/2})\;\; ,
\end{equation}
with (setting $r=|\BFp|$)
\begin{equation}
\co_{2l}(\ort,\theta):=\Int_0^{\infty}\Int_0^{2\pi}\Int_{\Omega} 
   \rho(\ort-\BFq)W(\BFp,\BFq)\; \ud q\, 
\cos(2l (\varphi-\theta))\, r\; \ud \varphi \,\ud r\;\; .
\end{equation}
The coefficients $\co_{2l}(\ort,\theta)$ can be further 
decomposed 
\begin{equation}\label{eq:co_split}
\begin{split}
\co_{2l}(\ort,\theta)&=\cos(2l \theta) \;
       \Int_0^{\infty}\Int_0^{2\pi}\Int_{\Omega} 
      \rho(\ort-\BFq)W(\BFp,\BFq)\; \ud q\, 
\cos(2l \varphi)\, r\;\ud \varphi\,\ud  r \\ 
&\quad +
 \sin(2l \theta) \;\Int_0^{\infty}\Int_0^{2\pi}\Int_{\Omega} 
      \rho(\ort-\BFq)W(\BFp,\BFq)\; \ud q\, 
\sin(2l \varphi)\, r\; \ud \varphi\,\ud  r\;\; .
\end{split}
\end{equation}
Recall that for an operator 
$\widehat{A}$ with Weyl symbol $A(\BFp,\BFq)$ the expectation 
value $\langle \psi , \widehat{A}\psi\rangle$ 
can be written as an integral over phase space 
of the symbol multiplied by the 
Wigner function of $\psi$, see e.g.\ \cite{Fol89}, 
\begin{equation}\label{eq:exp-value-repr}
\langle \psi , \widehat{A}\psi\rangle =\IInt W(\BFp,\BFq)A(\BFp,\BFq)
\; \ud p\,\ud q\;\; .
\end{equation}
Therefore the coefficients in \eqref{eq:co_split} can be interpreted as  
expectation values of certain operators
$\widehat{A}_{2l}(\ort)$, $\widehat{B}_{2l}(\ort)$ given as the Weyl quantizations 
of the functions 
\begin{equation}\label{eq:symbols}
A_{2l}(\BFp,\BFq):=\rho(\ort-\BFq)\cos(2l \varphi)\qquad 
B_{2l}(\BFp,\BFq)=\rho(\ort-\BFq)\sin(2l \varphi)\;\; , 
\end{equation}
respectively, 
\begin{align}
\Int_{\Omega}\Int_0^{2\pi}\Int_0^{\infty} W(\BFp,\BFq)\, 
\rho(\ort-\BFq)\cos(2l \varphi)\, r\;\ud  r \,\ud \varphi\, \ud q 
      &=\langle\psi,\widehat{A}_{2l}(\ort)\psi\rangle 
 \label{eq:QMexpA} \\
\Int_{\Omega}\Int_0^{2\pi}\Int_0^{\infty} W(\BFp,\BFq)\, \rho(\ort-\BFq)
\sin(2l \varphi)\, r\; \ud  r \,\ud \varphi\, \ud q 
      &=\langle\psi,\widehat{B}_{2l}(\ort)\psi\rangle \;\; .
 \label{eq:QMexpB} 
\end{align}
Note that the operators $\widehat{A}_{2l}(\ort)$ and $\widehat{B}_{2l}(\ort)$ 
depend on the parameter $\ort$. 
Since their symbols are smooth and homogeneous of 
degree zero in $\BFp$ 
they are classical pseudodifferential operators of order zero, see 
e.g.\ \cite{Fol89}
for the definition of pseudodifferential operators. 
So we  finally obtain the following 
general expansion of the autocorrelation function  
\begin{equation}\label{eq:auto_corr_expansion}
\begin{split}
\rC_{\rho}(\ort,\dort)&=
\langle\psi,\widehat{A}_0(\ort)\psi\rangle \, J_0(|\dort|)\\
&\qquad+2\sum_{l=1}^{\infty} (-1)^l 
[\langle\psi,\widehat{A}_{2l}(\ort)\psi\rangle \cos(2l \theta)+
\langle\psi,\widehat{B}_{2l}(\ort)\psi\rangle \sin(2l \theta)]\, J_{2l}(|\dort|)\\
&\qquad+O(E^{-1/2})\;\; ,
\end{split}
\end{equation}
in terms of the expectation values of a sequence of bounded 
operators given as Weyl quantizations of the symbols \eqref{eq:symbols}. 
Recall that the only 
approximation we have made was to insert 
for $|\BFp|$ in the exponent in eq.~\eqref{eq:approx_on_energy}
the value at the energy shell $\sqrt{E}$.

Since the Bessel functions have the property that $J_{2l}(|\dort|)\approx 0$ 
for $|\dort|\ll 2l$, 
this representation is an efficient 
expansion for small $|\dort|$; the larger $|\dort|$ becomes, the more 
terms of the sum have to be taken into account. Therefore it is desirable to 
have an estimate on the number of terms which have to be taken into
account for large $|\dort|$. 
The first, and largest,  maximum of $J_{2l}(r)$ lies around $r\sim 2l$, 
and close to it one has the expansion
\cite{AbrSte84}
\begin{equation}\label{eq:uniform-Bessel-asympt}
J_{2l}(2l-z l^{1/3})= \frac{1}{l^{1/3}} \Ai(z)+O(1/l)\,\, .
\end{equation}
So the first peak becomes broader with a rate $\sim l^{1/3}$ and 
therefore we have to take for large $r$ approximately 
\begin{equation}
m\sim \frac{r}{2}
+\frac{z}{2}\left(\frac{r}{2}\right)^{1/3}
\end{equation}
terms in the sum over $l$ into account; here $z$ determines the error 
term. We refer to  appendix \ref{app:Bessel} for a  more detailed discussion.

We would like to mention two works in which related results 
have been obtained. In \cite{VebRobLiu99} a similar expansion
was derived for the
case that the eigenfunction is concentrated on an ergodic 
component of the phase space 
of a classically mixed system,
however without extracting the 
Bessel function from the expectation values. 
For the case of a free 
particle on a surface of constant negative curvature
an expansion of the path correlation function in terms of
Legendre function was derived in \cite{AurSte93}.
In the special case of averaging over the whole billiard (i.e.\ $\rho=1$)
the path correlation function should
be for ergodic systems the same as the autocorrelation function.

The correlation length expansion \eqref{eq:auto_corr_expansion}
has various possible applications; some of them will be discussed
and illustrated in the next section.
In particular, the expansion leads to a prediction for 
the asymptotic limit
of the autocorrelation function in different situations.
More precisely, 
consider a subsequence of eigenfunctions $\{\psi_{n_j}\}_{j\in\N}$ 
for which 
the corresponding sequence of Wigner functions converges weakly
to a measure $\nu$ on phase space.
Such a measure $\nu$ is called a quantum limit, and it is an invariant measure
of the classical flow.

If a sequence of eigenfunctions 
$\{\psi_{n_j}\}_{j\in\N}$ converges to a quantum limit, the correlation 
length expansion for the autocorrelation function 
\eqref{eq:auto_corr_expansion} shows that the corresponding 
sequence of autocorrelation functions converges as well and their limit is 
obtained by substituting in \eqref{eq:auto_corr_expansion} 
the expectation values of
$\widehat{A}_{2l}(\ort)$ and $\widehat{B}_{2l}(\ort)$ 
by their corresponding classical limit.
Explicitly, this gives 
\begin{equation} \label{eq:auto-correl-limit}
 \rC_{\rho}^{\text{limit}}(\ort,\dort)=\overline{A_0} J_0(|\dort|)
  +2\sum_{l=1}^{\infty} (-1)^l 
  \bigl[\,\overline{A_{2l}}(\BFx) \cos(2l \theta)+
        \overline{B_{2l}}(\BFx) \sin(2l \theta)\bigr]\, J_{2l}(|\dort|)\;\; ,
\end{equation}
where
\begin{equation} \label{eq:classical-exp}
  \overline{A}:=\Int_{T^*\Omega} A \,\,\ud \nu  \;\;.
\end{equation}

As we will discuss in section~\ref{subsec:ergodic-systems},
for ergodic systems the terms $\overline{A_{2l}}$ and $\overline{B_{2l}}$
vanish for $E\to\infty$,
and with $\overline{A_{0}}=1$
we recover \eqref{eq:asymptotic-result-J-0}.

%%%%%%%%%%%%%%%%%%%%%%%%%%%%%%%%%%%%%%%%%%%%%%%%%%%%%%%%%%%%%%%%%%
\section{Applications of the correlation length expansion} 
\label{sec:application}
%%%%%%%%%%%%%%%%%%%%%%%%%%%%%%%%%%%%%%%%%%%%%%%%%%%%%%%%%%%%%%%%%%

%============================================================================
\subsection{Direct comparison}
%============================================================================

\newcommand{\einbild}[1]{
\PSImagx{stad_wfk_autocorrel_1907_direction_l_#1_analyt_test_expansion}{8.5cm}
}

\BILD{bth}
     {
     \begin{center}
       \hspace*{-0.75cm}\einbild{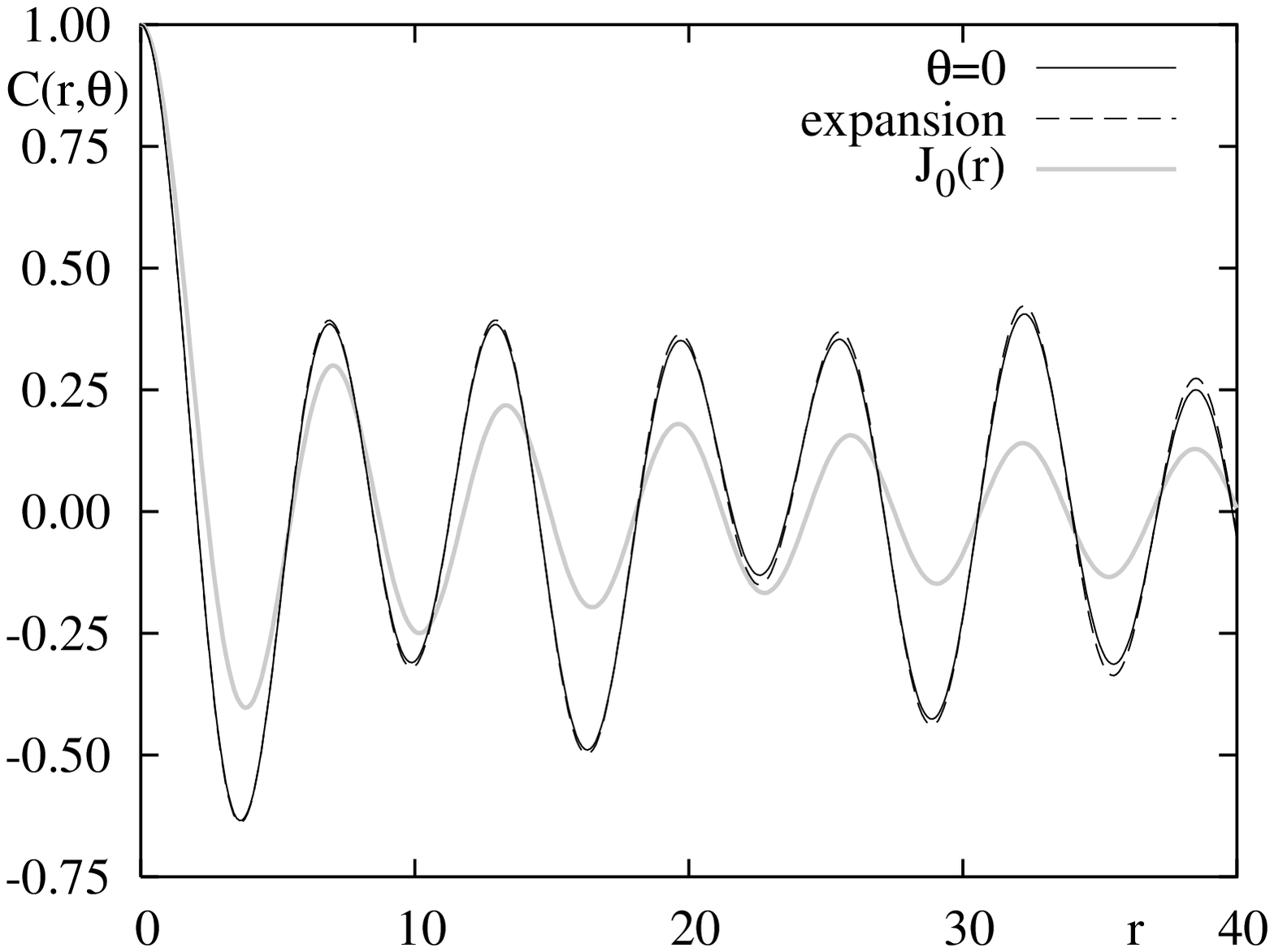}
       \einbild{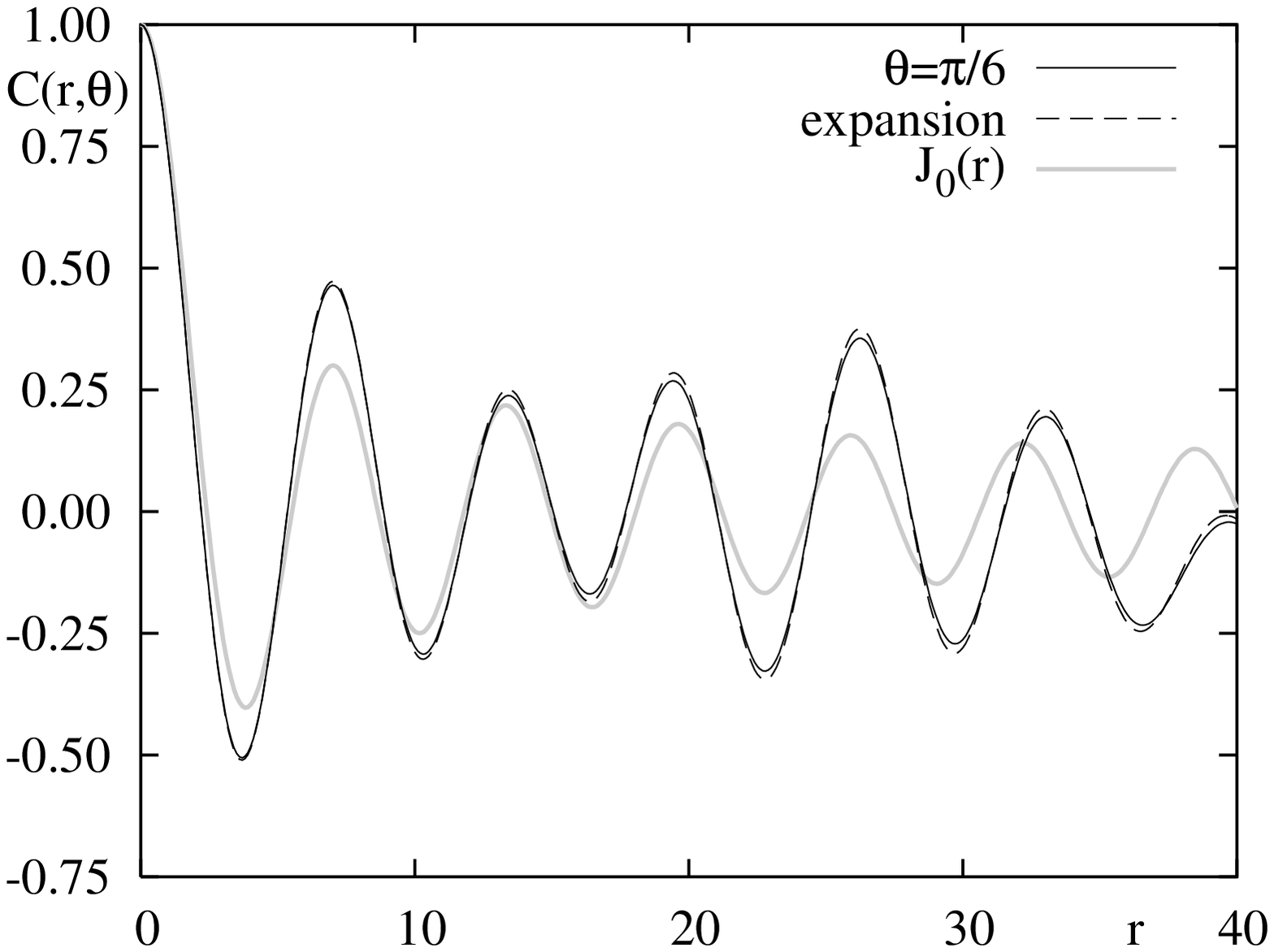}

       \hspace*{-0.75cm}\einbild{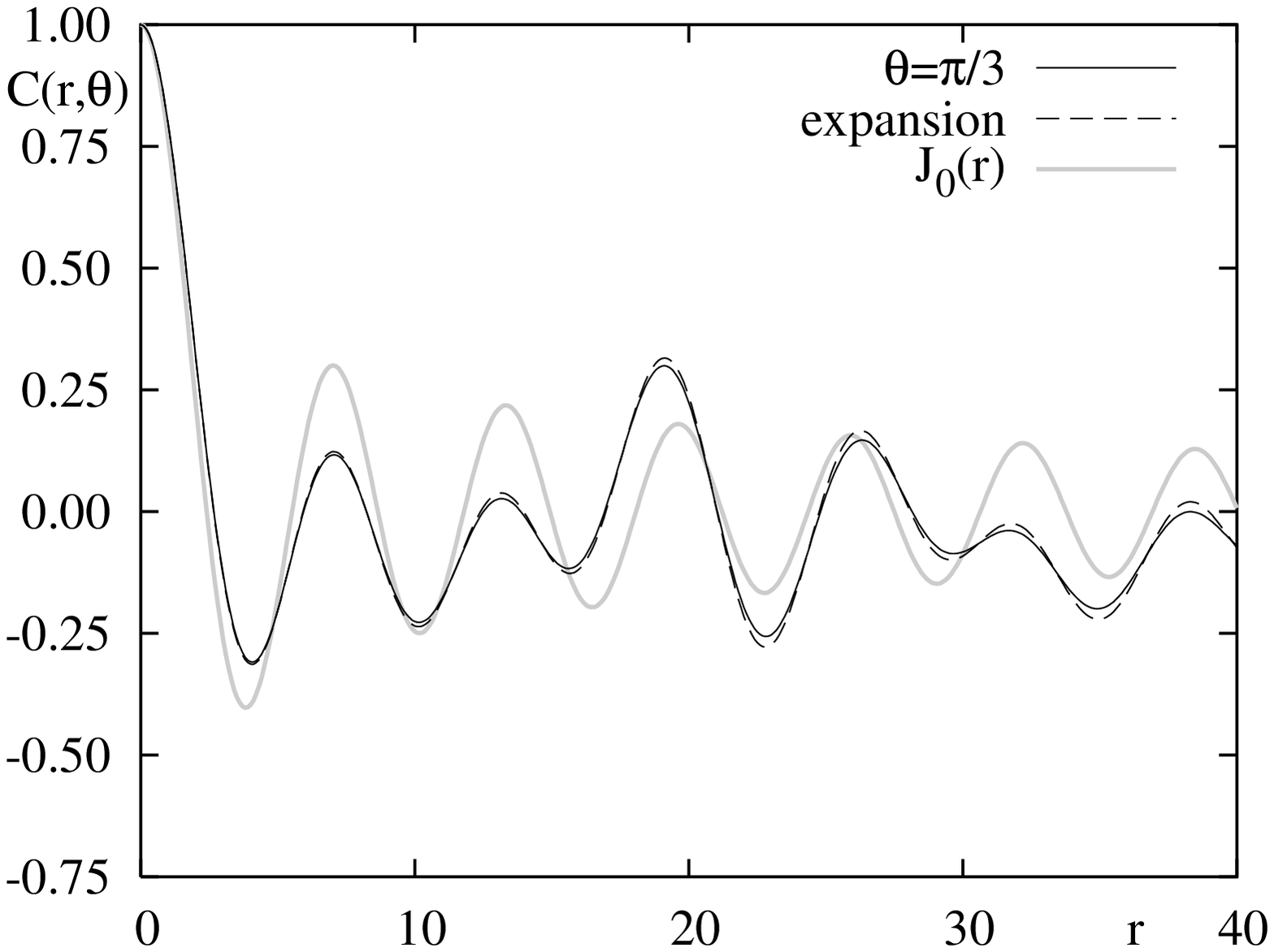}
       \einbild{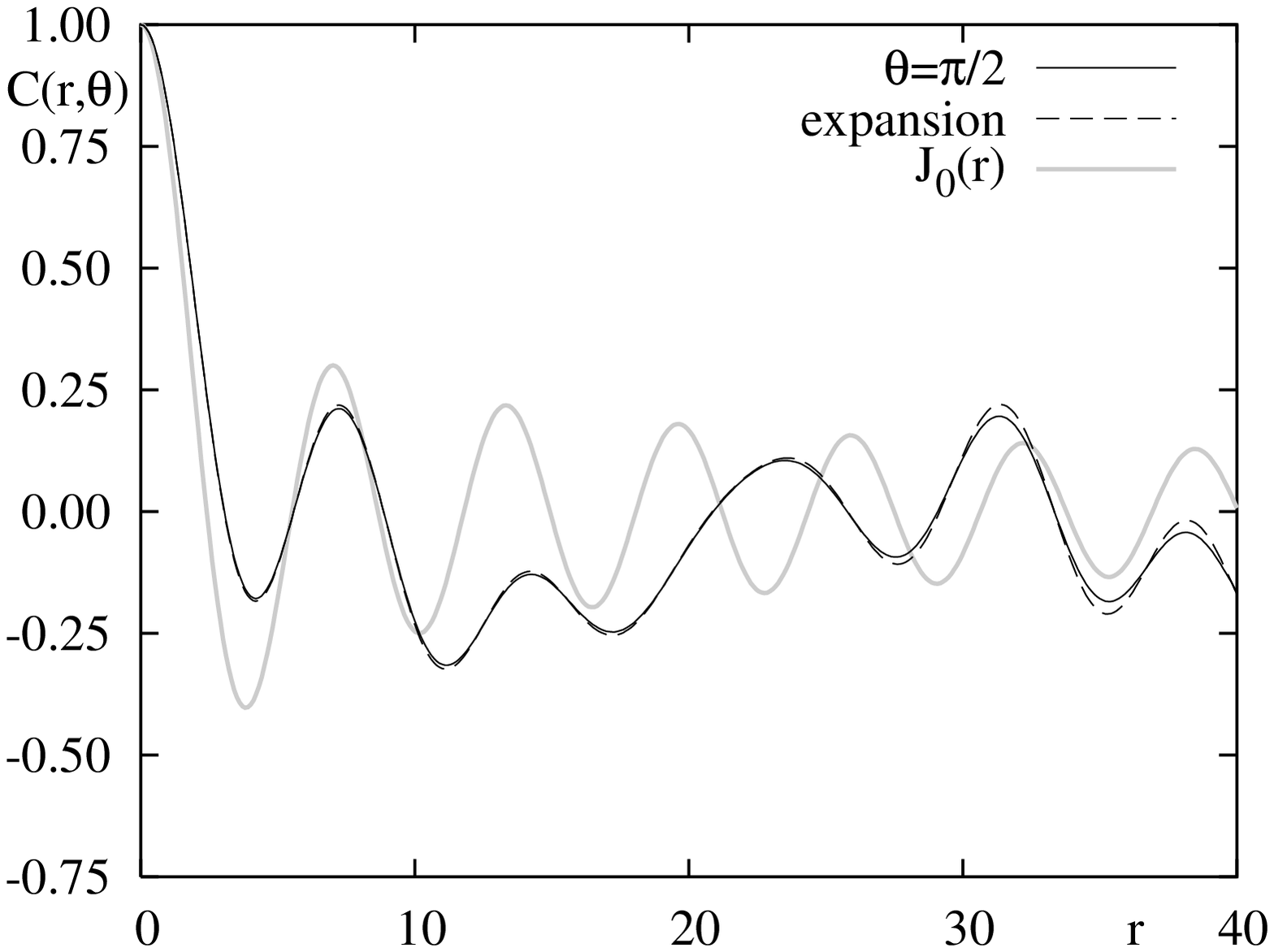}
     \end{center}
     }
     {Comparison of the autocorrelation function $C(r,\theta)$
      for $\psi_{1907}$ in the stadium billiard (full curve)
      with the expansion \eqref{eq:mean_auto_corr_expansion}.
      In particular for small $r$ the agreement is excellent,
      whereas for larger $r$ small differences become visible.
     }
     {fig:cardi-1907-direct-comparison}

\BILD{tbh}
     {
     \begin{center}
      \begin{minipage}{8cm}
        \begin{center}
           \vspace*{0.5cm}
           \PSImagx{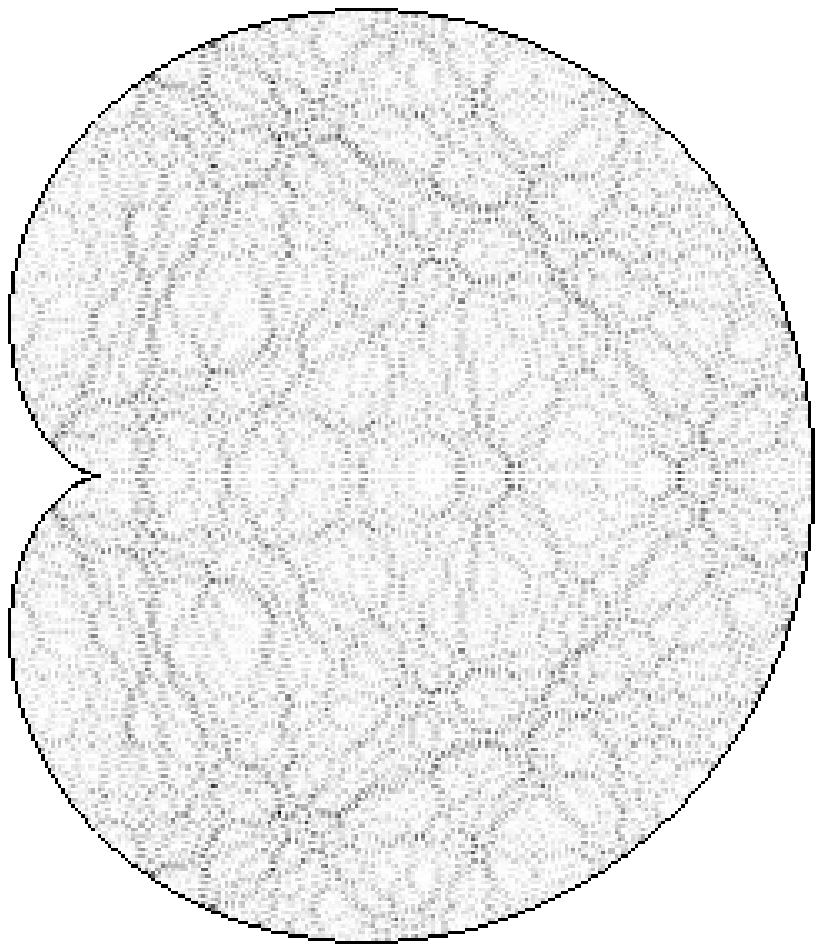}{4.5cm}
           \vspace*{0.5cm}
        \end{center}
      \end{minipage}
      \begin{minipage}{8cm}
        \PSImagx{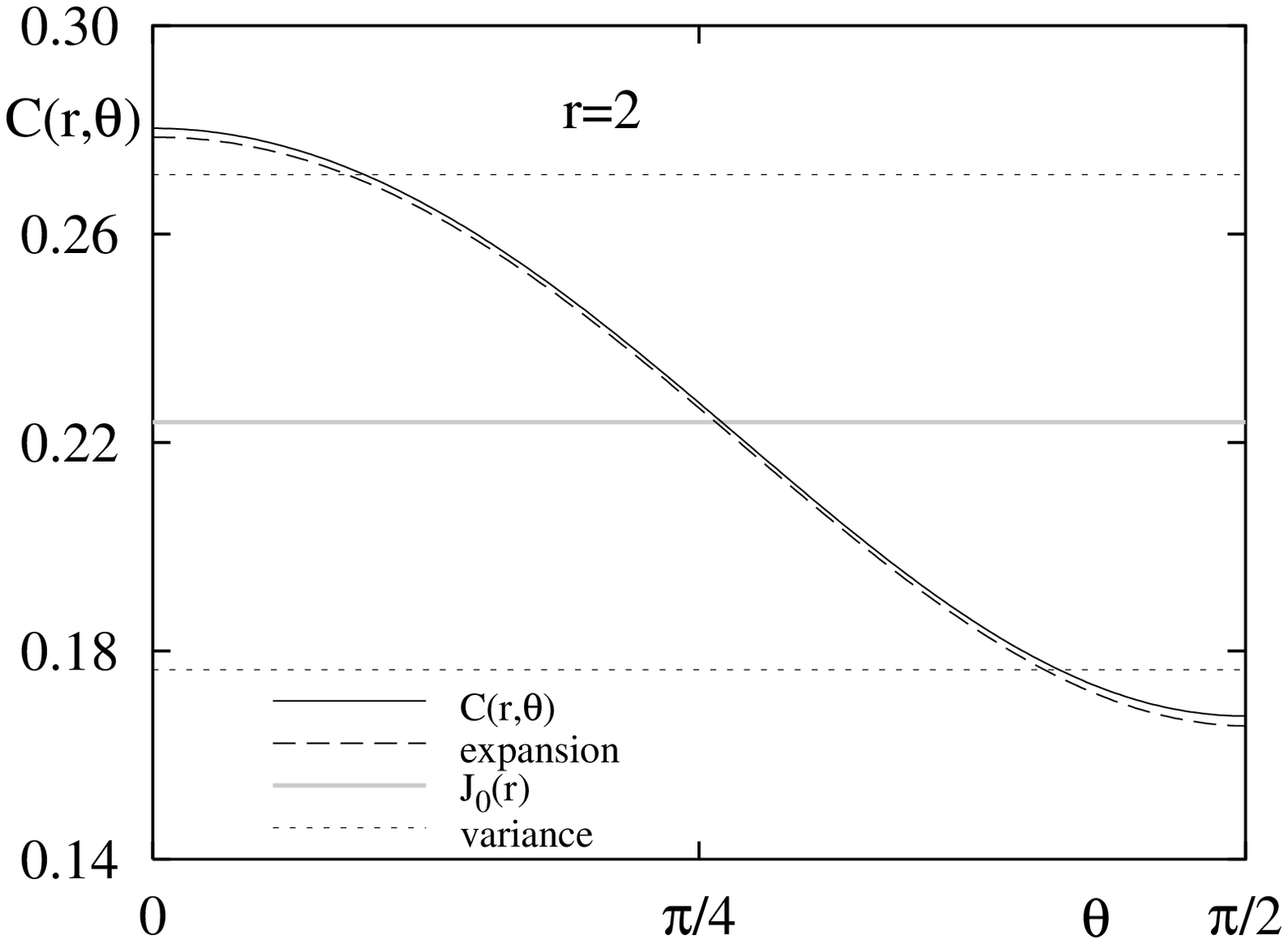}{8cm}
      \end{minipage}

\vspace*{1cm}                                       
        \PSImagx{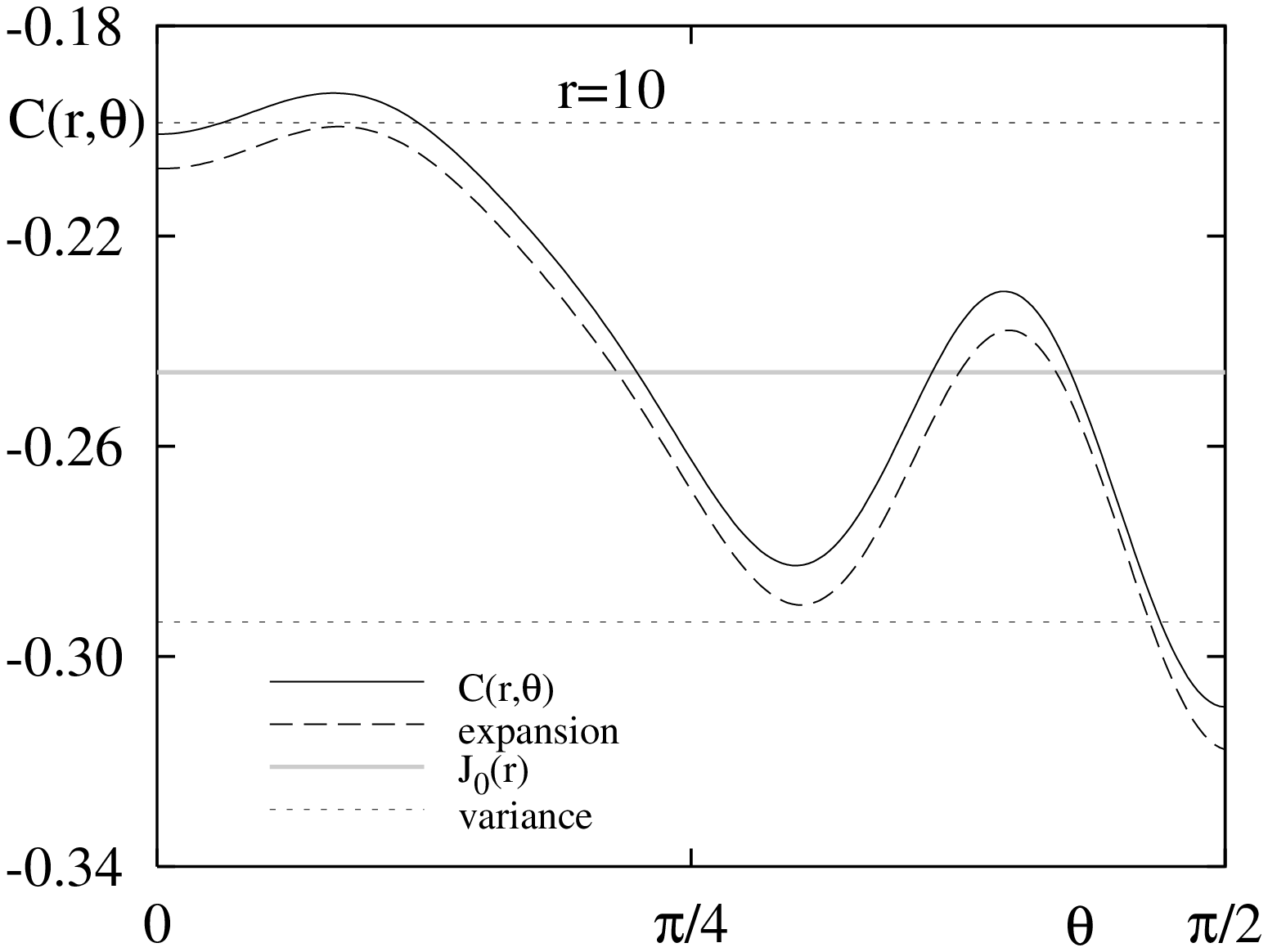}{8cm}
        \PSImagx{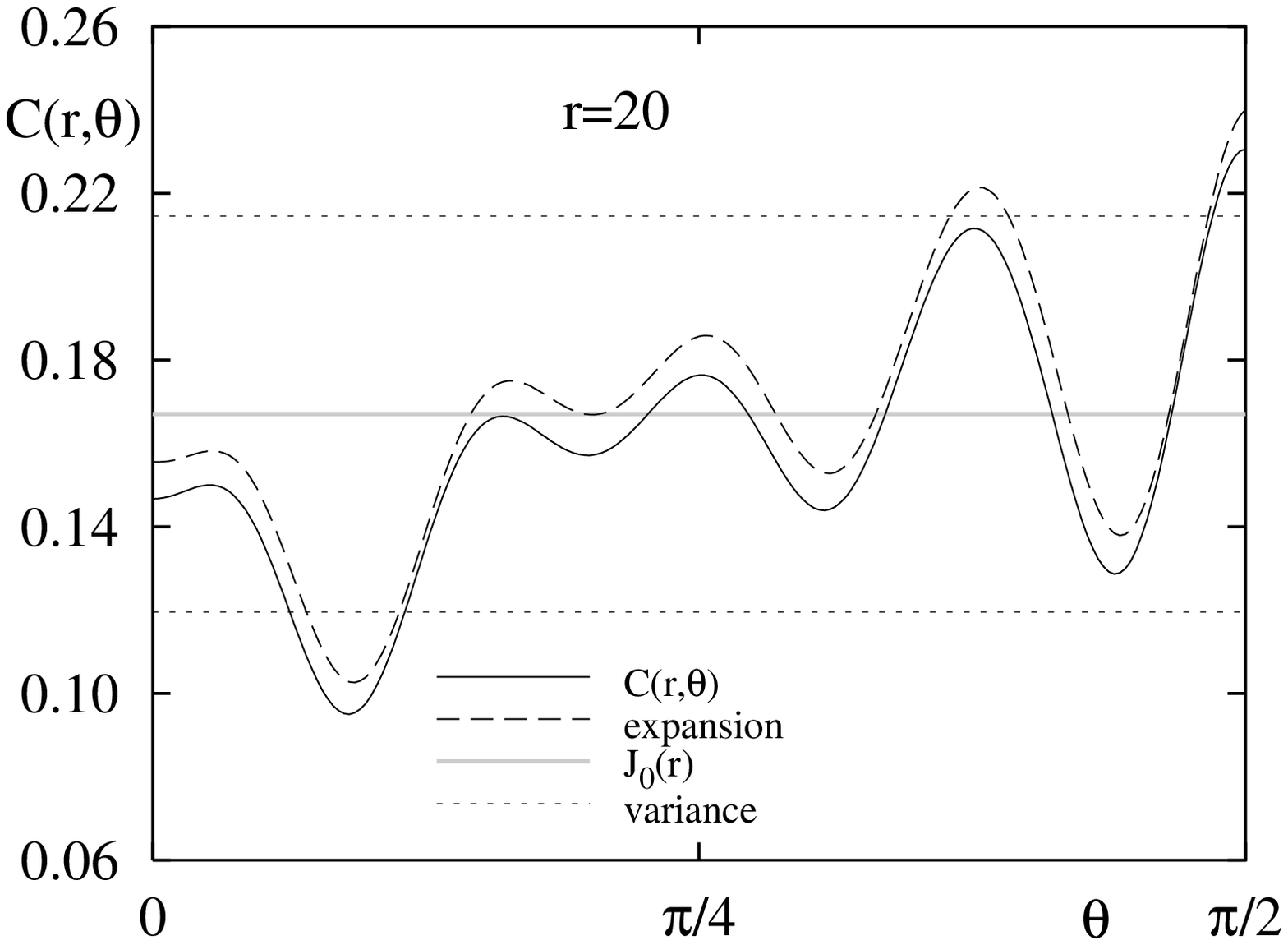}{8cm}

\vspace*{1cm}                                       
                                                 
        \PSImagx{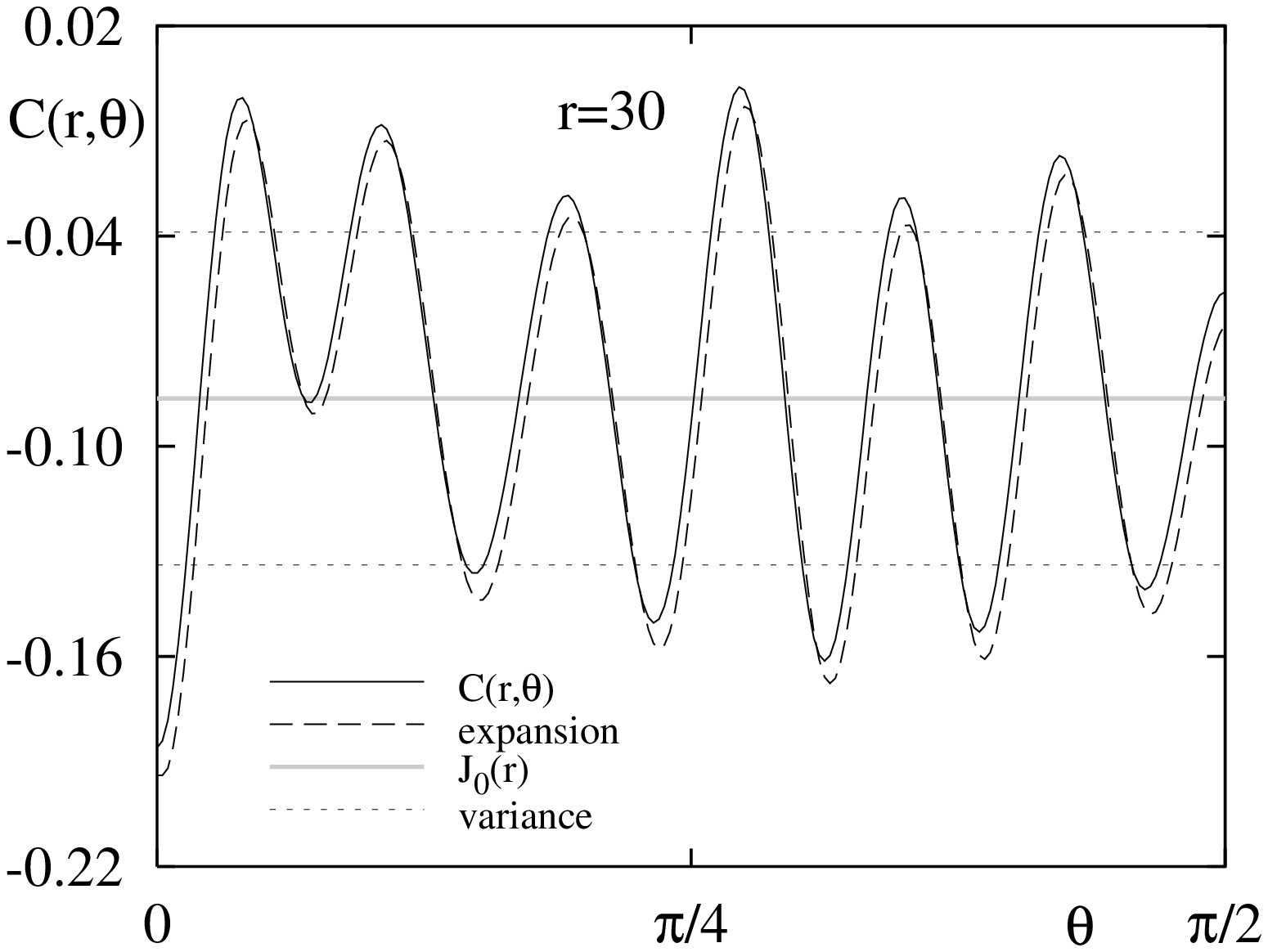}{8cm}
        \PSImagx{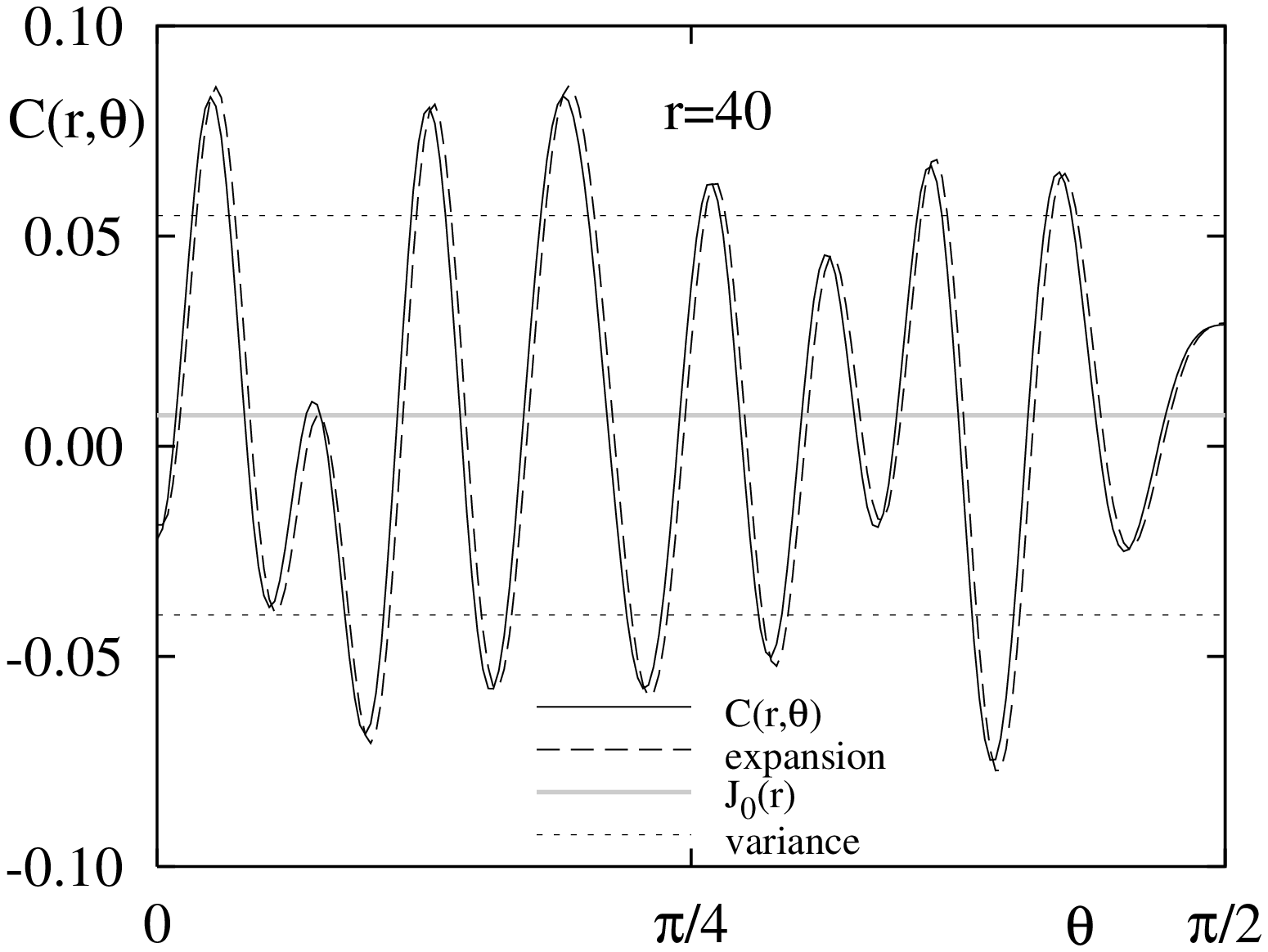}{8cm}

%      \end{minipage}
     \end{center}
     }
     {Angular dependence of the autocorrelation function $C(r,\theta)$  
      for different $r$. Shown are the results for $\psi_{6000}$ 
      in the  cardioid billiard with odd symmetry.
      The full line is the result for $C(r,\theta)$ using 
      \eqref{eq:autocorrel-via-boundary-integral}, the dashed
      line shows the result of the expansion 
      \eqref{eq:mean_auto_corr_expansion}, the full grey line 
      is the value of $J_0(r)$ and the dotted horizontal lines
      show the variance $J_0(r) \pm \Delta^{1/2}$,
      see eq.~\eqref{eq:variance-of-autocorrel-fluct}.
     }
     {fig:cardi-C-r-theta-A}

In the numerical examples we have studied the autocorrelation function 
in the case  $\rho=1$, which allows
for an exact computation of the autocorrelation function using
the representation 
\eqref{eq:autocorrel-via-boundary-integral},
which is much more efficient than a direct computation
of the autocorrelation function by its definition, 
eq.~\eqref{eq:def-autocorrel-rho1}.
In this case the general expansion
\eqref{eq:auto_corr_expansion} gives the representation 
\begin{equation}\label{eq:mean_auto_corr_expansion}
 \rC(r,\theta)= J_0(r)
  +2\pi \sum_{l=1}^{\infty} (-1)^l 
  \bigl[a_{2l}\cos(2l\theta)+b_{2l}\sin(2l\theta)\bigr] 
         \, J_{2l}(r)+O(E^{-1/2})\;\; ,
\end{equation}  
where the coefficients $a_{2l}$ and $b_{2l}$ are the 
Fourier coefficients 
\begin{equation}\label{eq:fourer-I-coeff}
a_{2l}=\frac{1}{\pi}\Int_0^{2\pi} 
    I(\varphi)\cos(2l \varphi)\; \ud\varphi
\qquad  \qquad
b_{2l}=\frac{1}{\pi}\Int_0^{2\pi} 
    I(\varphi)\sin(2l \varphi)\; \ud \varphi\;\; ,
\end{equation}
of the radially integrated momentum density \cite{Zyc92,BaeSch99}, 
\begin{equation} 
  I(\varphi):=\Int_0^{\infty}|\widehat{\psi}(r\BFe(\varphi))|^2r \; \ud r\;\;,
\end{equation}
where $\BFe(\varphi)=(\cos \varphi,\sin \varphi)$.
Also for $I(\varphi)$ a representation in terms
of a double integral of the normal derivative function
is available \cite{BaeSch99}. 
Taking the symmetries into account, one can show that for the 
odd eigenfunctions in the \limacon billiards and the odd-odd 
eigenfunctions in the stadium billiard all $b_{2l}$ 
vanish, so only the cosine terms remain in 
\eqref{eq:auto_corr_expansion} and \eqref{eq:mean_auto_corr_expansion}.

First we will test
the influence of the error term $O(E^{-1/2})$ in eq.~\eqref{eq:mean_auto_corr_expansion} for
computations at finite energies. To that end we use 
the exact quantum $I(\varphi)$ in eq.~\eqref{eq:fourer-I-coeff}. 
In fig.~\ref{fig:cardi-1907-direct-comparison} 
the autocorrelation function $C(r,\theta)$ for four different
angles $\theta$ 
is compared to \eqref{eq:mean_auto_corr_expansion}.
In particular for $r$ not too big the agreement is excellent.
Only for larger $r$ do small deviations become visible,
which go to zero for higher energies and $r$ fixed.
One should remark that for any $r>0$ the
effective integration region in eq.~\eqref{eq:def-autocorrel-rho1}
is reduced by the factor
\begin{equation} \label{eq:finite-area-correction}
  c(r,\theta):=\frac{\Vol\left( \Omega \cap \Omega(r/\sqrt{E},\theta) \right)}
             {\Vol(\Omega)} \;\;,
\end{equation}
where
$\Omega(r/\sqrt{E},\theta)$ is the set $\Omega$ shifted by
the vector $r/\sqrt{E} \; (\cos \theta,\sin \theta)$.
Incorporating this factor leads to an improvement in the agreement
of the expansion with the exact autocorrelation function
at larger $r$.

Instead of looking at the dependence of the autocorrelation function
$C(r,\theta)$ for fixed $\theta$ and varying $r$,
it is also interesting to keep $r$ fixed and consider the angular dependence.
For a ``chaotic'' eigenfunction in the cardioid billiard
some examples are shown in 
fig.~\ref{fig:cardi-C-r-theta-A}.
The result of the expansion \eqref{eq:mean_auto_corr_expansion}
is in good agreement with the exact result.
For larger $r$ the autocorrelation function
$C(r,\theta)$ oscillates more strongly around $J_0(r)$.
For even larger $r$ we observe clear deviations of the
expansion from the exact result (not shown).
For comparison the variance of the 
autocorrelation function around the  prediction $J_0(r)$ 
for a random wave model \cite{SreSti96} in leading order
\begin{equation} \label{eq:variance-of-autocorrel-fluct}
  \Delta^{1/2}=\left(\frac{16}{3\pi^{3/2}A}\right)^{1/2}\frac{1}{E^{1/4}}
\end{equation}
is shown and good agreement is found.

\FloatBarrier

%%%%%%%%%%%%%%%%%%%%%%%%%%%%%%%%%%%%%%%%%%%%%%%%%%%%%%%%%%%%%%%%%%%%%%%%%%%%%%
\subsection{Localized eigenfunctions}
%%%%%%%%%%%%%%%%%%%%%%%%%%%%%%%%%%%%%%%%%%%%%%%%%%%%%%%%%%%%%%%%%%%%%%%%%%%%%%

\BILD{b}
     {
     \begin{center}
     
     \begin{minipage}{8cm}
       \begin{center}
         \PSImagx{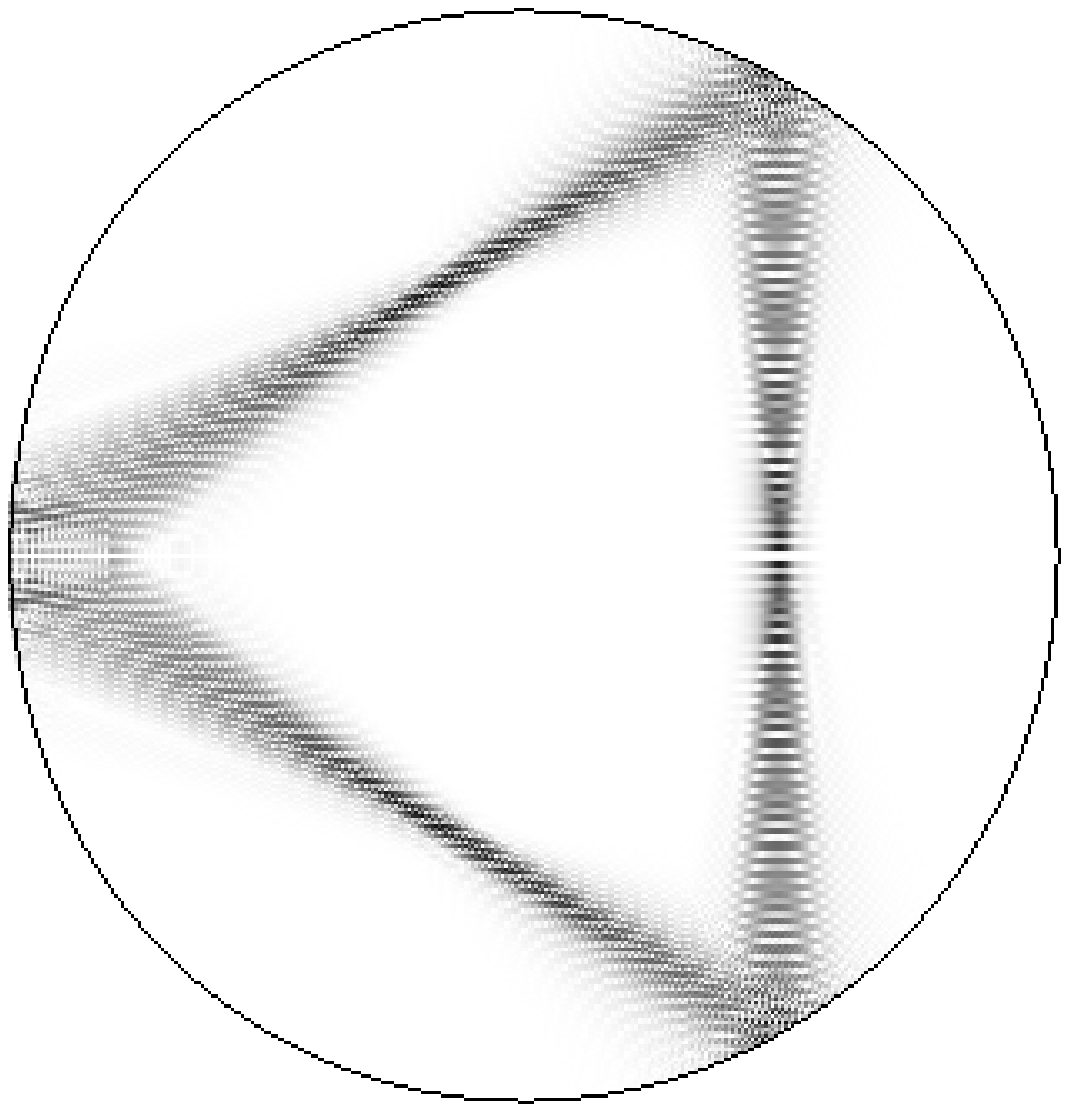}{4.5cm}
       \end{center}
     \end{minipage}
     \begin{minipage}{8cm}
\PSImagx{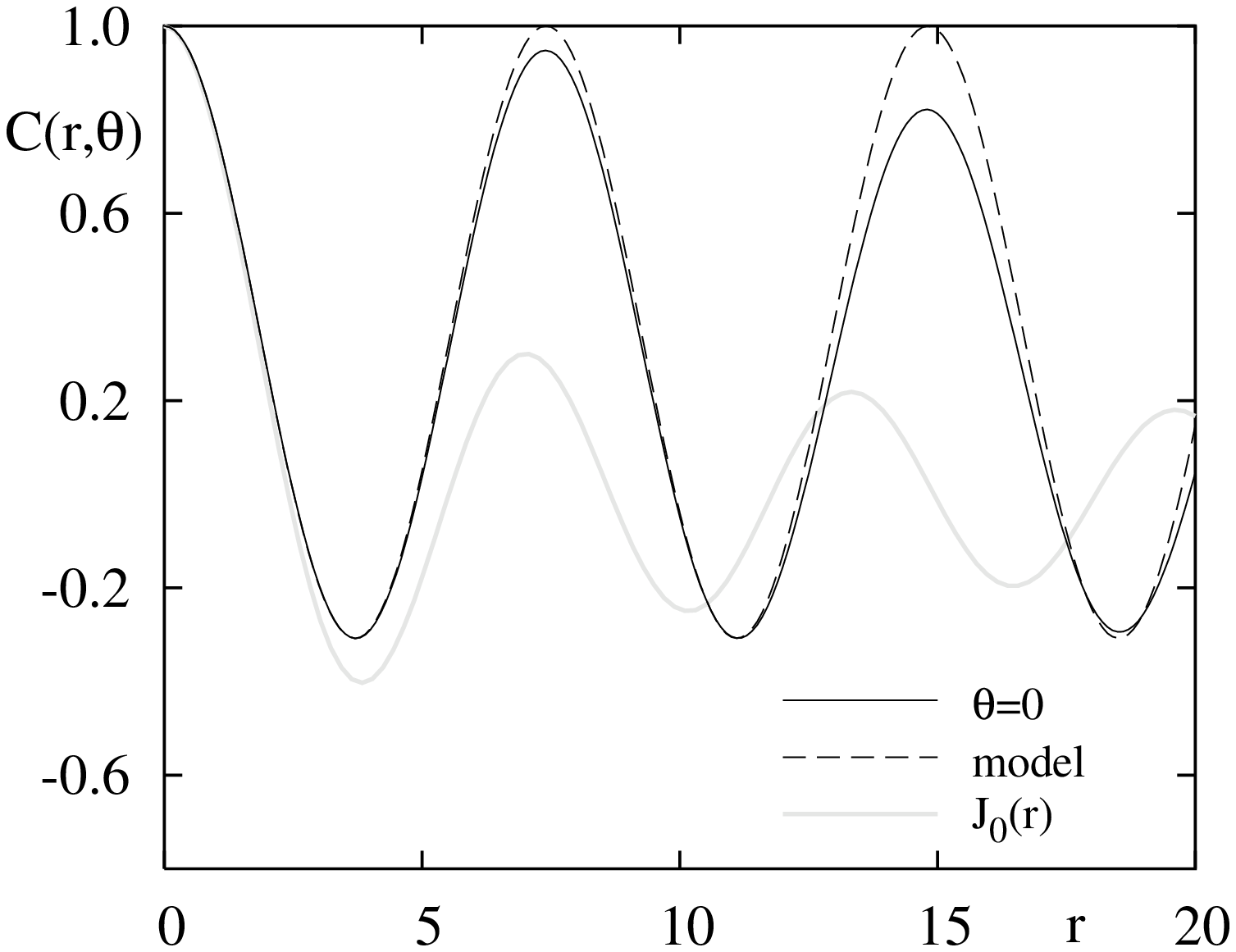}{8cm}
     \end{minipage}
                                                                       
\PSImagx{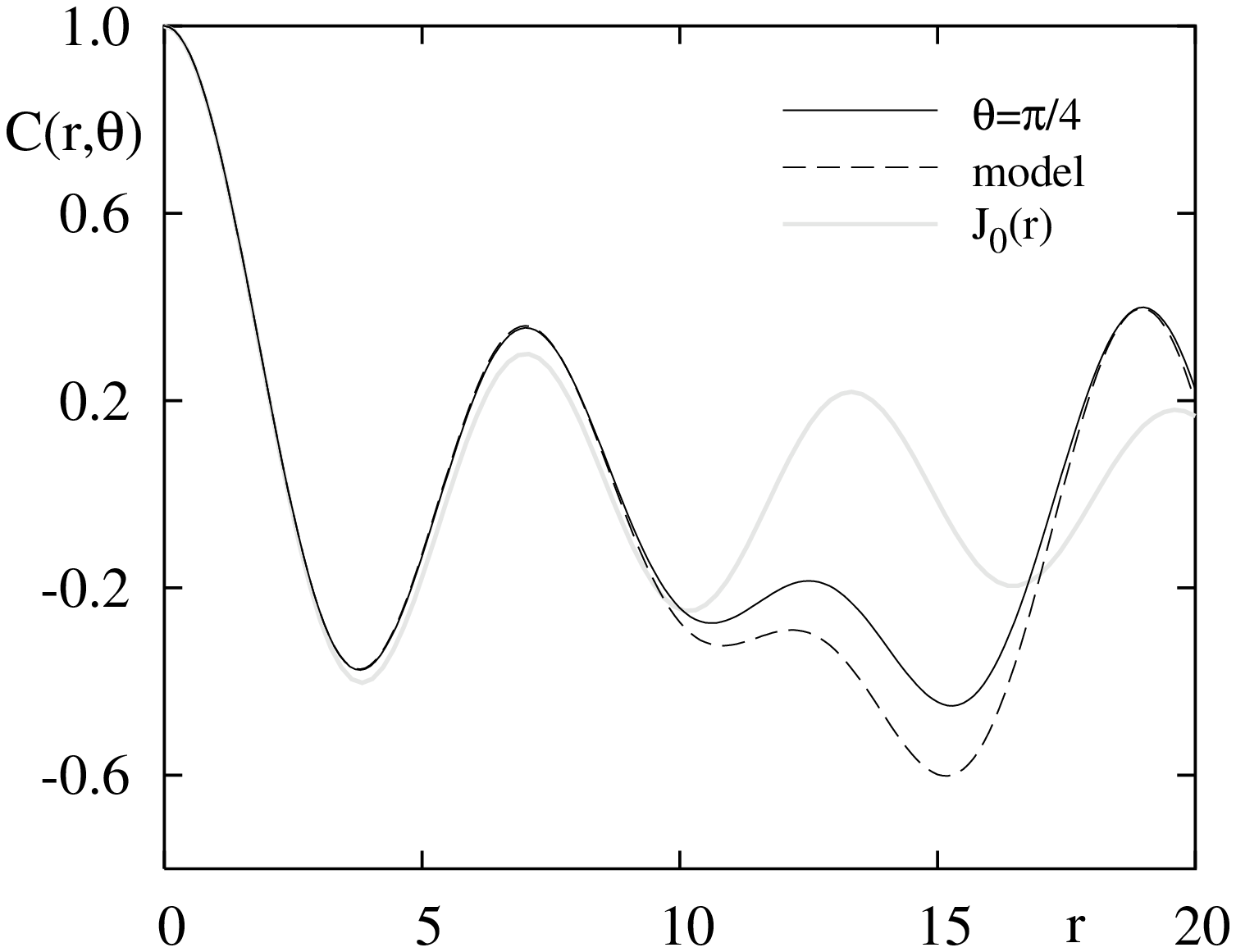}{8cm}
\PSImagx{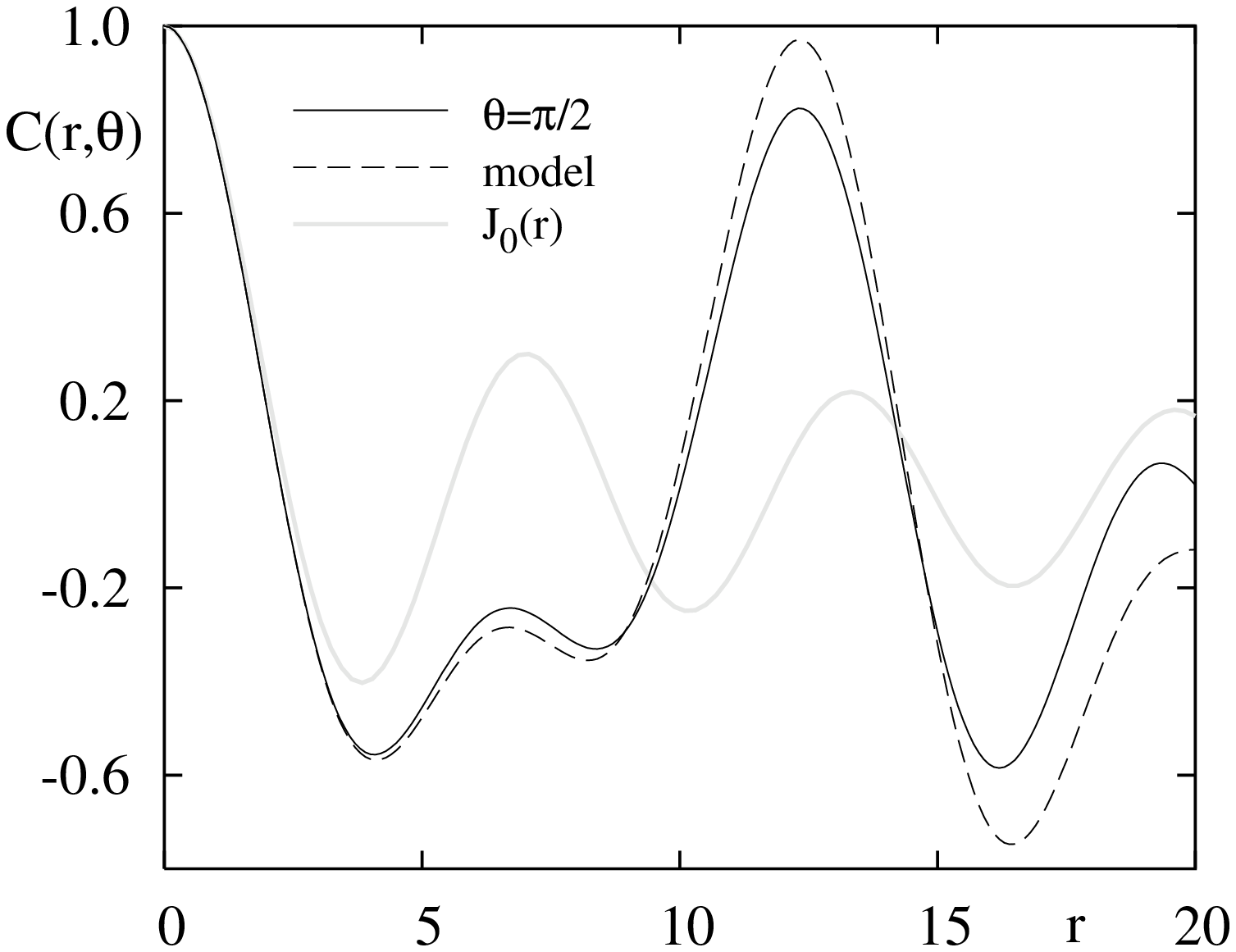}{8cm}

     \end{center}
     }
     {High lying eigenfunction 
      ($E=367984.82\ldots$, approx.\ 47788$^{\text{th}}$ eigenfunction
of odd symmetry) 
      in the \limacon billiard ($\varepsilon=0.3$),     
       which localizes on the
     stable orbit of triangular shape.
     The autocorrelation function for three different directions
     is compared with the 
     $\delta$-model, eq.~\eqref{eq:autocorr-limit-delta-model},
     shown as dashed line 
     using the directions of the stable orbit.
     }
     {fig:limacon-triangle}

\BILD{b}
     {
     \begin{center}

     \begin{minipage}{8cm}
       \begin{center}
         \PSImagx{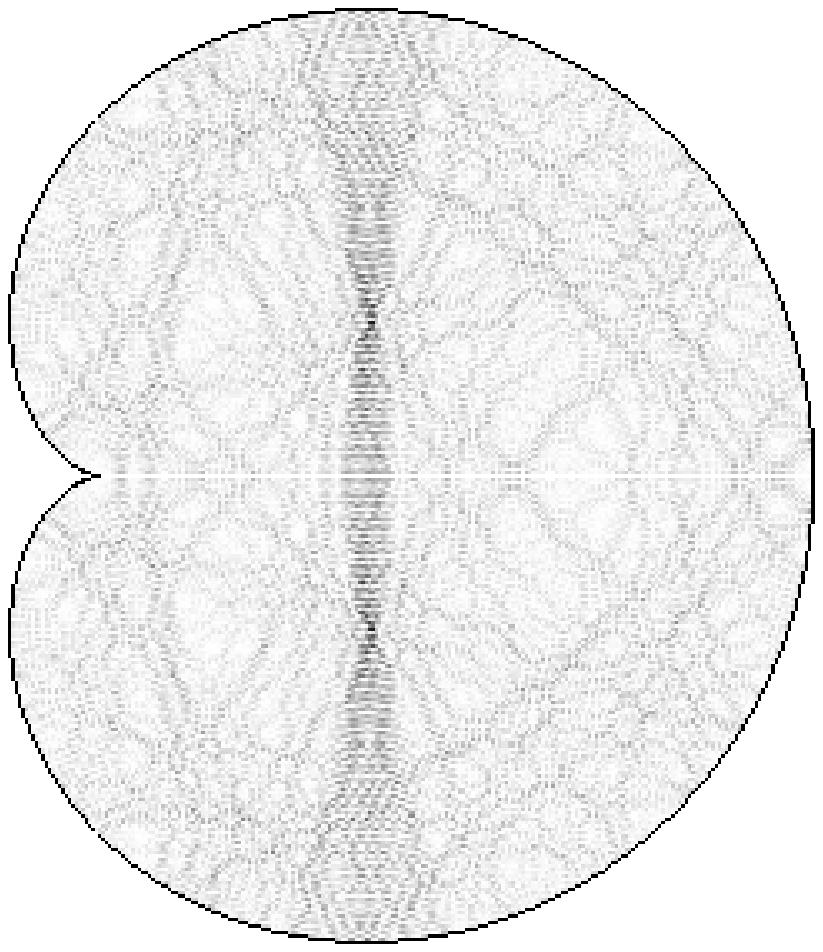}{4.5cm}
       \end{center}
     \end{minipage}
     \begin{minipage}{8cm}
        \PSImagx{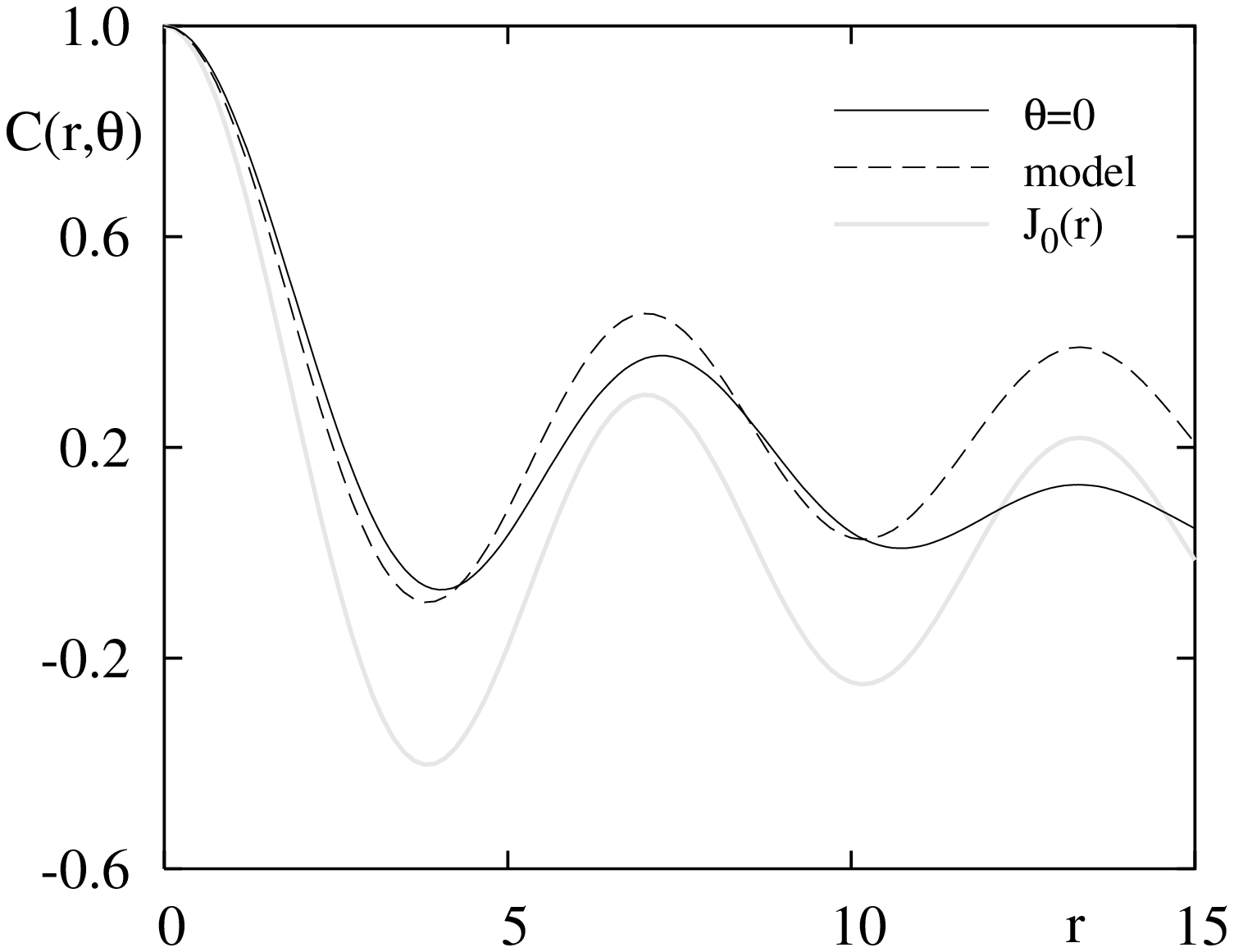}{8cm}
     \end{minipage}
                                                                       
      \PSImagx{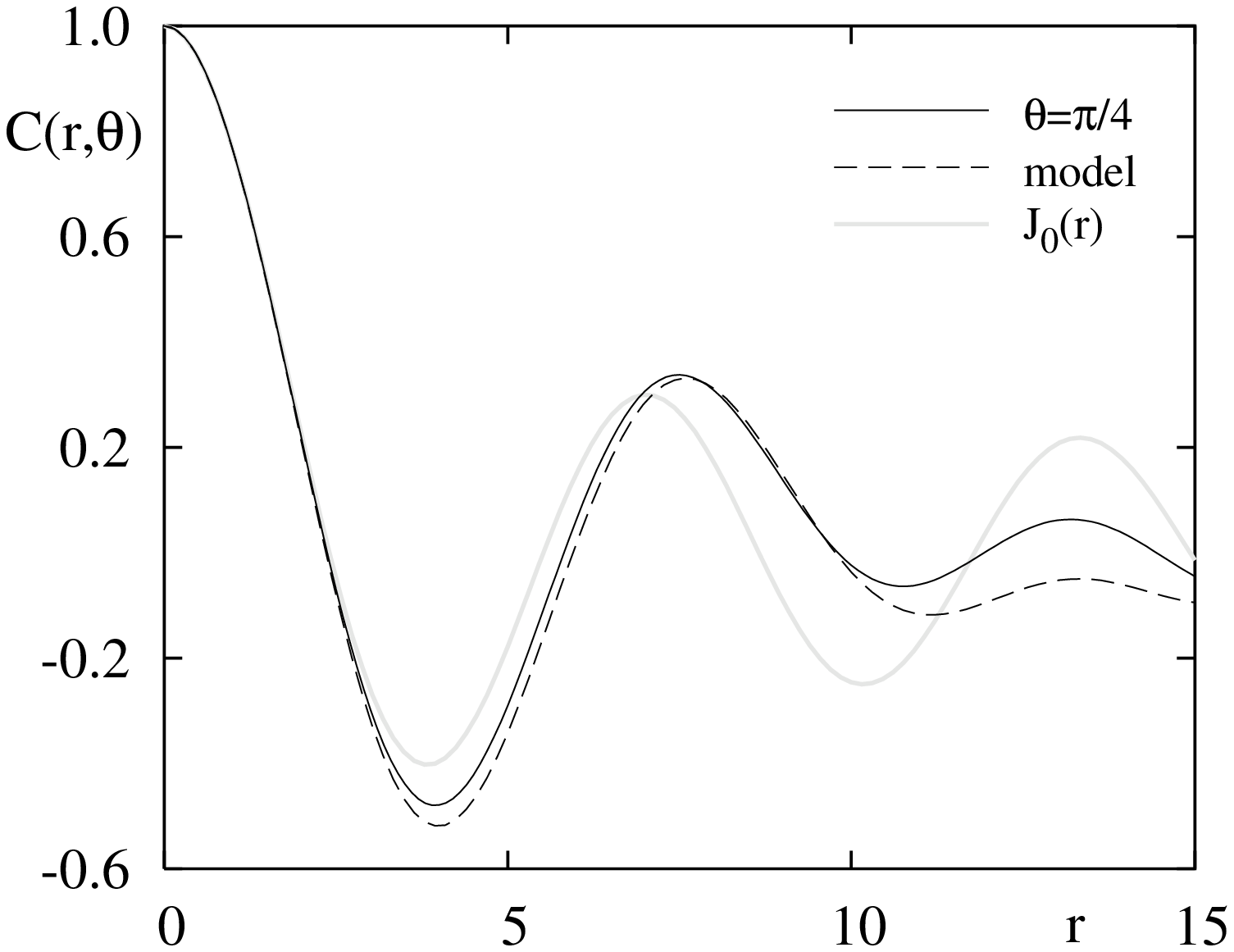}{8cm}
      \PSImagx{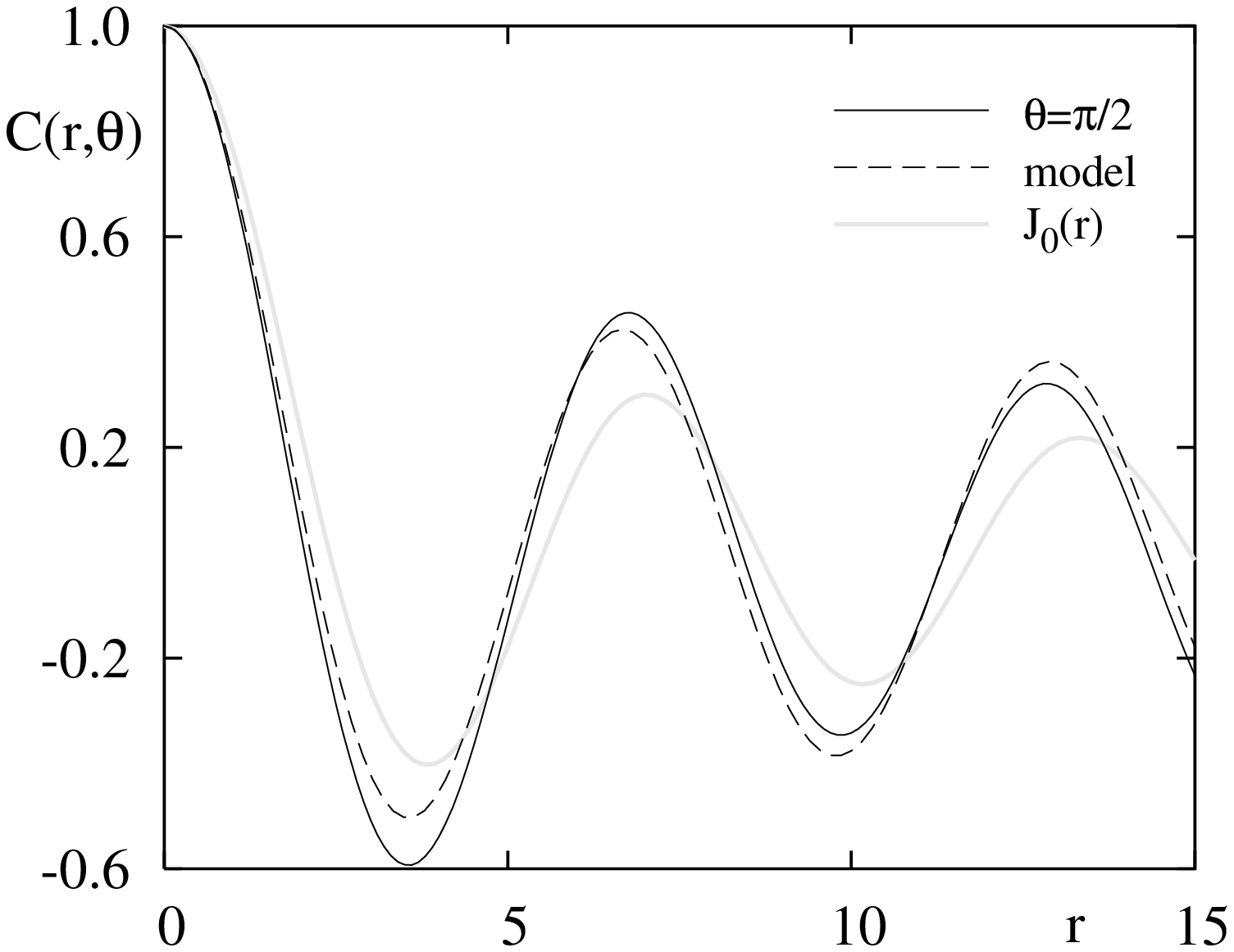}{8cm}

     \end{center}
     }
     {For a scarred state ($\psi_{7147}$ of odd symmetry)
      in the cardioid billiard the autocorrelation function
      is compared with the simple model \eqref{eq:scar-model-with-background}
      for $\alpha=0.22$. 
     }
     {fig:scar-cardioid}

For a state strongly localized on an periodic orbit of length $l_\gamma$
we have
(either in the semiclassical limit, or as a crude model at finite
energies)
\begin{equation}\label{eq:ellipt_I_model}
  I(\varphi)\sim \frac{1}{l_\gamma} 
     \sum l_{\gamma_i} \delta (\varphi-\varphi_i) \;\;,
\end{equation}
where $l_{\gamma_i}$ are the lengths of the segments
of the orbit with direction $\varphi_i$.
Thus we get
\begin{equation}
  a_{2l}=\frac{1}{\pi l_\gamma} \sum l_{\gamma_i} \cos(2 l \varphi_i)
  \qquad  \qquad
  b_{2l}=\frac{1}{\pi l_\gamma} \sum l_{\gamma_i} \sin(2 l \varphi_i) \;\;,
\end{equation}
which therefore  
using \eqref{eq:mean_auto_corr_expansion} gives
a prediction for $C(\dort)$ for such states, namely
\begin{equation} \label{eq:autocorr-limit-delta-model}
  C(\dort)\sim\frac{1}{l_{\gamma}}\sum_i l_{\gamma_i} 
   \cos(|\dort| \cos(\theta-\varphi_i))\;\; .
\end{equation}
Notice that in the presence of symmetries all
symmetry related directions have to be taken into account in
eq.~\eqref{eq:ellipt_I_model}.
For this simple model  
one can determine the autocorrelation function more directly by 
using \eqref{eq:corr-Wigner}
\begin{equation}
\begin{split}
  C(\dort)=\iint W(\BFp,\BFq)\ue^{\ui \BFp\dort}\; \ud p\,\ud q
  &=\Int |\widehat{\psi}(\BFp)|^2 \ue^{\ui \BFp\dort}\; \ud p\\
  &= \Int_0^{2\pi} I(\varphi)\cos(|\dort| \cos(\theta-\varphi))
  \; \ud \varphi+O(E^{-1/2})\;\; ;
\end{split}
\end{equation}
inserting \eqref{eq:ellipt_I_model} 
directly gives \eqref{eq:autocorr-limit-delta-model}.

\newcommand{\einbildabcd}[3]{
\begin{minipage}{#2.5cm}
  \begin{minipage}[t]{0.2cm}
  #3)

  \end{minipage}
  \begin{minipage}[t]{#2cm}
~

 \vspace*{-3ex}

    \PSImagx{#1}{#2cm}
  \end{minipage}
\end{minipage}
}

\BILD{t}
     {
     \begin{center}
       
        \vspace*{1cm}

        \einbildabcd{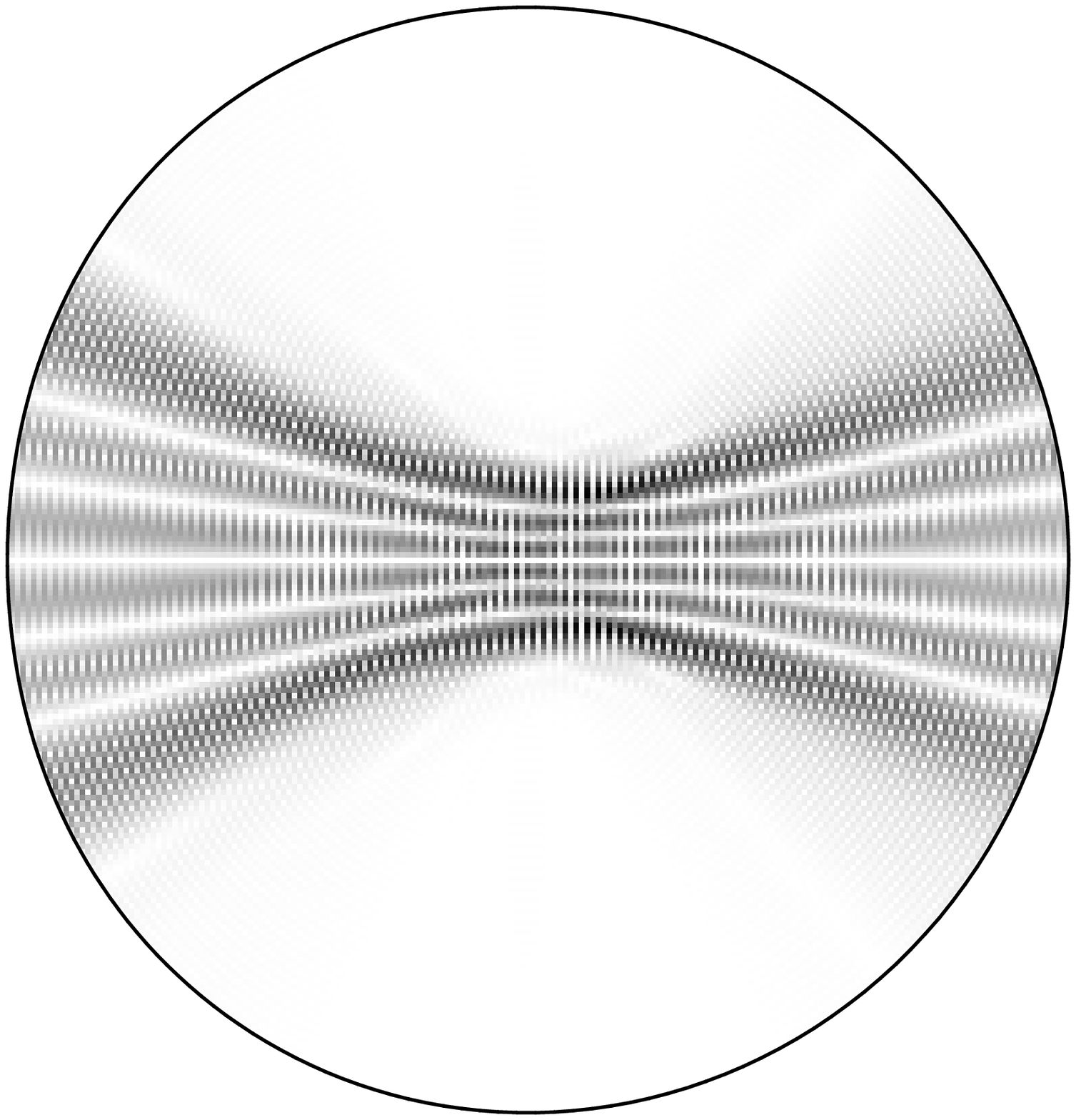}{5}{a}
        \einbildabcd{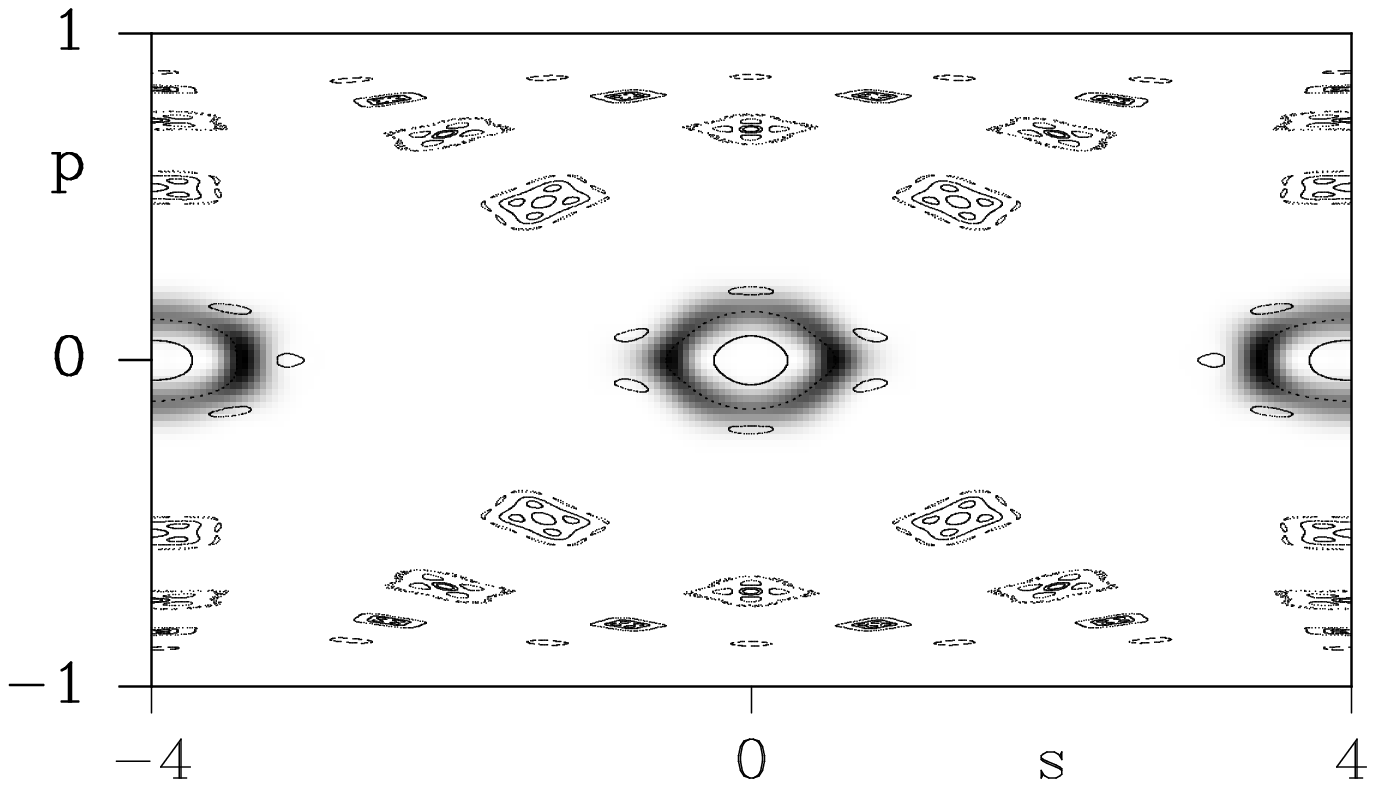}{10}{b}

        \vspace*{1cm}
 
        \hspace*{-0.5cm}\einbildabcd{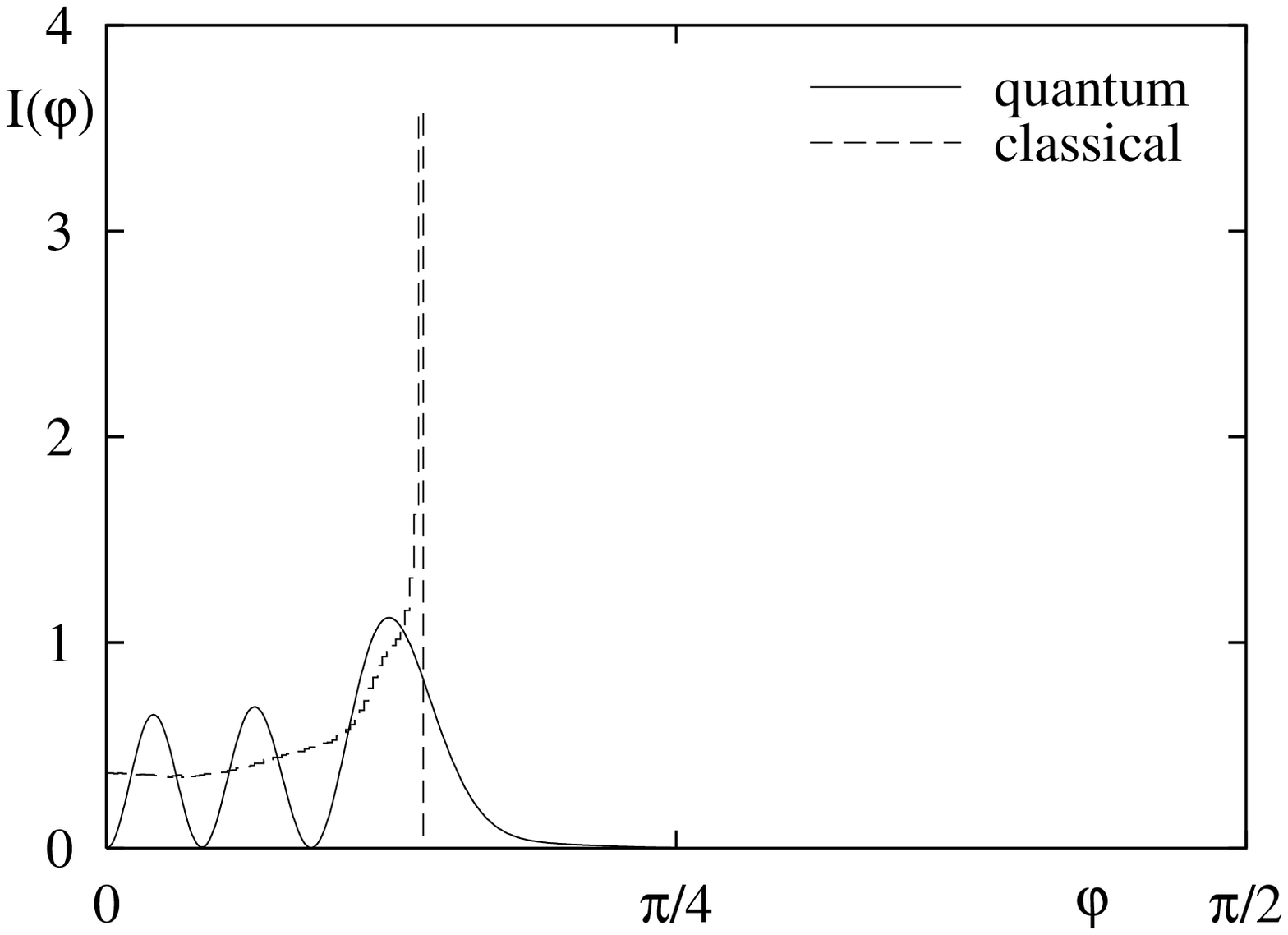}{8}{c}
        \hspace*{-0.25cm}\einbildabcd{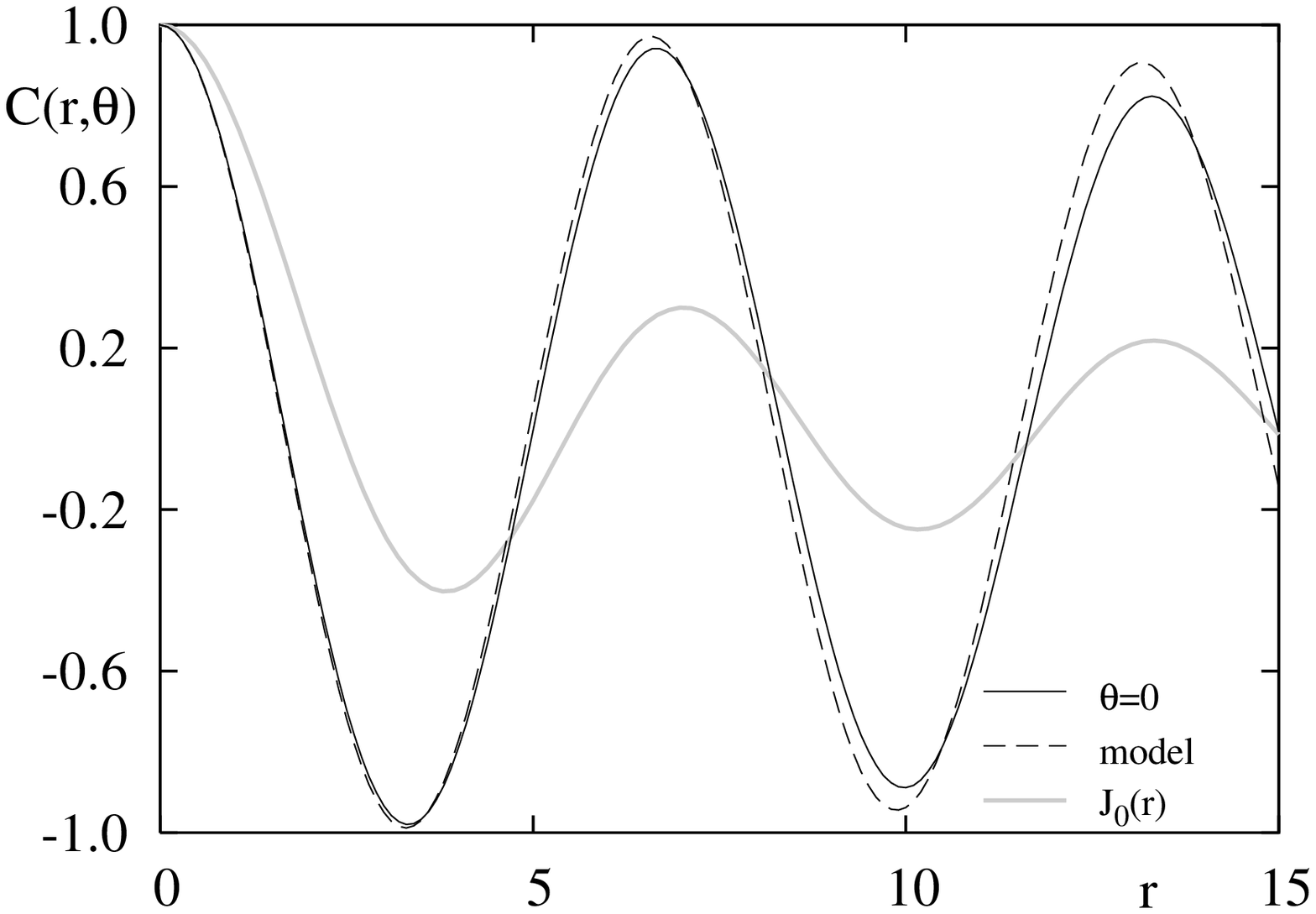}{8}{d}

        \vspace*{1cm}

        \hspace*{-0.5cm}\einbildabcd{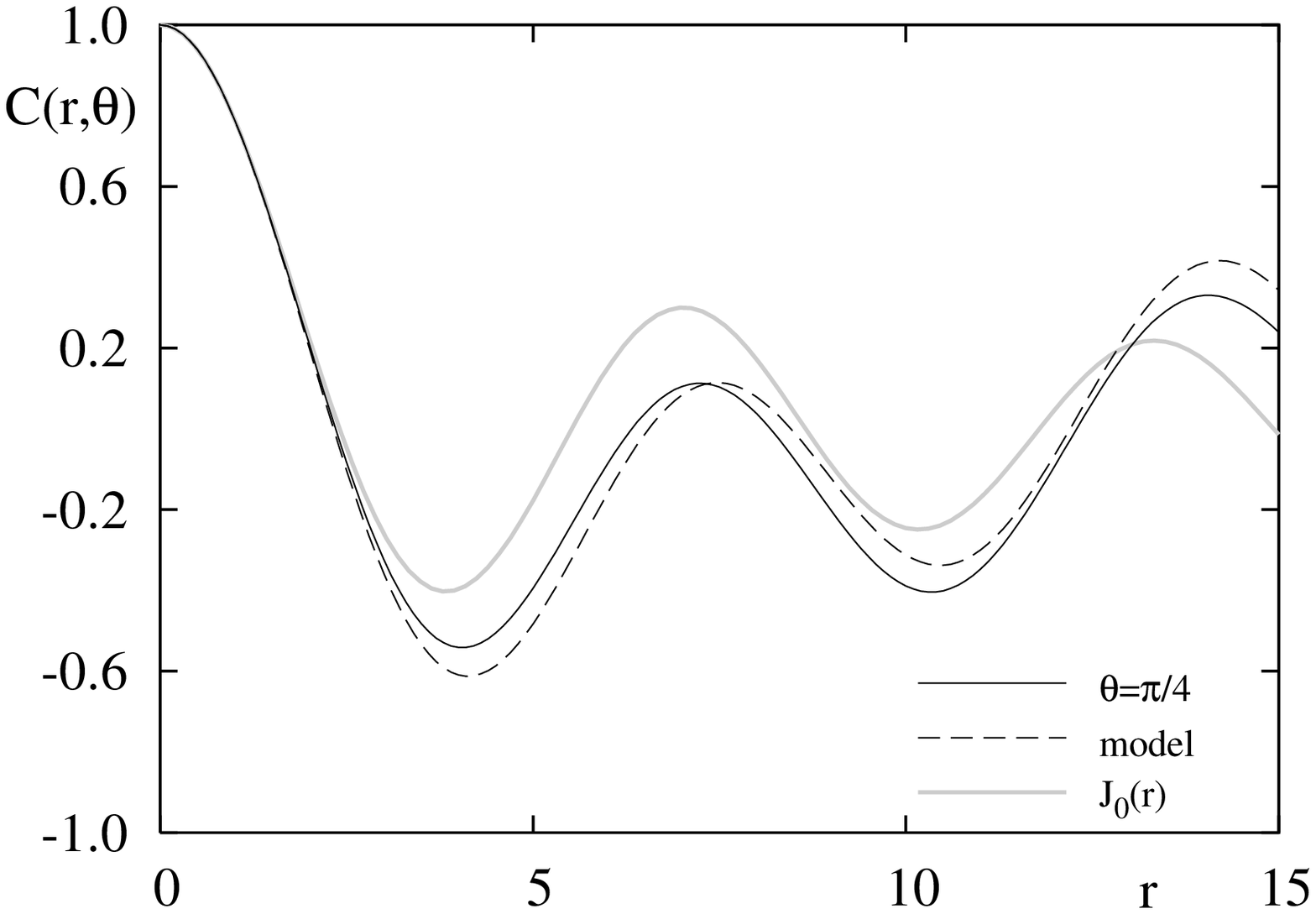}{8}{e}
        \hspace*{-0.25cm}\einbildabcd{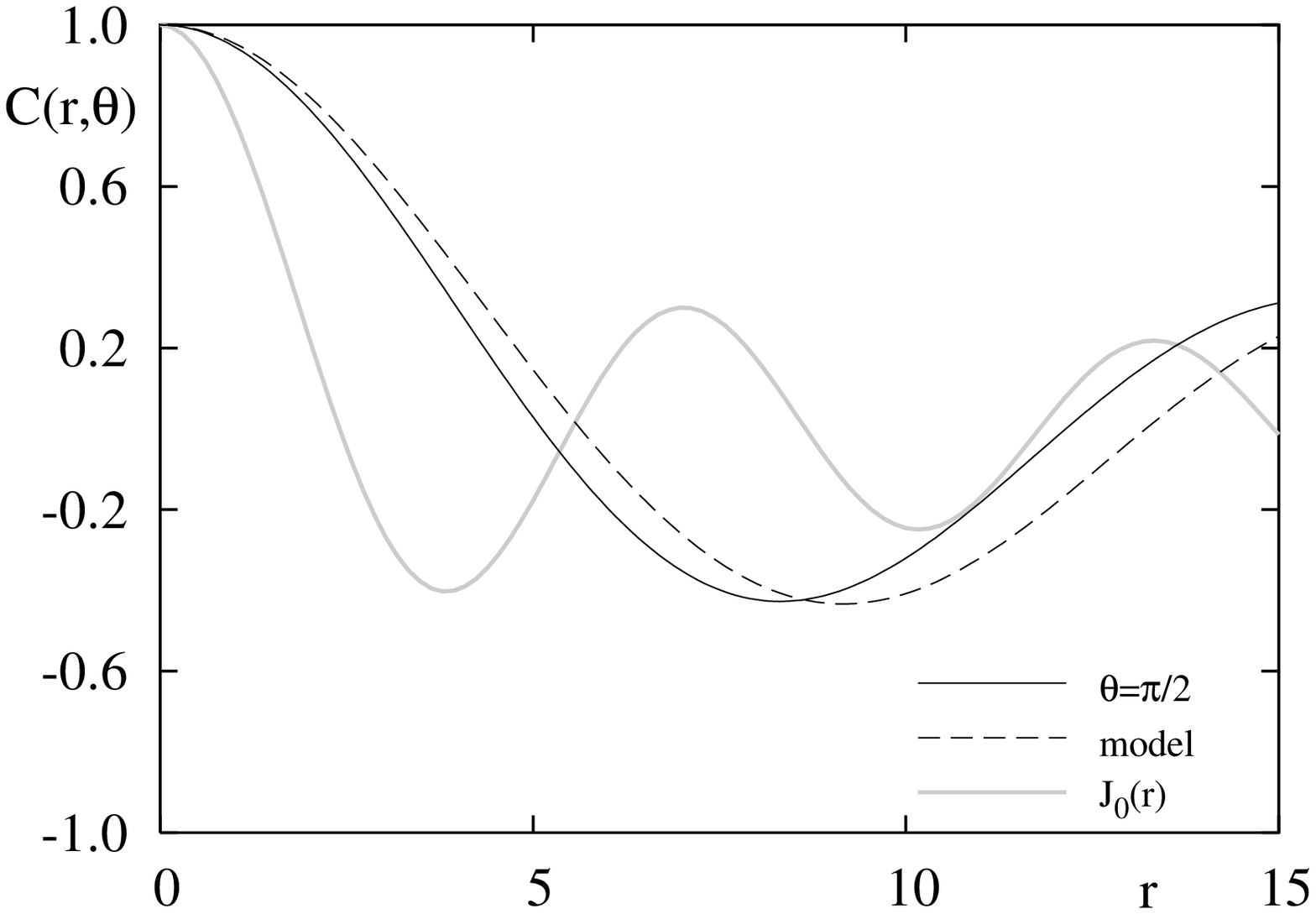}{8}{f}

     \end{center}
     }
     {Grey scale plot of $\psi_{3056}$ for the \limacon billiard with
      $\varepsilon=0.3$
      together with the corresponding Husimi plot, 
      for which in addition some orbits are shown.
      In c) the radially integrated momentum distribution 
      $I_{3056}(\varphi)$ and the corresponding classical distribution 
      $I^{\text{classical}}(\varphi)$
      for the torus are shown.
      In d)--f) the exact autocorrelation function is compared
      with the expansion of the autocorrelation function, 
      eq.~\eqref{eq:mean_auto_corr_expansion}, 
      using $I^{\text{classical}}(\varphi)$ 
      for different angles $\theta$.      
     }
     {fig:torus-limacon}

In fig.~\ref{fig:limacon-triangle} we
compare the limiting behavior \eqref{eq:autocorr-limit-delta-model} 
with the autocorrelation
function of a high lying eigenstate in the \limacon billiard.
The state localizes on the (stable) orbit of triangular shape.
Up to $r\approx 10$ the agreement is very good;
for larger $r$ the autocorrelation function of the eigenstate
shows deviations from the asymptotic behavior.
Notice that the state has a much higher energy
than the other examples.
At lower energies the agreement is not as good,
because the region in phase space on which the state localizes
is broader. This in turn implies that
its corresponding radially integrated momentum distribution
$I(\varphi)$ also has broad peaks, which  is not accounted 
for properly by the ansatz \eqref{eq:ellipt_I_model}.
However, when considering states of this type with increasing
energies, a clear trend to the asymptotic result 
\eqref{eq:autocorr-limit-delta-model} is observed.

This simple model has also been tested for a scarred state
in the cardioid. However, the agreement
is limited to a qualitative description for up to $r\approx 2$.
This is understandable in view of the observation (see \cite[fig.~8a)]{BaeSch99})
that for a scarred state the radially integrated
momentum distribution $I(\varphi)$
shows quite large fluctuations, 
and also in the considered case
the direction $\varphi=\pi/2$ is not clearly pronounced.
As these fluctuations essentially correspond to the random 
``background'' fluctuations of the state, 
a simple ansatz to model this behavior is
\begin{equation} \label{eq:scar-model-with-background}
  C(r,\theta)=(1-\alpha) J_0(r)
     +\alpha \frac{1}{l_{\gamma}}\sum_i l_{\gamma_i} 
             \cos(|\dort| \cos(\theta-\varphi_i))\;\; .
\end{equation}
It turns out that one can vary $\alpha$ such
that quite good agreement of this model with the exact
autocorrelation function is obtained, see fig.~\ref{fig:scar-cardioid}
where $\alpha=0.22$ (for all directions). Depending on the direction $\theta$
the ``optimal'' value for $\alpha$ does vary,
which already indicates the limitations of this simple model.
To get a better agreement a more precise description
of $I(\varphi)$ for scarred states is necessary.
In particular, this should also lead to an understanding of the
energy dependence of $\alpha$ which is expected
to go to zero in the semiclassical limit.
Notice, that the structure of the autocorrelation
function is quite similar to the one for $\psi_{1817}$ shown in 
fig.~\ref{fig:cardi-1817}.

Another case, for which we obtain much better agreement,
is for an eigenfunction localized on an invariant torus.
In such a case  
the expectation values, eqs.~\eqref{eq:QMexpA}, \eqref{eq:QMexpB},
tend to the mean of the classical observable 
over the torus, see eqs.~\eqref{eq:auto-correl-limit}, 
\eqref{eq:classical-exp}.
Fig.~\ref{fig:torus-limacon}a) shows for the \limacon billiard
the eigenfunction and the corresponding 
Husimi Poincar\'e representation \cite{TuaVor95,SimVerSar97};
see \cite{BaeSch2001b:p} for a more detailed discussion
and the formula which has beens used.
Also shown in the Husimi plot are the points of
some orbits.
Using an initial condition on the torus
we can determine the classical  
angular distribution $I^{\text{classical}}(\varphi)$.
As this has a singularity due to the caustic of the torus 
we show in fig.~\ref{fig:torus-limacon}c)
a binned distribution together with 
the corresponding quantum radially integrated momentum distribution
$I_{3056}(\varphi)$. There is qualitative agreement
between these two curves in the sense that smoothing 
$I^{\text{classical}}(\varphi)$ describes the mean behavior of the quantum
$I_{3056}(\varphi)$. 
Of course, the classical distribution cannot
describe the (quantum) oscillations visible for $I_{3056}(\varphi)$.
It turns out, see fig.~\ref{fig:torus-limacon}d)--f), 
that already this simple model
leads to surprisingly good agreement between the
exact autocorrelation function and the expansion 
\eqref{eq:mean_auto_corr_expansion} computed using 
$I^{\text{classical}}(\varphi)$.

\FloatBarrier
%%%%%%%%%%%%%%%%%%%%%%%%%%%%%%%%%%%%%%%%%%%%%%%%%%%%%%%%%%%%%%%%%%%%%%%%%%%%%%%
\subsection{Autocorrelation function of irregular states in mixed systems}
%%%%%%%%%%%%%%%%%%%%%%%%%%%%%%%%%%%%%%%%%%%%%%%%%%%%%%%%%%%%%%%%%%%%%%%%%%%%%%%

In classical systems with mixed phase space regions with regular and regions 
with stochastic behavior coexist. It is conjectured \cite{Per73} that 
correspondingly the  
quantum mechanical eigenfunctions split into 
regular and irregular ones, respectively, 
living semiclassically on the correponding parts of 
phase space. This has been confirmed numerically for several systems,
see e.g.\ \cite{BohTomUll90a,ProRob93b,LiRob95b,LiRob95,CarVerFen98}.
Consider now a sequence of eigenfunction $\psi_{n_j}$ which 
localize on some open 
ergodic domain $D$ in a system with mixed phase space, then 
almost all 
the expectation values $\langle\psi_{n_j}\widehat{A}\psi_{n_j}\rangle$ 
tend to the mean 
$\overline{A}^D$ of the corresponding classical observable
$A$ over this domain $D$ \cite{Sch2001:PhD}.
Therefore using \eqref{eq:mean_auto_corr_expansion} 
we get in the limit 
$E\to\infty$ for the autocorrelation function of such a sequence 
\begin{equation} \label{eq:autocorrel-limit-ergodic-subdomain}
 \rC_{\rho}^{\text{limit}}(\ort,\dort)=\overline{A_0}^D J_0(|\dort|)
  +2\sum_{l=1}^{\infty} (-1)^l 
  \bigl[\,\overline{A_{2l}}^D(\BFx) \cos(2l \theta)+
        \overline{B_{2l}}^D (\BFx) \sin(2l \theta)\bigr]\, 
J_{2l}(|\dort|)\;\; .
\end{equation}
Instead of computing $\overline{A_{2l}}^D$  and $\overline{B_{2l}}^D$
directly, we can also use a typical trajectory of the ergodic
component to determine the corresponding classical 
$I^{\text{classical}}(\varphi)$ via
\begin{equation} \label{eq:def-Iphi-classical}
  I^{\text{classical}}(\varphi) = 
      \lim_{l\to\infty} \frac{1}{l} \sum l_i \delta(\varphi-\varphi_i)
   \;\;,
\end{equation}
where $l$ is the total length of the trajectory and 
$\varphi_i$ is the direction of the $i$--th segment having length $l_i$.
Then we use \eqref{eq:mean_auto_corr_expansion} to get a prediction
for the autocorrelation function.

\renewcommand{\einbildabcd}[3]{
\begin{minipage}{#2.5cm}
  \begin{minipage}[t]{0.3cm}
  #3)

  \end{minipage}
  \begin{minipage}[t]{#2cm}
~

 \vspace*{-3ex}

    \PSImagx{#1}{#2cm}
  \end{minipage}
\end{minipage}
}

\BILD{tbh}
     {
     \begin{center}

         \hspace*{-0.25cm}\einbildabcd{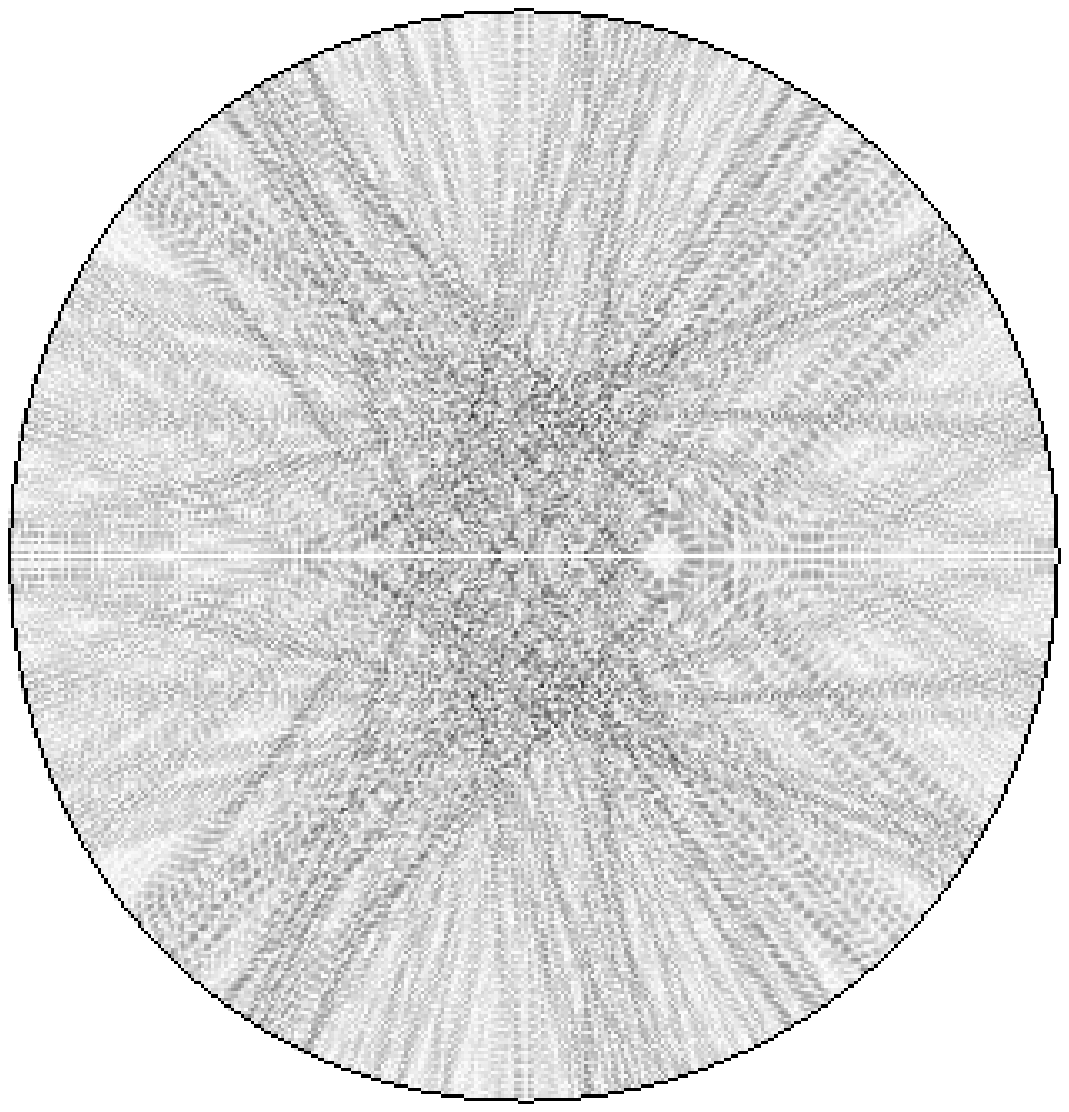}{5}{a}
        \hspace*{0.5cm}\einbildabcd{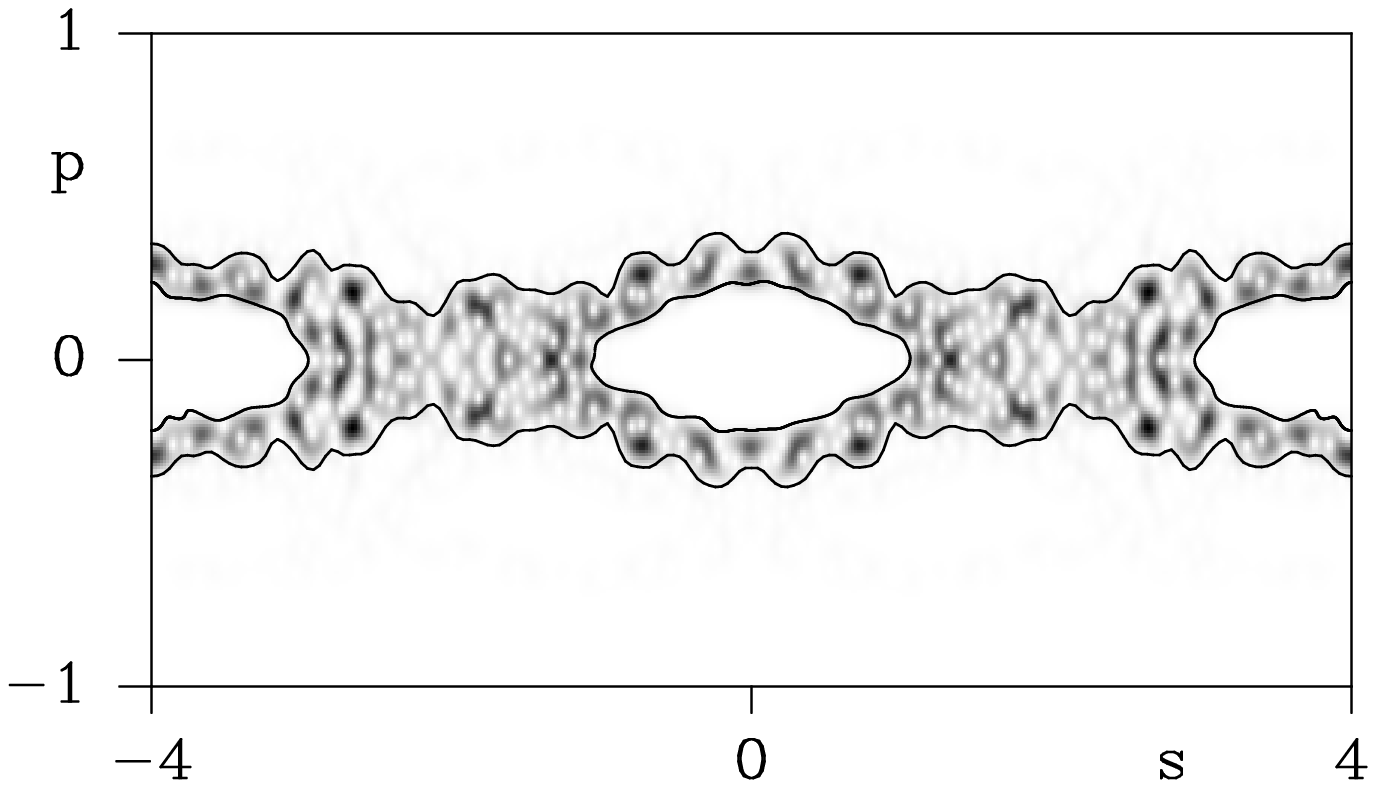}{10}{b}

\vspace*{0.5cm}

        \hspace*{-0.25cm}\einbildabcd{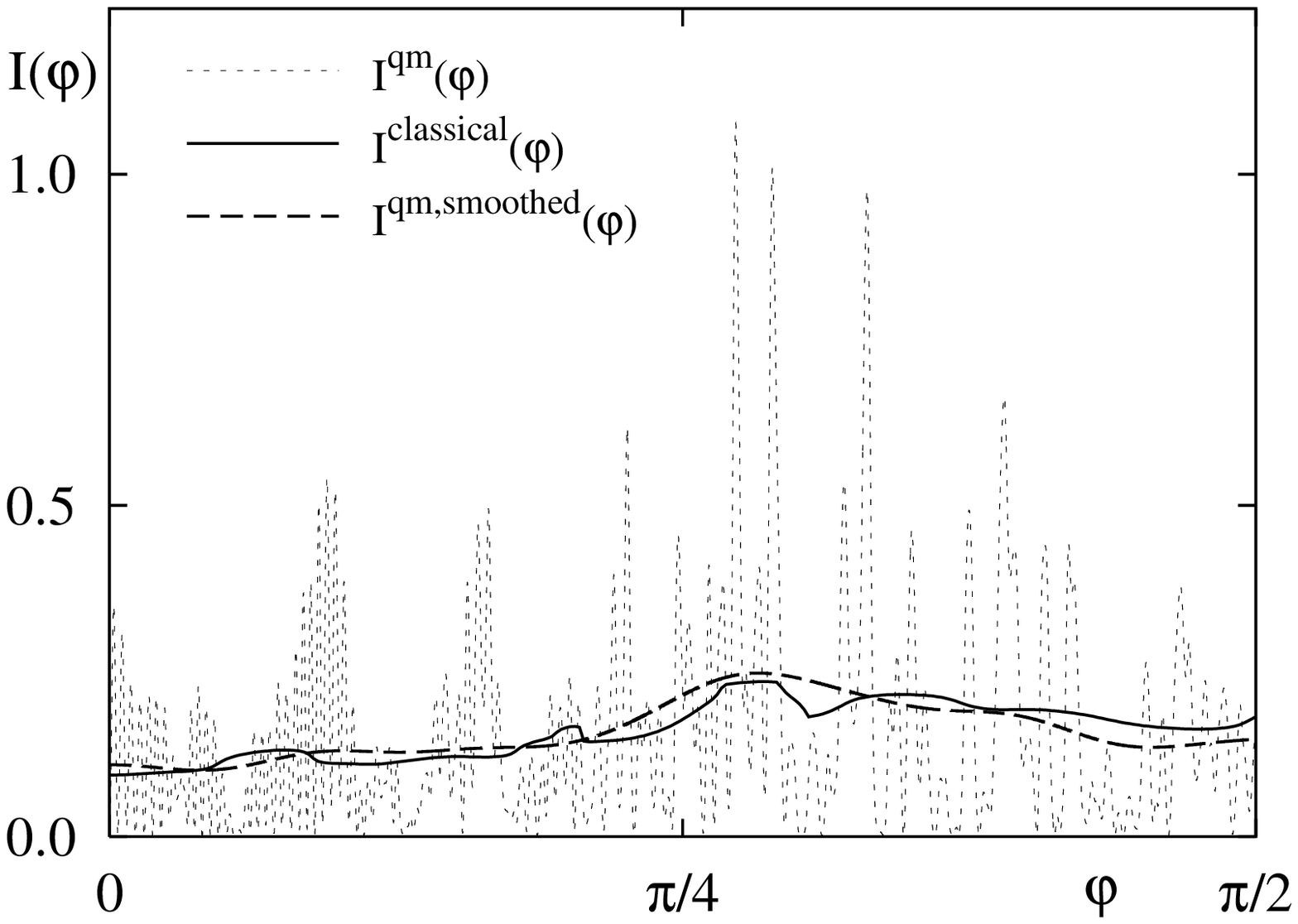}{8}{c}
        \hspace*{-0.5cm}\einbildabcd{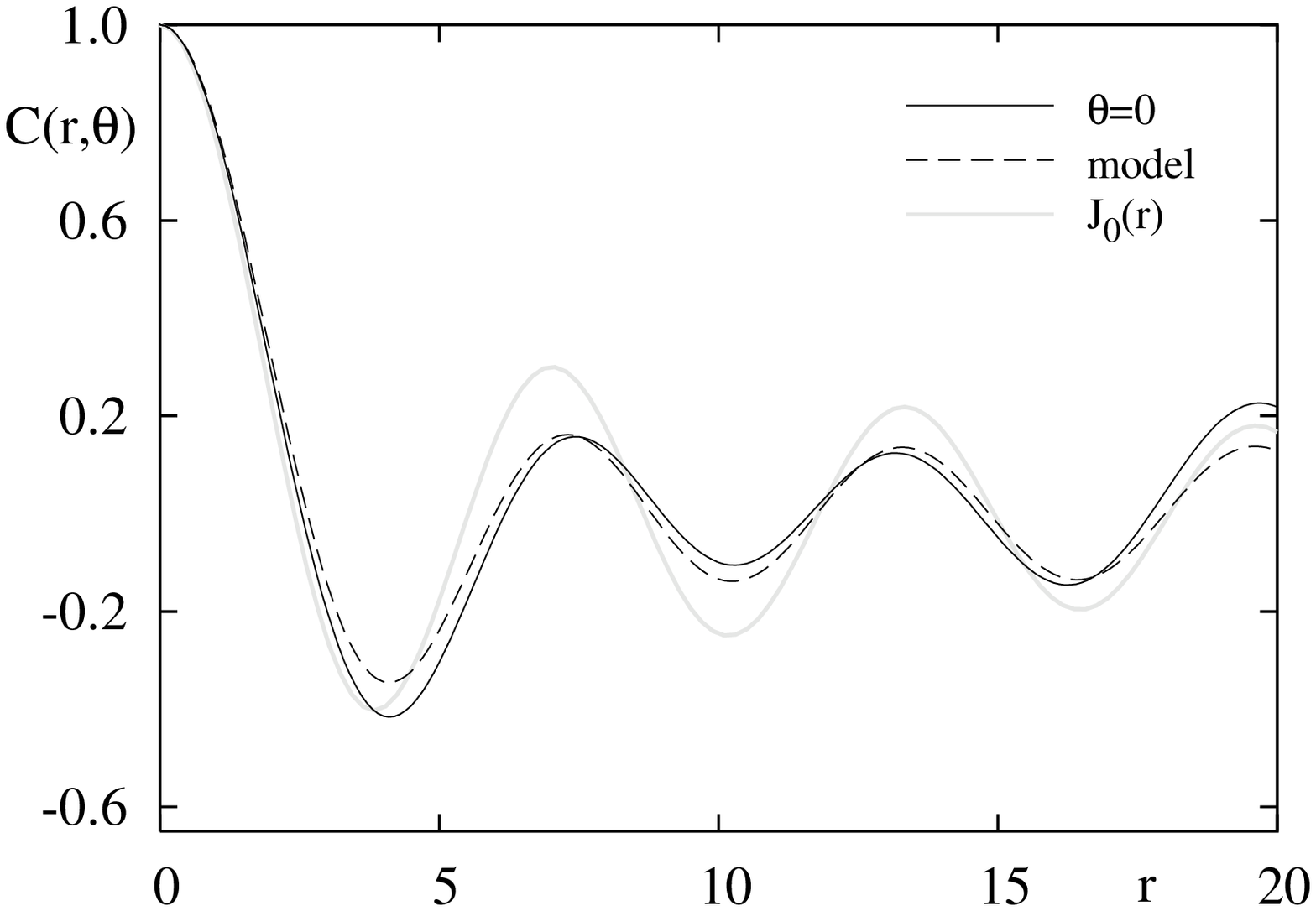}{8}{d}
                                     
\vspace*{0.5cm}

        \hspace*{-0.25cm}\einbildabcd{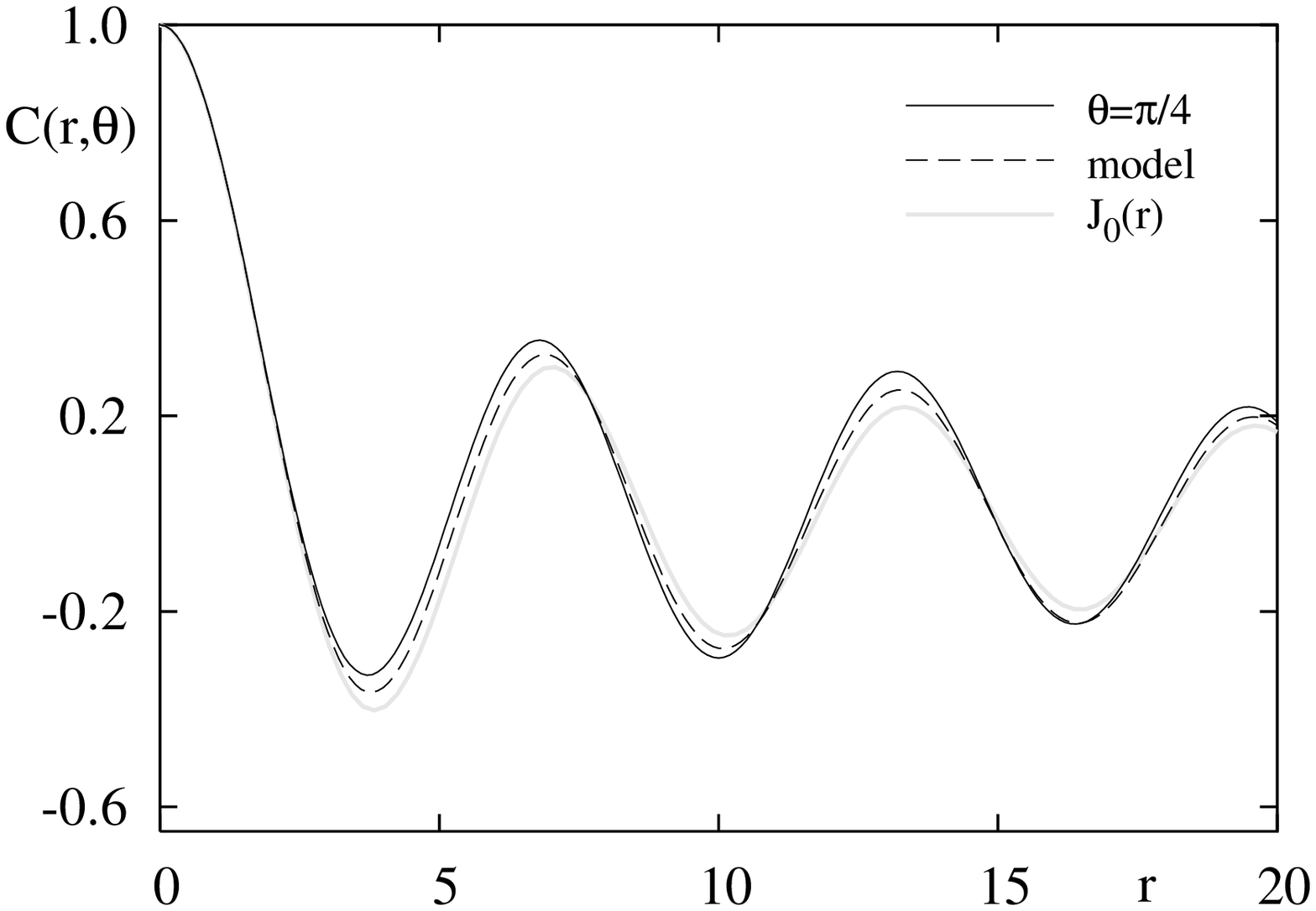}{8}{e}
        \hspace*{-0.5cm}\einbildabcd{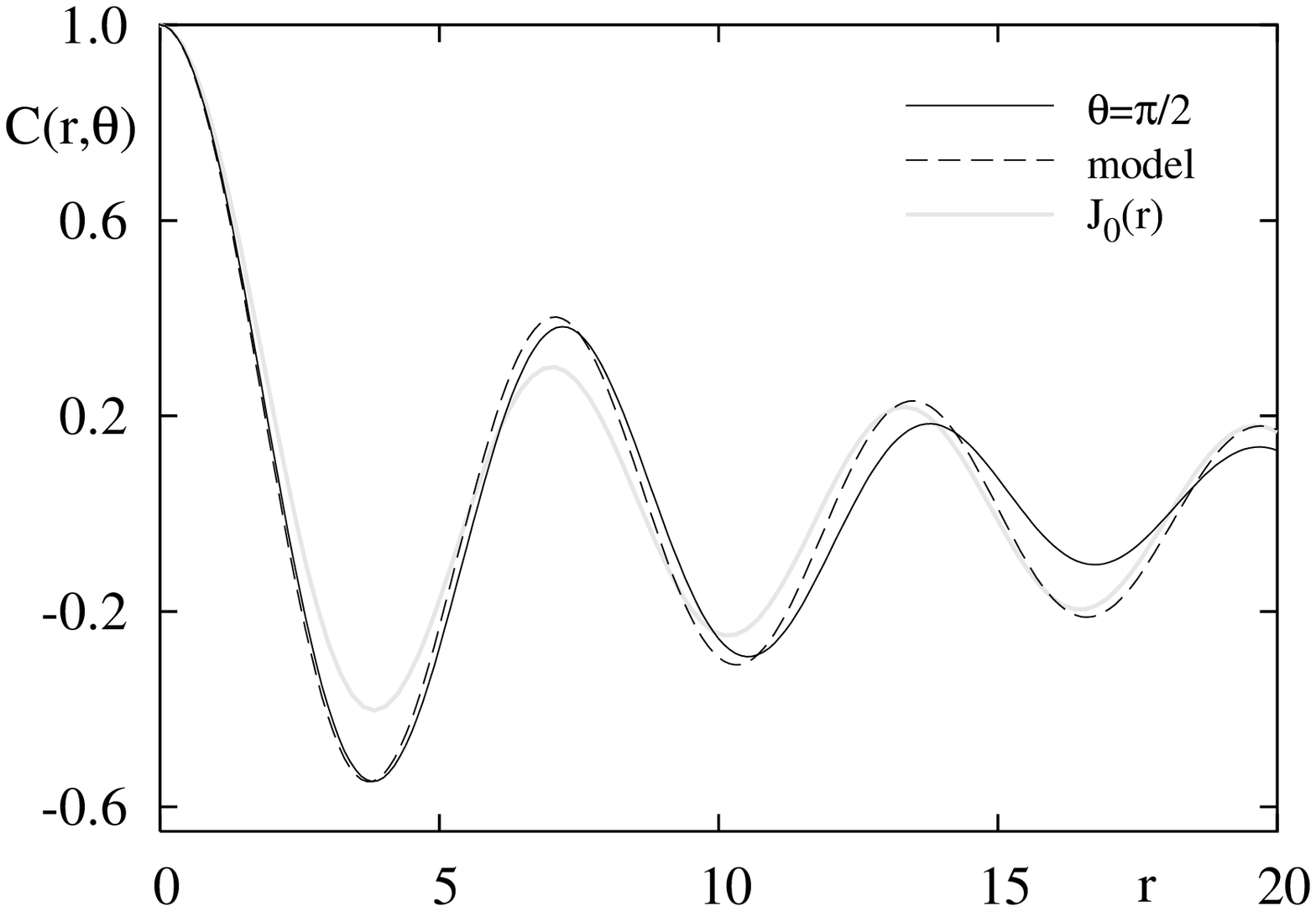}{8}{f}

            \end{center}
     }
     {Autocorrelation function for a high lying irregular
      state  ($E=1002754.70\ldots$, approx.\ 130516$^{\text{th}}$ eigenfunction
      of odd symmetry) in the \limacon billiard with
      $\varepsilon=0.3$.
      In b) the Husimi representation on the boundary is shown
      together with an approximate boundary (full line) of
      the region $\cD$ on which the state localizes.
      The resulting classical momentum distribution 
      $I^{\text{classical}}(\varphi)$
      is shown in c) as full line and compared with the
      radially integrated momentum distribution $I^{\text{qm}}(\varphi)$ of
      the state in a) and a smoothing of this,  
      $I^{\text{qm,smoothed}}(\varphi)$,
      shown as dashed line.
      In d)--f) the autocorrelation function $C(r,\theta)$ 
      of the eigenfunction is compared
      for three different directions with result of the expansion
      \eqref{eq:autocorrel-limit-ergodic-subdomain} using
      $I^{\text{classical}}(\varphi)$.
     }
     {fig:autocorrel-mixed-ergodic-state}

However, we observe that even quite high lying states
do not yet localize on the whole chaotic component.
Instead they are confined to smaller subregions 
due to partial barriers in phase space.
Fig.~\ref{fig:autocorrel-mixed-ergodic-state}a)
shows an example of a high lying state in the \limacon billiard
($\varepsilon=0.3$)
In fig.~\ref{fig:autocorrel-mixed-ergodic-state}b)
the corresponding Husimi function is plotted, which clearly shows
the localization on a chaotic subdomain (the whole irregular region
is much larger).
If $D$ is a open region in phase space, then
the corresponding classical  distribution of the momentum directions 
is given by
\begin{equation}
  I^{\text{classical}}(\varphi) = \frac{1}{\Vol(D)} \int 
      \chi_D(\BFp(\varphi),\BFq )\; \ud q\;\;,
\end{equation}
where $\BFp(\varphi)=(\cos \varphi,\sin \varphi).$
One can show that in terms of 
the projection $\cD$ of $D$ on the Poincar\'e section
this equation can be reduced to 
\begin{align}
   I^{\text{classical}}(\varphi) &= 
      \frac{\Int_{\cD} l(s,p) \delta(\varphi-\phi(s,p)) \; \ud s \,\ud p}
           {\Int_{\cD} l(s,p)  \; \ud s \,\ud p} \label{eq:Iphi-with-delta}\\
    &=   \frac{\Int_{\partial \Omega'(\varphi)} 
              l(s,p(s,\varphi)) \sqrt{1-p^2(s,\varphi)} 
               \chi_{\cD}(s,p(s,\varphi)) \; \ud s }
           {\Int_{\cD} l(s,p)  \; \ud s \,\ud p} \label{eq:Iphi}
  \;\;,
\end{align} 
where  $l(s,p)$
is the length of the orbit segment starting in the point $(s,p)\in\cP$
with direction $\phi(s,p)$ and in the second equation
$p(s,\varphi) = \BFp(\varphi) \Bft(s)$, with $\Bft(s)$ denoting the unit 
tangent vector to $\partial D$ in the point $s$. 
Furthermore $\partial \Omega'(\varphi):=\{s\in \partial \Omega \;|\; 
\BFp(\varphi)\BFn(s)\leq 0\}$, where $\BFn(s)$ denotes the outer normal vector 
to $\partial D$ in the point $s$, is the subset of $\partial \Omega$ where 
the vector $\BFp(\varphi)$ points inwards. 
For the numerical computation we have used \eqref{eq:Iphi}
because we just have to deal with a one-dimensional integral
to compute the $\varphi$ dependence, and also 
compared to \eqref{eq:Iphi-with-delta} no binning
of $I^{\text{classical}}(\varphi)$ is necessary.

After these general remarks on the computation of 
$I^{\text{classical}}(\varphi)$ let us describe how
we compute the relevant quantities to determine the
autocorrelation function for the state shown in 
fig.~\ref{fig:autocorrel-mixed-ergodic-state}a).
To describe the projection $\cD$ of the domain $D$ in phase space,
we use an approximation of the boundary of $\cD$ by a splines,
which are shown in the fig.~\ref{fig:autocorrel-mixed-ergodic-state}b)
as full curves. Then we use eq.~\eqref{eq:Iphi} to determine
the corresponding $I^{\text{classical}}(\varphi)$,
shown in c) as full curve.
Of course the radially integrated momentum distribution
$I^{\text{qm}}(\varphi)$ of the eigenstate shows strong
fluctuations, but the smoothing $I^{\text{qm,smoothed}}(\varphi)$
is well described by $I^{\text{classical}}(\varphi)$,
although the agreement is not perfect.
Using $I^{\text{classical}}(\varphi)$ we
employ the expansion \eqref{eq:autocorrel-limit-ergodic-subdomain}
to get a prediction for the autocorrelation function
for states localizing on $\cD$, which is compared in 
figs.~\ref{fig:autocorrel-mixed-ergodic-state}d--f)
with the exact autocorrelation function.
Up to $r\approx 10$ we get quite good agreement, whereas
for larger $r$ deviations become more visible.
This shows, that the eigenfunction has more structure
than accounted for by $I^{\text{classical}}(\varphi)$, i.e.\
it is not yet far enough in the semiclassical limit.

For higher energies the states tend to localize
on the full ergodic region, and then $I^{\text{classical}}(\varphi)$ 
can simply be computed using \eqref{eq:def-Iphi-classical}
by averaging of a typical trajectory in $D$.
One should emphasize that the agreement has to 
be compared with the agreement of the autocorrelation function
for ergodic systems with \eqref{eq:asymptotic-result-J-0}
as the prediction eq.~\eqref{eq:autocorrel-limit-ergodic-subdomain}
only takes into account the classical limit.
This has been studied in \cite{VebRobLiu99}
(in the case of averaging the local autocorrelation function
over a small disk), where in particular for
\cite[Fig.~13b]{VebRobLiu99} very good agreement has been found.

\FloatBarrier
%%%%%%%%%%%%%%%%%%%%%%%%%%%%%%%%%%%%%%%%%%%%%%%%%%%%%%%%%%%%%%%%%%%%%%%%%%%%%%%
\subsection{Ergodic systems and the rate of quantum ergodicity}
      \label{subsec:ergodic-systems}
%%%%%%%%%%%%%%%%%%%%%%%%%%%%%%%%%%%%%%%%%%%%%%%%%%%%%%%%%%%%%%%%%%%%%%%%%%%%%%%

If the classical system is ergodic and $\psi_{n_j}$ is a quantum 
ergodic sequence of eigenfunctions, then for $j\to\infty$
\begin{align}
  \langle\psi_{n_j},\widehat{A}_{2l}(\ort)\psi_{n_j}\rangle 
        &\sim \overline{A_{2l}}=\delta_{l0}\\
 \langle\psi_{n_j},\widehat{B}_{2l}(\ort)\psi_{n_j}\rangle 
        &\sim \overline{B_{2l}}=0 \;\;.
\end{align}
Thus using the expansion \eqref{eq:auto_corr_expansion}
we again get \eqref{eq:asymptotic-result-J-0} for $E\to\infty$.
Deviations from this universal behavior are then determined by the 
rate of quantum ergodicity. 
In order see this
it will be convenient to remove the angular dependence by  
taking the mean over all angular-directions in $\rC(\dort)$. 
Since by eq.~\eqref{eq:mean_auto_corr_expansion} 
\begin{equation}
  \frac{1}{2\pi}\Int_0^{2\pi} \rC_n(r,\theta)\; \ud \theta 
        =J_0(r)+O(E^{-1/2})\;\; ,
\end{equation}
we consider the second moment, 
\begin{equation} \label{eq:sigma2-autocorrel}
\VAR_n(r):=\
   \frac{1}{2\pi}\Int_0^{2\pi} [\rC_n(r,\theta)-J_0(r)]^2\; \ud \theta\;\; , 
\end{equation}
where $\rC_n(r,\theta)$ denotes the autocorrelation function of 
$\psi_n$.
Inserting the expansion \eqref{eq:mean_auto_corr_expansion} of 
$\rC_n(\dort)$ leads to 
\begin{equation} \label{eq:sigma2-autocorrel-expansion}
\VAR_n(r)= 2\pi^2
   \sum_{l=1}^{\infty}(a_{2l,n}^2+b_{2l,n}^2)
       [J_{2l}(r)]^2\,(1+O(E^{-1/2}))\;\; .
\end{equation}

\BILD{b}
     {
     \begin{center}
      \PSImagx{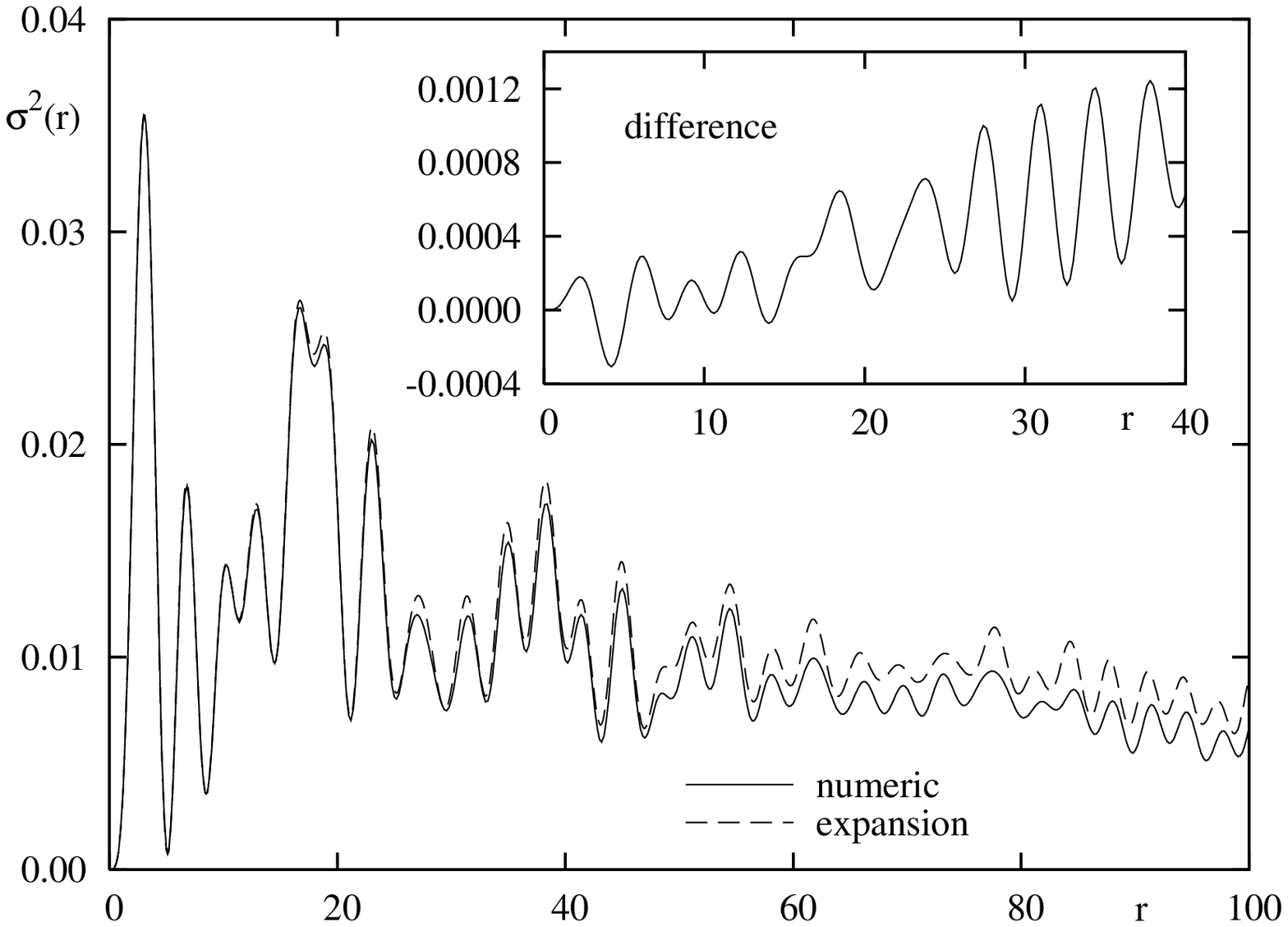}{15cm}
     \end{center}
     }
     {Comparison of the second moment $\VAR_{1907}(r)$
      of the autocorrelation function,
      eq.~\eqref{eq:sigma2-autocorrel},
      with the expansion \eqref{eq:sigma2-autocorrel-expansion}
      for the stadium billiard.
      The inset shows the difference for $r\in[0,20]$.
     }
     {fig:rate}

In fig.~\ref{fig:rate} we compare $\VAR(r)$ for
an eigenfunction in the stadium billiard with the expansion
\eqref{eq:sigma2-autocorrel-expansion}.
For small $r$ we get excellent agreement and some deviations become
visible in the plot for $r>20$.
The inset shows a plot of the difference up to $r=20$.
It is surprising that even though for large $r$ the amplitudes do not 
match anymore, still the expansion gives the right oscillatory structure.

If we take the mean of \eqref{eq:sigma2-autocorrel-expansion}
over all eigenfunctions up to energy $E$, we get
\begin{align}
 \barVAR(E,r)&:=\frac{1}{N(E)}\sum_{E_n\leq E} \VAR_n(r)  
   \label{eq:sum-sigma2-def}\\
 &=2\pi^2\sum_{l=1}^{\infty}\frac{1}{N(E)}
  \sum_{E_n\leq E}(a_{2l,n}^2+b_{2l,n}^2)\,\, [J_{2l}(r)]^2\,(1+O(E^{-1/2}))
\label{eq:sum-sigma2}
\end{align} 
Remarkably, together with eqs.~\eqref{eq:auto_corr_expansion} and 
\eqref{eq:sigma2-autocorrel-expansion} this shows that
the rate of quantum ergodicity can 
be studied in terms of the autocorrelation function.
Particularly interesting 
is that the observables in the expansion 
\eqref{eq:auto_corr_expansion} become more and more oscillatory with 
increasing $l$, so by varying $|\dort|$ one can determine 
the rate of quantum ergodicity on different length scales. 

A prediction for the behavior of $\barVAR(E,r)$
follows from \cite{EckFisKeaAgaMaiMue95}, 
where it is shown that (under suitable conditions on the system)
in the mean 
\begin{equation}\label{eq:indiv_rate}
\frac{1}{N(E)}\sum_{E_n\leq E}  
[\langle\psi_n,\widehat{A}\psi_n\rangle -\overline{A}]^2
      \sim \frac{4\VARcl (A)}{\Vol \Omega}\, 
  \frac{1}{\sqrt{E}}
\end{equation}
for any pseudodifferential operator $\widehat{A}$ of order zero with 
symbol $A$. Here $\overline{A}$ denotes the mean value of $A$, and 
$\VARcl (A)/\sqrt{T}$ is the variance of the fluctuations of 
\begin{equation}
  \frac{1}{T}\Int_0^T A(\BFp(t),\BFq(t))\;\; \ud t
\end{equation}
around $\overline{A}$. 
So if we insert  \eqref{eq:indiv_rate} into \eqref{eq:sum-sigma2} we obtain
\begin{equation} \label{eq:semiclassical-expectation-Sigma2}
\barVAR(E,r)\sim\frac{8\pi^2}{\Vol \Omega}
   \sum_{l=1}^{\infty}
    \bigl[\VARcl(A_{2l})+\VARcl(B_{2l})\bigr] \, 
       [J_{2l}(r)]^2\, \frac{1}{\sqrt{E}}\;\; .
\end{equation}

A detailed study of the rate of quantum ergodicity in terms
of the autocorrelation function, i.e.\ 
via eq.~\eqref{eq:sum-sigma2-def},
and a comparison with the semiclassical expectation
\eqref{eq:semiclassical-expectation-Sigma2}
will be given in a separate paper.

%%%%%%%%%%%%%%%%%%%%%%%%%%%%%%%%%%%%%%%%%%%%%%%%%%%%%%%%%%%%%%%%%%%%%%%%%%%%%%%
\section{Summary}
%%%%%%%%%%%%%%%%%%%%%%%%%%%%%%%%%%%%%%%%%%%%%%%%%%%%%%%%%%%%%%%%%%%%%%%%%%%%%%%

We have discussed the autocorrelation function for eigenstates of 
quantum mechanical systems, and its relation  
to the behavior of the classical system. 
For billiards we have derived a formula 
for the autocorrelation function of an eigenfunction 
in terms of the normal derivative on the boundary 
\eqref{eq:autocorrel-via-boundary-integral}, which
enables an efficient numerical computation. 

Our main result  is the correlation length expansion 
of the autocorrelation function \eqref{eq:auto_corr_expansion}, 
which provides an efficient expansion for small 
correlation length, where only a small number of terms enter.
Moreover, it provides a tool for understanding the 
behavior of the autocorrelation function for different types of
eigenfunctions in terms of their semiclassical limit. 

The coefficients in the correlation length expansion 
\eqref{eq:auto_corr_expansion} can be computed in terms of the 
radially integrated momentum density.
Even though it is based on an approximation,  
our numerical study shows very good agreement with the corresponding
exact results; only for large correlation lengths
deviations become visible.
As the expansion coefficients have to be determined
just once for a given eigenfunction, this is also a numerically
efficient method to compute the autocorrelation function.
Similar expansions can be derived in higher dimension
and for more general systems (e.g.\ systems with potential and
magnetic field), but then the Bessel functions
have to be modified in order to reflect the structure of the energy shell 
of the classical system.

We applied the expansion of the autocorrelation function
to different types of eigenfunctions, 
and showed that it provides a good tool for 
the understanding of their autocorrelation functions. 
In systems with mixed phase space  regular states concentrated 
on tori and irregular states have been successfully treated. 
For chaotic system the fluctuations of 
the autocorrelation functions around the leading term 
are shown to be connected with the rate of quantum ergodicity. 
Moreover by varying the correlation length the autocorrelation function 
is shown to be an interesting new tool to measure the 
rate of quantum ergodicity on different length scales.

\vspace{1cm}

{\bf Acknowledgements}

\vspace{0.25cm}

We would like to thank Professor Jonathan Keating for useful comments
on the \mbox{manuscript}. 
A.B.\ acknowledges support by the 
Deutsche Forschungsgemeinschaft under contract No. DFG-Ba 1973/1-1.
R.S.\ acknowledges support by the 
Deutsche Forschungsgemeinschaft under contract No. DFG-Ste 241/7-3.

\section*{Appendix}
\appendix
%%%%%%%%%%%%%%%%%%%%%%%%%%%%%%%%%%%%%%%%%%%%%%%%%%%%%%%%%%%%%%%%%%%%%%%%%%%%%%%
\section{Autocorrelation function in terms of normal derivatives on 
the boundary} \label{sec:appendix-autocorrel-via-u_n}
%%%%%%%%%%%%%%%%%%%%%%%%%%%%%%%%%%%%%%%%%%%%%%%%%%%%%%%%%%%%%%%%%%%%%%%%%%%%%%%

We will give a derivation of the formula 
\eqref{eq:autocorrel-via-boundary-integral}
which provides an expression of the autocorrelation function $C(\dort)$
in terms of the normal derivative.
Let $\psi (\BFq)$ be a solution of the Helmholtz equation with 
Dirichlet boundary condition on $\partial\Omega$, 
\begin{equation}
(\Delta +k^2 )\psi (\BFq)=0\; ,\quad\psi (\BFq)=0\quad\text{for}\;  
\BFq\in\partial\Omega\;\; ,
\end{equation} 
where we have defined $k=\sqrt{E}$, and let 
\begin{equation}
u(s):=\BFn (s)\nabla \psi (\BFq(s))
\end{equation}
be the outer normal derivative of $\psi$ on $\partial\Omega$, where $s$ 
parameterizes $\partial\Omega$ in arclength.  
It is well known that 
\begin{equation}
-\frac{1}{4}\Int_{\partial\Omega} Y_0(k|\BFq-\BFq(s)|) u(s)\; \ud s
=\begin{cases} \psi(\BFq)   &\text{for}\; \BFq\in \overset{\circ}{\Omega} \\
                  0         &\text{for}\; \BFq\notin \Omega
\end{cases}
\end{equation}
and furthermore 
\begin{equation}\label{eq:J_vanish}
\Int_{\partial\Omega} J_0(k|\BFq-\BFq(s)|) u(s)\; \ud s =0\;\; .
\end{equation}
Let $\rho(t)$ be a smooth cut-off function with 
\begin{equation}\label{eq:cutoff}
\rho(kt)=\begin{cases} 1 &\text{for}\quad t\leq 2\diam(\Omega) \\
                      0 &\text{for}\quad t\geq 3\diam( \Omega)  
        \end{cases} \;\;,
\end{equation}
where  $\diam( \Omega)$ denotes the diameter of $\Omega$.  
Then we have for $\BFq$ in some
neighborhood of $\Omega$
\begin{equation}
\psi(\BFq) =
-\frac{1}{4}\Int_{\partial\Omega} \rho (k|\BFq-\BFq(s)|)
Y_0(k|\BFq-\BFq(s)|) u(s)\; \ud s 
\end{equation}
and obtain 
\begin{equation}
C(\dort)=\Int_{\R^2}\psi^*(\BFq)\psi(\BFq+\dort)\; \ud q
=\IInt_{\partial\Omega\times\partial\Omega } K_{\rho}(\dort,s,s') u^*(s)u(s')
\; \ud s\,\ud s'\;\; ,
\end{equation}
with 
\begin{equation}\label{eq:Krho}
\begin{split}
K_{\rho}(\dort,s,s')
&=\frac{1}{16}\Int_{\R^2}\rho (k|\BFq-\BFq(s)|)
Y_0(k|\BFq-\BFq(s)|)Y_0(k|\BFq-\BFq(s')+\dort|)\; \ud q\\
&=\frac{1}{16}\Int_{\R^2}\rho (k|\BFq|)
Y_0(k|\BFq|)Y_0(k|\BFq+\BFq(s)-\BFq(s')+\dort|)\; \ud q\;\; .
\end{split}
\end{equation}
Due to the factor $\rho (k|\BFq-\BFq(s)|)$ this integral is absolutely 
convergent. We now use Grafs addition theorem \cite{AbrSte84}
\begin{equation} \label{eq:Grafs-addition}
Y_0(k|\BFq+\Delta\BFq|) =
\begin{cases}   \sum_{l\in\Z} Y_l(k|\Delta\BFq|) J_l(k|\BFq|) \cos (l\varphi) 
&\text{for}\; |\BFq|<|\Delta\BFq| \\
\sum_{l\in\Z} Y_l(k|\BFq|) J_l(k|\Delta\BFq|) \cos (l\varphi) 
&\text{for}\; |\BFq|>|\Delta\BFq| \;\;,
\end{cases} 
\end{equation}
where $\Delta\BFq=\BFq(s)-\BFq(s')+\dort$ and $\varphi$ is the angle 
between $\Delta\BFq$ and $\BFq$. Introducing polar coordinates in
the integral in \eqref{eq:Krho}  and using \eqref{eq:Grafs-addition}
  %Grafs addition theorem 
gives 
\begin{equation}
\begin{split}
K_{\rho}(\dort,s,s')  = &
\frac{\pi}{8}\Int_0^{|\Delta\BFq|}Y_0(kr)J_0(kr)r\; 
     \ud r \, Y_0(k|\Delta\BFq|)\\
&+\frac{\pi}{8}\Int_{|\Delta\BFq|}^{\infty}
     \rho (kr)Y_0(kr)Y_0(kr)r\; \ud r\,  J_0(k|\Delta\BFq|) \;\;,
\end{split}
\end{equation}
where we have furthermore used that $\rho (kr)=1$ for $r\leq |\Delta\BFq|$ 
by \eqref{eq:cutoff}.  
The first integral is 
\begin{equation}
  \Int_0^{|\Delta\BFq|}Y_0(kr)J_0(kr)r\; \ud r
    =\frac{|\Delta\BFq|^2}{2}
      \bigl[Y_0(k|\Delta\BFq|)J_0(k|\Delta\BFq|)
           +Y_1(k|\Delta\BFq|)J_1(k|\Delta\BFq|)\bigr]\,\, ,
\end{equation}
see e.g. \cite{AbrSte84}, and for the second one partial integration  gives 
\begin{equation}\label{eq:part-int}
\begin{split}
\Int_{|\Delta\BFq|}^{\infty}\rho (kr)Y_0(kr)Y_0(kr)r\; \ud r
  &=-\frac{|\Delta\BFq|^2}{2} \bigl[Y_0(k|\Delta\BFq|)Y_0(k|\Delta\BFq|)
  +Y_1(k|\Delta\BFq|)Y_1(k|\Delta\BFq|)\bigr]  \\
 &\qquad 
 -\frac{k}{2}
  \Int_{|\Delta\BFq|}^{\infty}\rho' (kr) 
    \bigl[Y_0(kr)Y_0(kr)+Y_1(kr)Y_1(kr)\bigr] r^2\; \ud r
  \;\;.
\end{split}
\end{equation}
Note that since $\rho' $ has compact support the second integral is 
over a finite interval, and for $s,s'\in\partial\Omega$, $\dort\in \Omega$ 
 the lower limit of the integral, 
 $|\Delta\BFq|$,  is outside the support of $\rho' $, hence the 
second term on the right hand side of  eq.~\eqref{eq:part-int} is constant. 
So we get 
\begin{equation}
K_{\rho}(\dort,s,s')=K(\dort,s,s')+R_{\rho}(\dort,s,s')
\end{equation}
with 
\begin{equation}
K(\dort,s,s')=\frac{\pi |\Delta\BFq|^2}{16}
  \bigl[Y_1(k|\Delta\BFq|)J_1(k|\Delta\BFq|) Y_0(k|\Delta\BFq|)
       -Y_1(k|\Delta\BFq|)Y_1(k|\Delta\BFq|)J_0(k|\Delta\BFq|) \bigr]
\end{equation}
and 
\begin{equation}
R_{\rho}(\dort,s,s')=C J_0(k|\Delta\BFq|)
\end{equation}
with $C$ constant and by \eqref{eq:J_vanish} this term gives no 
contribution to 
$C(\dort)$. Using a Wronsky determinant of Bessel functions \cite{AbrSte84} 
we can simplify 
$K(\dort,s,s')$ further 
\begin{equation}
\begin{split}
 K(\dort,s,s')&=\frac{\pi |\Delta\BFq|^2}{16} 
       Y_1(k|\Delta\BFq|) \bigl[J_1(k|\Delta\BFq|)Y_0(k|\Delta\BFq|)
                               -Y_1(k|\Delta\BFq|)J_0(k|\Delta\BFq|)\bigr]\\
  &=\frac{\pi |\Delta\BFq|^2}{16} Y_1(k|\Delta\BFq|)\frac{2}{\pi k|\Delta\BFq|}
= \frac{ |\Delta\BFq|}{8k} Y_1(k|\Delta\BFq|)
\end{split}
\end{equation}
which gives the final result.  

%%%%%%%%%%%%%%%%%%%%%%%%%%%%%%%%%%%%%%%%%%%%%%%%%%%%%%%%%%%%%%%%%%%%%%%%%%%%%%%
\section{Remainder estimate} \label{sec:appendix:estimate}
%%%%%%%%%%%%%%%%%%%%%%%%%%%%%%%%%%%%%%%%%%%%%%%%%%%%%%%%%%%%%%%%%%%%%%%%%%%%%%%
 
In this appendix we sketch the derivation of the remainder estimate in 
equation \eqref{eq:approx_on_energy}. We start by representing
the integral as an expectation value, see \eqref{eq:exp-value-repr}, 
\begin{equation}
\Int_0^{\infty}\Int_0^{2\pi}\Int_{\Omega} 
    \rho(\ort-\BFq)W(\BFp,\BFq)\; \ud q'
\, \ue^{\ui r |\dort|\cos(\varphi-\theta)/\sqrt{E}}\, r\;
\ud \varphi \,\ud r
=\langle \psi, A\psi \rangle\;\; 
\end{equation}
where $A$ is the Weyl quantization of the symbol 
\begin{equation}
a(\BFp,\BFq):=
\rho(\ort-\BFq)\ue^{\ui \frac{|\BFp|}{\sqrt{E}}|\dort|\cos(\varphi-\theta)}\;\; .
\end{equation}
The basic idea is to find a decomposition of the operator $A$, 
\begin{equation}\label{eq:quant-Taylor}
A=A_0+(\sqrt{-\Delta}-\sqrt{E})A_1 +R
\end{equation}
where $A_0$ has Weyl symbol 
\begin{equation}
a_0(\BFp,\BFq)=\rho(\ort-\BFq)\ue^{\ui |\dort|\cos(\varphi-\theta)}
\end{equation}
and the remainder $R$ satisfies 
\begin{equation}\label{eq:est:R}
||R||\leq C\, E^{-1/2}\;\; .
\end{equation}
%%
%%
%Supposing  we have 
If we assume the decomposition \eqref{eq:quant-Taylor} and take
the expectation value of both sides, one gets
\begin{equation} \label{eq:desired-result}
\langle \psi, A\psi \rangle
     =\langle \psi, A_0\psi \rangle+\langle \psi, R\psi \rangle \;\;,
\end{equation}
where $(\sqrt{-\Delta}-\sqrt{E})\psi=0$  has been used.
In terms of the symbols eq.~\eqref{eq:desired-result}
is the desired result, see \eqref{eq:approx_on_energy}, 
\begin{equation}
\begin{split}
\Int_0^{\infty}\Int_0^{2\pi}\Int_{\Omega} 
    &W(\BFp,\BFq)\rho(\ort-\BFq)\; \ud q'
\, \ue^{\ui r |\dort|\cos(\varphi-\theta)/\sqrt{E}}\, r\;
\ud \varphi \,\ud r \\
&=
\Int_0^{\infty}\Int_0^{2\pi}\Int_{\Omega} 
    W(\BFp,\BFq) \rho(\ort-\BFq)\; \ud q'
\, \ue^{\ui  |\dort|\cos(\varphi-\theta)/}\, r\;
\ud \varphi \,\ud r +O(E^{-1/2}) \;\;.
\end{split}
\end{equation}

Let us now show, that 
the decomposition \eqref{eq:quant-Taylor} is basically a quantization of 
the Taylor expansion of the symbol $a(\BFp,\BFq)$ around $|\BFp|=\sqrt{E}$, 
\begin{equation}
a(\BFp,\BFq)=a_0(\BFp,\BFq)+\bigl(|\BFp|-\sqrt{E}\bigr) a_1(\BFp,\BFq)\;\; .
\end{equation}
Quantizing this classical decomposition yields \eqref{eq:quant-Taylor} with 
$R$ given as Weyl quantization of 
\begin{equation}
r(\BFp,\BFq)=\bigl(|\BFp|-\sqrt{E}\bigr) a_1(\BFp,\BFq)-\bigl(|\BFp|-\sqrt{E}\bigr) 
    \# a_1(\BFp,\BFq)
\end{equation}
since the Weyl symbol of $(\sqrt{-\Delta}-\sqrt{E})A_1$ is 
$(|\BFp|-\sqrt{E}) \# a_1(\BFp,\BFq)$ with $\#$ denoting the 
symbol product, see e.g.\ \cite{Fol89}. 
Since $E$ is a constant we have 
\begin{equation}
r(\BFp,\BFq)=|\BFp|a_1(\BFp,\BFq)-|\BFp|\# a_1(\BFp,\BFq)\;\; ,
\end{equation}
and this is a function which is bounded and of order $O(E^{-1/2})$, 
and all its derivatives are bounded and of order $O(E^{-1/2})$, too. 
So by the Calderon Vallaincourt theorem \cite{Fol89} the estimate 
\eqref{eq:est:R} follows. 

%%%%%%%%%%%%%%%%%%%%%%%%%%%%%%%%%%%%%%%%%%%%%%%%%%%%%%%%%%%%%%%%%%%%%%%%%%%%%%%
\section{Estimating the Bessel sum} \label{app:Bessel}
%%%%%%%%%%%%%%%%%%%%%%%%%%%%%%%%%%%%%%%%%%%%%%%%%%%%%%%%%%%%%%%%%%%%%%%%%%%%%%%

In this appendix we determine how many terms in the sum 
\eqref{eq:auto_corr_expansion} have to be taken into account
such that the remainder is smaller than some given error $\delta$.
From \eqref{eq:QMexpA} and \eqref{eq:QMexpB} it follows
that for fixed $\psi$
\begin{equation}
|\langle\psi,\widehat{A}_{2l}(\ort)\psi\rangle \cos(2l \theta)+
\langle\psi,\widehat{B}_{2l}(\ort)\psi\rangle \sin(2l \theta)|\leq C \;\;.
\end{equation}
Thus if we  split the sum 
\begin{equation}
\begin{split}
\sum_{l=1}^{\infty} &  (-1)^l
[\langle\psi,\widehat{A}_{2l}(\ort)\psi\rangle \cos(2l \theta)+
\langle\psi,\widehat{B}_{2l}(\ort)\psi\rangle \sin(2l \theta)]\, J_{2l}(|\dort|)\\
&=\sum_{l=1}^{m-1} (-1)^l 
[\langle\psi,\widehat{A}_{2l}(\ort)\psi\rangle \cos(2l \theta)+
\langle\psi,\widehat{B}_{2l}(\ort)\psi\rangle \sin(2l \theta)]\, J_{2l}(|\dort|)+R_m(|\dort|)   \;\; ,
\end{split}
\end{equation}
we get for the  remainder
\begin{equation}
|R_m(r)|\leq C \sum_{l=m}^{\infty} |J_{2l}(r)|\;\; .
\end{equation}

Therefore we have to estimate the sum over Bessel functions
\begin{equation} \label{eq:sum-over-Bessels}
\sum_{l=m}^{\infty}|J_{2l}(r)|\,\, ,
\end{equation}
and determine its dependence on $m$ and $r$. 
The asymptotics in the transition region 
\begin{equation}
J_{2l}(2l-z(2l)^{1/3})\sim \frac{1}{l^{1/3}} \Ai(2^{1/3}z)\,\, ,
\end{equation}
gives that $J_{2l}(r)$ is monotonically increasing for 
$r<2l$, such that for $r<2m$
\begin{equation}
\sum_{l=m}^{\infty}|J_{2l}(r)|=
\sum_{l=m}^{\infty}\frac{1}{l^{1/3}} \Ai\bigg(\frac{2l-r}{l^{1/3}}\bigg)
+O(m^{-1})\,\, .
\end{equation}
Defining $z$ by 
\begin{equation}\label{eq:rzm-relation}
r=2m-zm^{1/3}\,\, ,
\end{equation}
we obtain 
\begin{equation}
\begin{split}
\sum_{l=m}^{\infty}\frac{1}{l^{1/3}} \Ai\bigg(\frac{2l-r}{l^{1/3}}\bigg)
&=\sum_{l=m}^{\infty}\frac{1}{l^{1/3}} \Ai\bigg(\frac{2(l-m)}{l^{1/3}}
+z\bigg(\frac{m}{l}\bigg)^{1/3}\bigg)\\
&=\sum_{l=0}^{\infty}\frac{1}{(l+m)^{1/3}} \Ai\bigg(\frac{2l}{(l+m)^{1/3}}
+z\bigg(\frac{m}{l+m}\bigg)^{1/3}\bigg)\\
&=\sum_{l=0}^{\infty}\frac{1}{m^{1/3}} \Ai\bigg(\frac{2l}{m^{1/3}}
+z\bigg)+O(m^{-1/3})
\end{split}
\end{equation}
where we have furthermore used that for large $m$ only the terms 
with $l \ll m$ contribute, because the Airy function is exponentially
decreasing for positive arguments. The Euler McLaurin formula 
then gives 
\begin{equation}
\begin{split}
\sum_{l=0}^{\infty}\frac{1}{m^{1/3}} \Ai\bigg(\frac{2l}{m^{1/3}}
+z\bigg)&=\Int_0^{\infty}\frac{1}{m^{1/3}}\Ai\bigg(\frac{2l}{m^{1/3}}
+z\bigg)\,\, \ud l+O(m^{-1/3})\\
&=\frac{1}{2}\Int_z^{\infty}\Ai(x)\,\, \ud x+O(m^{-1/3})\;\; ,
\end{split}
\end{equation}
\BILD{t}
     {
     \begin{center}
       \PSImagx{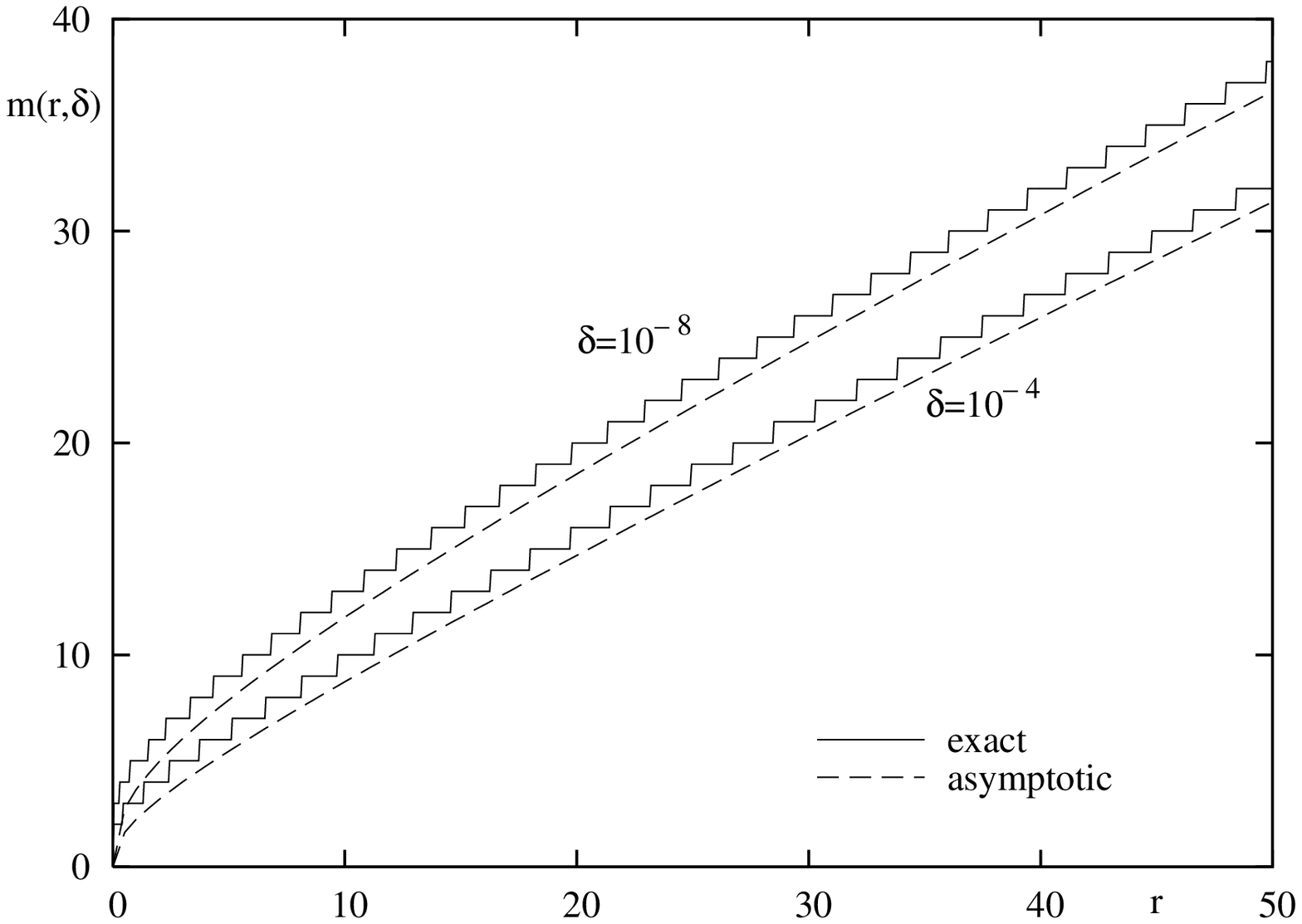}{12cm}
     \end{center}
     }
     {For the bounds $\delta=10^{-4}$ 
      and $\delta=10^{-8}$ of the sum over Bessel functions 
      \eqref{eq:sum-over-Bessels}
      the result of the exact computation of $m(r,\delta)$ 
      (full curves) and the asymptotic 
      result \eqref{eq:asympt-m} are compared.
      The asymptotic results approaches the exact one slowly from below with 
      a rate $O(r^{-1/3})$.
     }
     {fig:remainder}
And so finally  we arrive at 
\begin{equation}
\sum_{l=m}^{\infty}|J_{2l}(r)|=\frac{1}{2}\Int_z^{\infty}\Ai(x)\; \ud x+O(m^{-1/3})\;\; .
\end{equation}
The function $\int_{z}^{\infty} \Ai(x)\,\, \ud x$ is monotonically decreasing, 
so for a given $\delta>0$ we can define a $z(\delta)$ by 
\begin{equation}\label{eq:Airy-delta-z}
\frac{1}{2}\Int_{z(\delta)}^{\infty} \Ai(x)\,\, \ud x=\delta\;\; ,
\end{equation}
and then \eqref{eq:rzm-relation} defines together with 
\eqref{eq:Airy-delta-z} a function $m(r,\delta)$ such that 
\begin{equation}\label{eq:remainder-delta}
\sum_{l=[m(r,\delta)+1]}^{\infty} |J_{2l}(r)|
=\delta+O(r^{-1/3})\;\; .
\end{equation}
By solving \eqref{eq:rzm-relation} for large $r$, we see that 
we have to take approximately 
\begin{equation}\label{eq:asympt-m}
m(r,\delta)\sim \frac{r}{2}
+\frac{z}{2}\left(\frac{r}{2}\right)^{1/3}
\end{equation}
terms in the sum \eqref{eq:auto_corr_expansion} over $l$ into account such
that the error is $\delta +O(r^{-1/3})$. 

For instance, if we require $\delta=10^{-4}$, then \eqref{eq:Airy-delta-z} 
gives $z(\delta)=4.359\dots$;
for $\delta=10^{-8}$ one gets $z(\delta)=7.925\dots$.
In fig.~\ref{fig:remainder} we show for these choices of $z$ 
the asymptotic result \eqref{eq:asympt-m} compared to
the exact computation, corresponding to \eqref{eq:sum-over-Bessels}.
The asymptotic results approaches the exact one slowly from below;
in the plotted region a constant offset by two compared to 
\eqref{eq:asympt-m} gives a good bound for $m(r,\delta)$.


\begin{thebibliography}{10}\parskip0.75ex

\bibitem{Ber77b}
M.~V. Berry: {\em Regular and irregular semiclassical wavefunctions\/}, J.
  Phys. A {\bf 10} (1977) ~2083--2091.

\bibitem{McDKau88}
S.~W. McDonald and A.~N. Kaufmann: {\em Wave chaos in the stadium: Statistical
  properties of short-wave solutions of the Helmholtz equation\/}, Phys. Rev. A
  {\bf 37} (1988) ~3067--3086.

\bibitem{AurSte93}
R.~Aurich and F.~Steiner: {\em Statistical properties of highly excited quantum
  eigenstates of a strongly chaotic system\/}, Physica D {\bf 64} (1993)
  ~185--214.

\bibitem{LiRob94}
B.~Li and M.~Robnik: {\em Statistical properties of high-lying chaotic
  eigenstates\/}, J. Phys. A {\bf 27} (1994) ~5509--5523.

\bibitem{Rob98}
D.~Robert: {\em Semi-classical approximation in quantum mechanics. {A} survey
  of old and recent mathematical results\/}, Helv. Phys. Acta {\bf 71} (1998) 1
  ~44--116.

\bibitem{Shn74}
A.~I. Shnirelman: {\em Ergodic properties of eigenfunctions {\rm (in
  Russian)}\/}, Usp. Math. Nauk {\bf 29} (1974) ~181--182.

\bibitem{Zel87}
S.~Zelditch: {\em Uniform distribution of eigenfunctions on compact hyperbolic
  surfaces\/}, Duke. Math. J. {\bf 55} (1987) ~919--941.

\bibitem{CdV85}
Y.~{Colin de Verdi\`ere}: {\em Ergodicit\'e et fonctions propres du laplacien
  {\rm (in French)}\/}, Commun. Math. Phys. {\bf 102} (1985) ~497--502.

\bibitem{HelMarRob87}
B.~Helffer, A.~Martinez and D.~Robert: {\em Ergodicit\'e et limite
  semi-classique {\rm (in French)}\/}, Commun. Math. Phys. {\bf 109} (1987)
  ~313--326.

\bibitem{GerLei93}
P.~G\'erard and E.~Leichtnam: {\em Ergodic properties of eigenfunctions for the
  Dirichlet problem\/}, Duke Math. J. {\bf 71} (1993) ~559--607.

\bibitem{ZelZwo96}
S.~Zelditch and M.~Zworski: {\em Ergodicity of eigenfunctions for ergodic
  billiards\/}, Commun. Math. Phys. {\bf 175} (1996) ~673--682.

\bibitem{BaeSchSti98}
A.~B\"acker, R.~Schubert and P.~Stifter: {\em Rate of quantum ergodicity in
  Euclidean billiards\/}, Phys. Rev. E {\bf 57} (1998) ~5425--5447, erratum
  ibid. {\bf 58} (1998) 5192.

\bibitem{Vor76}
A.~Voros: {\em Semi--classical approximations\/}, Annales de l'Institut Henri
  Poincar\'e A {\bf 24} (1976) ~31--90.

\bibitem{Vor77}
A.~Voros: {\em Asymptotic $\hbar$--expansions of stationary quantum states\/},
  Annales de l'Institut Henri Poincar\'e A {\bf 26} (1977) ~343--403.

\bibitem{Ber83}
M.~V. Berry: {\em Semiclassical mechanics of regular and irregular motion\/},
  in: {\em Comportement Chaotique des Syst{\`e}mes D{\'e}terministes ---
  Chaotic Behaviour of Deterministic Systems\/} (Eds. G.~Iooss, R.~H.~G.
  Hellemann and R.~Stora),  171--271, {N}orth-{H}olland, {A}msterdam,  (1983).

\bibitem{EckDoeKuhStoe99}
B.~Eckhardt, U.~D\"orr, U.~Kuhl and H.-J. St\"ockmann: {\em Correlations of
  electromagnetic fields in chaotic cavities\/}, Europhys. Lett. {\bf 46}
  (1999) ~134--140.

\bibitem{SreSti96}
M.~Srednicki and F.~Stiernelof: {\em Gaussian fluctuations in chaotic
  eigenstates\/}, J. Phys. A {\bf 29} (1996) ~5817--5826.

\bibitem{Sre96b}
M.~Srednicki: {\em Gaussian random eigenfunctions and spatial correlations in
  quantum dots\/}, Phys. Rev. E {\bf 54} (1996) ~954--955.

\bibitem{HorSre98a}
S.~Hortikar and M.~Srednicki: {\em Correlations in chaotic eigenfunctions at
  large separation\/}, Phys. Rev. Lett. {\bf 80} (1998) ~1646--1649.

\bibitem{HorSre98}
S.~Hortikar and M.~Srednicki: {\em Random matrix elements and eigenfunctions in
  chaotic systems\/}, Phys. Rev. E {\bf 57} (1998) ~7313--7316.

\bibitem{ShaGoe84}
M.~Shapiro and G.~Goelman: {\em Onset of chaos in an isolated energy
  eigenstate\/}, Phys. Rev. Lett. {\bf 53} (1984) ~1714--1717.

\bibitem{ShaRonBru88}
M.~Shapiro, J.~Ronkin and P.~Brumer: {\em Scaling laws and correlation length
  of quantum and classical ergodic states\/}, Chem. Phys. Lett. {\bf 148}
  (1988) 2,3 ~177--182.

\bibitem{VebRobLiu99}
G.~Veble, M.~Robnik and J.~Liu: {\em Study of regular and irregular states in
  generic systems\/}, J. Phys. A {\bf 32} (1999) ~6423--6444.

\bibitem{Sch2001:PhD}
R.~Schubert: {\em Semiclassical localization in phase space\/}, Ph.D. thesis,
  Abteilung Theoretische Physik, Universit\"at Ulm,  (2001).

\bibitem{Bun74}
L.~A. Bunimovich: {\em On ergodic properties of certain billiards\/}, Funct.
  Anal. Appl. {\bf 8} (1974) ~254--255.

\bibitem{Bun79}
L.~A. Bunimovich: {\em On the ergodic properties of nowhere dispersing
  billiards\/}, Commun. Math. Phys. {\bf 65} (1979) ~295--312.

\bibitem{Rob83}
M.~Robnik: {\em Classical dynamics of a family of billiards with analytic
  boundaries\/}, J. Phys. A {\bf 16} (1983) ~3971--3986.

\bibitem{Rob84}
M.~Robnik: {\em Quantising a generic family of billiards with analytic
  boundaries\/}, J. Phys. A {\bf 17} (1984) ~1049--1074.

\bibitem{Woj86}
M.~Wojtkowski: {\em Principles for the design of billiards with nonvanishing
  Lyapunov exponents\/}, Commun. Math. Phys. {\bf 105} (1986) ~391--414.

\bibitem{Sza92}
D.~Sz\'asz: {\em On the K-property of some planar hyperbolic billiards\/},
  Commun. Math. Phys. {\bf 145} (1992) ~595--604.

\bibitem{Mar93}
R.~Markarian: {\em New ergodic billiards: exact results\/}, Nonlinearity {\bf
  6} (1993) ~819--841.

\bibitem{PrivComProRob}
T.~Prosen and M.~Robnik:  private communication.

\bibitem{ProRob93a}
T.~Prosen and M.~Robnik: {\em Energy level statistics in the transition region
  between integrability and chaos\/}, J. Phys. A {\bf 26} (1993) ~2371--2387.

\bibitem{Rid79}
R.~J. {Riddel Jr.}: {\em Boundary-distribution solution of the Helmholtz
  equation for a region with corners\/}, J. Comp. Phys. {\bf 31} (1979)
  ~21--41.

\bibitem{BerWil84}
M.~V. Berry and M.~Wilkinson: {\em Diabolical points in the spectra of
  triangles\/}, Proc. R. Soc. London Ser. A {\bf 392} (1984) ~15--43.

\bibitem{Bae98:PhD}
A.~B\"acker: {\em Classical and Quantum Chaos in Billiards\/}, Ph.D. thesis,
  Abteilung Theo\-re\-ti\-sche Physik, Universit\"at Ulm,  (1998).

\bibitem{VerSar95}
E.~Vergini and M.~Saraceno: {\em Calculation of highly excited states of
  billiards\/}, Phys. Rev. E {\bf 52} (1995) ~2204--2207.

\bibitem{Hel84}
E.~J. Heller: {\em Bound-state eigenfunctions of classically chaotic
  Hamiltonian systems: Scars of periodic orbits\/}, Phys. Rev. Lett. {\bf 53}
  (1984) ~1515--1518.

\bibitem{BaiHosSteTay85}
Y.~Y. Bai, G.~Hose, K.~Stefa\'nski and H.~S. Taylor: {\em Born-Oppenheimer
  adiabatic mechanism for regularity of states in the quantum stadium
  billiard\/}, Phys. Rev. A {\bf 31} (1985) ~2821--2826.

\bibitem{ConHel88}
P.~W. O'Connor and E.~J. Heller: {\em Quantum localization for a strongly
  classically chaotic system\/}, Phys. Rev. Lett. {\bf 61} (1988) ~2288--2291.

\bibitem{Tan97}
G.~Tanner: {\em How chaotic is the stadium billiard? A semiclassical
  analysis\/}, J. Phys. A {\bf 30} (1997) ~2863--2888.

\bibitem{BaeSchSti97a}
A.~B\"acker, R.~Schubert and P.~Stifter: {\em On the number of bouncing-ball
  modes in billiards\/}, J. Phys. A {\bf 30} (1997) ~6783--6795.

\bibitem{AbrSte84}
M.~Abramowitz and {I. A. Stegun (eds.)}: {\em Pocketbook of Mathematical
  Functions\/}, Verlag Harri Deutsch, Thun -- Frankfurt/Main, abridged edn.,
  (1984).

\bibitem{Fol89}
G.~B. Folland: {\em Harmonic Analysis in Phase Space\/}, vol. 122 of {\em
  Annals of Mathematics Studies\/}, Princeton University Press, Princeton,
  (1989).

\bibitem{Zyc92}
K.~\.Zyczkowski: {\em Classical and quantum billiards, integrable,
  nonintegrable, and pseudo-integrable\/}, Acta Physica Polonica B {\bf 23}
  (1992) ~245--270.

\bibitem{BaeSch99}
A.~B\"acker and R.~Schubert: {\em Chaotic eigenfunctions in momentum space\/},
  J. Phys. A {\bf 32} (1999) ~4795--4815.

\bibitem{TuaVor95}
J.~M. Tualle and A.~Voros: {\em Normal modes of billiards portrayed in the
  stellar (or nodal) representation\/}, Chaos, Solitons and Fractals {\bf 5}
  (1995) ~1085--1102.

\bibitem{SimVerSar97}
F.~P. Simonotti, E.~Vergini and M.~Saraceno: {\em Quantitative study of scars
  in the boundary section of the stadium billiard\/}, Phys. Rev. E {\bf 56}
  (1997) ~3859--3867.

\bibitem{BaeSch2001b:p}
A.~B\"acker and R.~Schubert: {\em Amplitude distribution of eigenfunctions in
  mixed systems\/}, preprint, 13 pages  (2001).

\bibitem{Per73}
I.~C. Percival: {\em Regular and irregular spectra\/}, J. Phys. B {\bf 6}
  (1973) ~L229--L232.

\bibitem{BohTomUll90a}
O.~Bohigas, S.~Tomsovic and D.~Ullmo: {\em Dynamical quasidegeneracies and
  separation of regular and irregular quantum levels\/}, Phys. Rev. Lett. {\bf
  64} (1990) ~1479--1482.

\bibitem{ProRob93b}
T.~Prosen and M.~Robnik: {\em Survey of the eigenfunctions of a billiard system
  between integrability and chaos\/}, J. Phys. A {\bf 26} (1993) ~5365--5373.

\bibitem{LiRob95b}
B.~Li and M.~Robnik: {\em Separating the regular and irregular energy levels
  and their statistics in a Hamiltonian system with mixed classical
  dynamics\/}, J. Phys. A {\bf 28} (1995) ~4843--4857.

\bibitem{LiRob95}
B.~Li and M.~Robnik: {\em Geometry of high-lying eigenfunctions in a plane
  billiard system having mixed-type classical dynamics\/}, J. Phys. A {\bf 28}
  (1995) ~2799--2818.

\bibitem{CarVerFen98}
G.~Carlo, E.~Vergini and A.~Fendrik: {\em Numerical verification of Percival's
  conjecture in a quantum billiard\/}, Phys. Rev. E {\bf 57} (1998)
  ~5397--5403.

\bibitem{EckFisKeaAgaMaiMue95}
B.~Eckhardt, S.~Fishman, J.~Keating, O.~Agam, J.~Main and K.~M\"uller: {\em
  Approach to ergodicity in quantum wave functions\/}, Phys. Rev. E {\bf 52}
  (1995) ~5893--5903.

\end{thebibliography}
\end{document}